\documentclass[12pt,a4paper]{article}
\usepackage{amsmath,geometry}
\usepackage{amsmath}
\usepackage{amsfonts}
\usepackage{amssymb}
\usepackage{verbatim}
\usepackage{graphicx}
\usepackage{multirow}
\usepackage{tablefootnote}
\usepackage{empheq}
\usepackage[all]{xy} 
\usepackage{color}
\usepackage{hyperref}
\usepackage{indentfirst}
\usepackage{cite}
\usepackage{array}

\usepackage[ruled,vlined]{algorithm2e}
\SetKw{Continue}{continue}
\SetKw{Break}{break}
\SetKw{And}{and}
\newcommand{\nc}{\newcommand}
\nc{\beq}{\begin{equation}}
\nc{\eeq}{\end{equation}}
\nc{\bea}{\begin{eqnarray}}
\newcommand{\bal}{\begin{aligned}}   \newcommand{\eal}{\end{aligned}}
\nc{\eea}{\end{eqnarray}}

\def\cO{{\cal O}}
\def\IP{\mathbb{P}}
\def\cO{\mathcal{O}}

\def\cI{\mathcal{I}}

\def\ra{\rightarrow}
\def\fto{\longrightarrow}

\def\clap#1{\hbox to 0pt{\hss#1\hss}}

\newcolumntype{P}[1]{>{\centering\arraybackslash}p{#1}}

\newdimen\csize\csize=1.5ex
\def\young#1{\tiny\vcenter{\hbox{\vrule\vtop{\hrule
  \offinterlineskip\halign{&\vbox
  {\hbox to\csize {\strut\hss##\hss\vrule}\hrule}\cr#1 \crcr}}}}}

\newcommand{\eq}[1]{\begin{equation}
                     \begin{split} #1 \end{split}
                     \end{equation}}

\def\fnote#1#2{\begingroup\def\thefootnote{#1}\footnote{#2}
     \addtocounter{footnote}{-1}\endgroup}
     
     \def\IP{\mathbb{P}}

\def\IZ{\mathbb{Z}}

\def\cA{\mathcal{A}}

\def\cI{\mathcal{I}}

\def\cN{\mathcal{N}}
\def\cO{\mathcal{O}}

\def\dburl{\url{www.rossealtman.com/tcy}}

\begin{document}

\vspace*{-1.5cm}
\begin{flushright}
  {\small
 % MPP-2013-...\\
  }
\end{flushright}

\vspace{1.5cm}
\begin{center}
%  {\LARGE    Divisor Involutions and Free Actions \\\vskip0.4cm in   Toric Orientifold Calabi-Yau Threefolds}
{\LARGE  Orientifold Calabi-Yau Threefolds with  \\\vskip0.4cm  Divisor Involutions and  String Landscape}

\end{center}

\vspace{0.75cm}
\begin{center}
Ross Altman$^*$, ${}{}$  Jonathan Carifio$^*$, ${}{}$ Xin Gao$^{\dagger,\ddagger}$, ${}{}$ and Brent D.~Nelson$^*$
 
   \end{center}

\vspace{0.1cm}
\begin{center} {\small 
 {\small 
 $^{\dagger}${\it College of Physics, Sichuan University, Chengdu, 610065, China}\\
 {\it $^{*}$ Department of Physics, Northeastern University, Boston, MA 02115, USA}}\\
 $^{\ddagger}${\it Institute for Theoretical Physics, Heidelberg University,\\ Philosophenweg 19, 69120, Heidelberg, Germany}
 }\\

\fnote{}{
 \hskip-0.5cm knowbodynos@gmail.com, \,
 jon.carifio@gmail.com, \,
xingao@scu.edu.cn, \,
 b.nelson@neu.edu}
\end{center}

\vspace{0.2cm}

\vspace{1cm}

\begin{abstract}
\noindent

We establish an  orientifold Calabi-Yau threefold database for $h^{1,1}(X) \leq 6$ by considering  non-trivial $\mathbb{Z}_{2}$ divisor exchange involutions, using a toric  Calabi-Yau database (\dburl).  We first determine the topology for each individual  divisor (Hodge diamond), then identify and classify the proper involutions which are globally consistent across all disjoint phases of the K\"ahler cone for each unique  geometry.  Each of the proper involutions will result in an orientifold Calabi-Yau manifold. Then we clarify all possible fixed loci under the proper involution, thereby determining the locations of different types of $O$-planes. It is shown that under the proper involutions, one typically ends up with a system of $O3/O7$-planes, and most of these will further admit naive Type IIB string vacua.
%For those geometries with $O3/O7$-plane, we identify the naive string vacua by checkding the $D3$-tadpole cancelation condition.   
The geometries with freely acting involutions are also determined. We further determine the splitting of the Hodge numbers into odd/even parity in the orbifold limit. The final result is a class of orientifold Calabi-Yau threefolds with non-trivial odd class cohomology ($h^{1,1}_{-}(X/\sigma^*) \neq 0$). 
\end{abstract}

\clearpage

\tableofcontents

\pagebreak

%%%%%%%%%%%%%%%%%%%%%%%%%%%%%%%%%%%%%%%%%%%%%%%%%%%%%%%%
%%%%%%%%%%%%%%%%%%%%%%%%%%%%%%%%%%%%%%%%%%%%%%%%%%%%%%%%
%%%%%%%%%%%%%%%%%%%%%%%%%%%%%%%%%%%%%%%%%%%%%%%%%%%%%%%%
\section{Introduction}
\label{sec:intro}

String compactification is crucial for high dimensional string theory to describe the four dimensional real world.
%Consistent perturbative superstring theories are required to have an underlying ten dimensional spacetime, which must be compactified down to four dimensions in order to describe the real world. 
The methods for doing so are best understood for supersymmetric compactifications, where without turning on more general fluxes, the compactification manifold must be a Calabi-Yau threefold, $X$. 
Compactifying a Type~IIA or Type~IIB string theory on a Calabi-Yau threefold results in an $\cN = 2$ supersymmetric theory in four dimensions. In order to break the theory further to an $\cN = 1$ supersymmetric theory in four dimensions, one must perform an orientifold projection on one of the two gravitinos of the theory.  As a result, among the vast number of Calabi-Yau threefolds, those that permit such an orientifold under some proper $\mathbb{Z}_{2}$ involution, $\sigma$, are of great phenomenological interest.  In this paper, we extend and improve on previous work \cite{Gao:2013pra} classifying non-trivial $\mathbb{Z}_{2}$ divisor exchange involutions in  Calabi-Yau threefolds for $h^{1,1}(X)\leq 6$, using a database \cite{Altman:2014bfa} (\dburl) constructed from the Kreuzer-Skarke dataset of reflexive four-dimensional polyhedra \cite{Kreuzer:2000xy}.  For a general review on flux compactification on these orientifold Calabi-Yau threefolds and the landscape of string vacua, see \cite{Grana:2005jc, Jockers:2005pn, Douglas:2006es,
Lust:2006zg, Denef:2008wq, Blumenhagen:2008zz, Ibanez:2012zz, Hebecker:2020aqr}. 

In this work, we focus on Type IIB orientifold geometries where an orientifold projection $\cO$ is generally composed of two parts. One is the  worldsheet parity $\Omega_p$,  and  another one is a diffeomorphism  map  $\sigma$ acting on the internal manifold, i.e, the involution. The involution $\sigma$ is a non-trivial  $\IZ_2$ action on the Calabi-Yau space such that $\sigma^2=1$. In order for the orientifold action to still preserve some supersymmetry, the involution must be isometric and holomorphic \cite{Acharya:2002ag, Brunner:2003zm}. These conditions require that the pullback $\sigma^{*}$ of the involution must always map $(p,q)$-forms to $(p,q)$-forms on $X$. In particular, the K\"ahler $(1,1)$-form $J$ must be preserved and the unique holomorphic $(3,0)$-form $\Omega$ must be an eigenform of $\sigma^{*}$ with eigenvalues $\pm 1$.
%\bea\label{eq:KInvol}
%\sigma^{*}J=J\hspace{5mm}\text{and}\hspace{5mm}\sigma^{*}\Omega_{3}= \pm \, \Omega_{3}.
%\eea

 The geometry may contain some fixed loci under the involution $\sigma$, which will correspond to orientifold planes ($O$-planes).
 The precise structure of the orientifold projection $\cO$ in Type IIB is determined according to the different $O$-plane systems relevant in D-brane constructions:
\begin{eqnarray}
\label{eq:orientifold}
 {\cal O}= \begin{cases}
                       \Omega_p\, \sigma \qquad &{\rm with} \quad
                       \sigma^*(J)=J\,,\quad  \sigma^*(\Omega_3)=+\Omega_3,  \quad  $O5/O9$\,\, {\rm system} ,\\[0.1cm]
                       (-1)^{F_L}\,\Omega_p\, \sigma\qquad & {\rm with} \quad
        \sigma^*(J)=J\,, \quad \sigma^*(\Omega_3)=-\Omega_3, \quad    $O3/O7$\,\, {\rm system},
\end{cases}
\end{eqnarray}
where $F_L$ is the left-moving fermion number.

If there is no fixed locus and the orientifold is smooth, such an involution describes a freely-acting $\IZ_2$. This, however, is non-trivial since under the orientifold action, the Calabi-Yau threefold may acquire new singularities. In general, the involution $\sigma$ splits the cohomology groups $H^{p,q}(X/\sigma^{*})$ into eigenspaces of even and odd parity:
 \bea
 H^{p,q}(X/\sigma^{*})=H^{p,q}_{+}(X/\sigma^{*})\oplus H^{p,q}_{-}(X/\sigma^{*}).
 \eea

In particular, Calabi-Yau orientifolds with non-trivial odd equivariant cohomology class $H^{1,1}_{-}(X/\sigma^{*})$ play an important role in string phenomenology. In Type~IIB orientifold compactifications with $O3/O7$-planes, when $h^{1,1}_{-}(X/\sigma^{*})>0$, there are non-trivial involutively odd moduli $(b^{a}, c^{a})$, $a=1,\cdots, h^{1,1}_-(X/\sigma^{*})$, in the bosonic closed spectrum coming from the R-R and NS-NS two-form fields $C_{2}$ and $B_{2}$ respectively.  These odd moduli can be stabilized by D-term conditions~\cite{Jockers:2005zy} or  by F-term conditions~\cite{Gao:2013rra, Cicoli:2021tzt}.  In the Type IIB orientifold compactifications of both the KKLT~\cite{Kachru:2003aw} and Large Volume Scenario~\cite{Balasubramanian:2005zx},  these axion-like odd moduli  can drive inflation~\cite{Grimm:2007hs, McAllister:2008hb,   Flauger:2009ab, Hebecker:2011hk,  Arends:2014qca, Blumenhagen:2014gta, 
 Marchesano:2014mla, Hebecker:2014eua, %McAllister:2014mpa, 
 Ben-Dayan:2014zsa, Long:2014dta, Gao:2014uha,  Ben-Dayan:2014lca,  Shiu:2015xda, Escobar:2015ckf, Blumenhagen:2016bfp, Landete:2017amp, Blumenhagen:2017cxt,   Hebecker:2018yxs}  in the natural, monodromy, and aligned inflationary models, and play an important role in the recently discussed Weak Gravity Conjecture and Swampland Conjecture (first proposed in \cite{Vafa:2005ui, ArkaniHamed:2006dz} and see \cite{Brennan:2017rbf, Obied:2018sgi, Palti:2019pca} for a detailed review). In addition to these moduli stabilization and inflationary models, many efforts have been made to combine the global issues with local issues, such as constructing global string compactifications that can simultaneously describe  D-brane models of particle physics. One of the most serious obstacles to this is the tension between chirality and moduli-fixing by non-perturbative effects \cite{Blumenhagen:2007sm}. Several potential solutions, however, have already been proposed  by carefully analyzing the Freed-Witten anomaly and K\"ahler cone condition \cite{Collinucci:2008sq}, tuning on fluxed-instantons on the divisors \cite{Grimm:2011dj,Kerstan:2012cy} or putting the D-brane at the singularity \cite{Cicoli:2011qg, Balasubramanian:2012wd,Cicoli:2012vw,Cicoli:2013mpa, Cicoli:2016xae, Cicoli:2017shd, Cicoli:2017axo, Cicoli:2021dhg}. In most of these approaches, finding Calabi-Yau threefolds with a non-trivial odd cohomology $h^{1,1}_{-}(X/\sigma^{*})$ is a crucial ingredient.

A great deal of progress has been made in understanding the statistical structure of the moduli in many classes of Calabi-Yau threefolds, by brute force calculation and scans, without considering the orientifold involution explicitly, and how this relates to properties such as the axion landscape or Swiss cheese structure \cite{Gray:2012jy, Long:2014fba, He:2015fif, Galvez:2016qll, Long:2016jvd,  Altman:2017vzk,  Demirtas:2018akl, Halverson:2019cmy}. Recently, in the context of Complete Intersection Calabi-Yau 3-folds (CICYs) embedded in products of projective spaces \cite{Candelas:1987kf}, a landscape of orientifold vacua has been constructed~\cite{Carta:2020ohw, Carta:2021uwv} from the most favorable description of the CICY 3-folds database \cite{Anderson:2017aux}.  More general free quotients have been classified and studied in the case of CICY 3-~and 4-folds~\cite{Braun:2010vc, Gray:2013mja, Candelas:2015amz, Constantin:2016xlj}. Free quotients in the toric case for $h^{1,1}(X)\leq 3$ were systematically studied  in~\cite{Braun:2017juz} (See \cite{He:2020bfv}  for a recent review on these Calabi-Yau database).

 It would be a great step forward to provide, as an extension of our database of Calabi-Yau threefolds, a concrete classification of Calabi-Yau orientifold data, explicitly counting the even and odd moduli, determining the types and numbers of fixed O-plane loci. This is a primary motivating factor for this paper.
In a previous work~\cite{Gao:2013pra}, divisor involutions in Calabi-Yau threefolds up to $h^{1,1}(X)=4$ with maximal triangulations were classified. In this paper we will extend and improve this work in several respects:
\begin{enumerate}
\item
We push our  classification upper bound to $h^{1,1}(X)=6$ in the Calabi-Yau database~\cite{Altman:2014bfa} constructed from the Kreuzer-Skarke list~\cite{Kreuzer:2000xy}.
\item
Instead of limiting ourselves to Calabi-Yau hypersurfaces in unique, fully desingularized toric varieties, we expand our analysis to hypersurfaces in all possible maximal projective crepant partial (MPCP) desingularizations. The number of toric triangulations we will analyze increases by two orders of magnitude from 2,968 \cite{Gao:2013pra} to { 646,903} independent phases.
\item
We determine all individual topologies of divisors in each of the toric Calabi-Yau threefolds.  Furthermore, we improve our algorithm to determine the proper involution and identify the involutions which are globally consistent across all disjoint phases of the K\"ahler cone for each unique Calabi-Yau geometry.
\item
We identify all possible fixed loci under non-trivial actions, thereby determining the location of different types of O-planes relevant in D-brane constructions.  We then further classify the involutions as freely or non-trivially acting depending on whether there exists a fixed locus on the Calabi-Yau threefold under the involution. 
\item In the orientifold Calabi-Yau threefolds with an $O3/O7$-system, we then further classify the so-called \lq\lq naive orientifold Type IIB string vacua" by considering the D3 tadpole cancelation condition when putting eight $D7$-branes on top of the $O7$-plane. 
\item
We determine the equivariant cohomology (Hodge number splitting) under these involutions in the $\IZ_2$-orbifold limit. 
\end{enumerate}

Since we only consider favorable Calabi-Yau threefolds, we will not include coordinate reflections such as  $\sigma: x_i \leftrightarrow -x_i$ in this paper, as the corresponding divisor involution $\sigma^{*}: D_i \leftrightarrow D_i$ is manifestly trivial and will not contribute to $h^{1,1}_-(X/\sigma^{*})$. 
%However, the fix locus on the Calabi-Yau hypersurface under these  reflections are still of interest for phenomenology reason, we leave  the discussion in a future work. 
When the geometry is unfavorable,  it contains a divisor with disconnected pieces like $\IP^n \cup \dots \cup \IP^n$, $dP_n \cup \dots \cup dP_n$ or others. Under the reflection, these pieces exchange to one another and split to $h^{0,0}_+(D)$ and $h^{0,0}_-(D)$, which will contribute to the odd equivariant cohomology $h^{1,1}_-(X/\sigma^{*})$ \cite{Blumenhagen:2012kz, Gao:2013pra}.

The paper is organized as follows: In Section~\ref{subsec:favorable}, we briefly review the construction of Calabi-Yau threefolds as hypersurfaces in toric varieties.
Then, in Section~\ref{subsec:topology} we  show how to compute the Hodge numbers of all toric divisors on a given Calabi-Yau threefold, and classify these divisors according to topology, in a manner similar to~\cite{Gao:2013pra}. 
We next identify all pairs of ``Non-trivial Identical Divisors'' (NID) and present the proper divisor involutions in Section~\ref{subsec:involutions}. All fixed-point loci are then identified in Section~\ref{subsec:fixedloci}, and this information is used to classify the involutions as either non-trivially or freely acting. As this is the heart of the current work, we take the time to describe our procedures in detail. We additionally provide a pseudocode description of the fixed point algorithm in Appendix \ref{appendix:pseudocode} for those interested in the algorithm's implementation.
The cohomology class splitting under these involutions is then determined in Section~\ref{subsec:splitting}.
We illustrate the procedures via detailed examples in Section~\ref{sec:example} and summarize our results in Section~\ref{sec:orientifolddiscuss}.

%%%%%%%%%%%%%%%%%%%%%%%%%%%%%%%%%%%%%%%%%%%%%%%%%%%%%%%%
%%%%%%%%%%%%%%%%%%%%%%%%%%%%%%%%%%%%%%%%%%%%%%%%%%%%%%%%
%%%%%%%%%%%%%%%%%%%%%%%%%%%%%%%%%%%%%%%%%%%%%%%%%%%%%%%%
\section{Constructing Calabi-Yau Orientifolds}
\label{sec:involutions}

%%%%%%%%%%%%%%%%%%%%%%%%%%%%%%%%%%%%%%%%%%%%%%%%%%%%%%%%
%%%%%%%%%%%%%%%%%%%%%%%%%%%%%%%%%%%%%%%%%%%%%%%%%%%%%%%%
\subsection{Polytopes, Geometries, and Triangulations}
\label{subsec:favorable}

It is a well-known result of Batyrev~\cite{Batyrev:1993} that Calabi-Yau threefolds can generically be obtained by taking the anticanonical hypersurface in an ambient four-dimensional Gorenstein toric Fano variety. In a previous work~\cite{Altman:2014bfa}, the procedure for computationally extracting the topology of such a toric variety, as well as its restriction to the anticanonical hypersurface, from combinatorical information encoded in a 4-dimensional reflexive lattice polytope $\Delta$, was outlined. In fact, a complete enumeration of all 4-dimensional reflexive polytopes exists due to Kreuzer and Skarke~\cite{Kreuzer:2000xy}.

In the context of toric geometry, when the toric divisor classes on the Calabi-Yau hypersurface $X$ are all descended from the ambient space $\cA$, we say that it is a favorable geometry. Consider the short exact sequence and its dual sequence
 \bea
  0 \ra TX \ra T\cA|_X \ra \cN_{X/\cA} \ra 0, \\\nonumber
    0 \ra \cN^*_{X/\cA} \ra T^*\cA|_X \ra T^*X \ra 0.
 \eea
This induces the long exact sequence in sheaf cohomology
 {\small
\bea\label{eq_generalizedkoszulwithdiviso}
      \parbox{0.1cm}{\xymatrix{
          &
           \hspace*{-1cm} \cdots \fto H^{1} (X, \cN^*_{X/\cA})   \xrightarrow{\quad\quad\alpha\quad\quad}    &
         {H^{1} (X, T^*\cA|_{X})}  \xrightarrow{\quad\quad\quad}  &
      H^1(X, T^*X) \ar`[rd]`[l]`[dlll]`[d][dll] &
         \\ &
           \quad  H^{2} (X, \cN^*_{X/\cA})   \xrightarrow{\quad\quad\beta\quad\quad}     &
               {H^{2} (X, T^*\cA|_{X})}  \xrightarrow{\quad\quad\quad}  &
             H^2(X, T^*X)  \fto \cdots\,. &
        }}%^\big._\big.
    \eea
}

By Dolbeault's theorem, $H^1(X, T^{*}X) \cong H^{1,1}(X) \cong {\rm coker }(\alpha) \oplus {\rm ker} (\beta)$. It has two contributions, the cokernel of the map $\alpha$ describes the descent of the K\"ahler moduli on $\cA$ to K\"ahler moduli on X, while the kernel of the map $\beta$ describes some new K\"ahler moduli on $X$ which do not descend from $\cA$. 
 If the kernel part is zero ${\rm ker}(\beta) = 0$, the only divisors on $X$ are those descending from ${\cal A}$ (possibly with additional linear relations) and we say the geometry is  \lq\lq favorable"\footnote{There exits a stronger notion of \lq\lq K\"ahler favourability" where  K\"ahler cones on $X$ descend  from an ambient space  in which they are embedded \cite{Anderson:2017aux}.  This involve a careful set of arguments on the descent of the effective, nef and ample cones of divisors which we refer the reader to see \cite{Oguiso,Oguiso2}. In some cases, the \lq\lq favorable" geometry are not \lq\lq K\"ahler
favorable" since the K\"ahler cone of $X$ is actually larger than the positive orthant one we calculated following \cite{Altman:2014bfa}. }.  In this case   $h^{1,1}(X) ={\rm  dim}(H^{1,1} (X)) \cong {\rm dim}({\rm Pic}(\cA))$.  The simplest case of a favorable  geometry is when $h^{2} (X, \cN|_{X}^*) = 0$ (or equivalently, when $h^{1} (X, \cN|_{X}) = 0$).    In this paper, we restrict us to study these so-called favorable geometry and leave those unfavorable cases for future work.

If we restrict ourselves to smooth manifolds, we must at least partially desingularize the ambient toric variety $\cA$ by blowing up enough of its singular points that $X$ is generically smooth, but without adding any discrepancies to its cohomology class. A method for doing such a maximal projective crepant partial (MPCP) desingularization involves the triangulation of the polar dual reflexive polytope $\Delta^{*}$, which contains at least one  fine, star, regular triangulation (FSRT).

When the Calabi-Yau hypersurfaces of two or more desingularizations share certain key topological
invariants, then it can be shown that they are topologically equivalent and can be considered
representations of the same Calabi-Yau threefold. In this case, the K\"ahler form of this Calabi-Yau
threefold is allowed to reside within the K\"ahler cone of either representation, and we refer to these
disjoint K\"ahler cone chambers as its phases. In order to allow the K\"ahler form to smoothly vary over its full range, the phases of the K\"ahler cone must be glued together in an appropriate manner. In practice, we use
Wall’s theorem \cite{Wall} to glue together the various phases of the complete K\"ahler cone corresponding to a
distinct Calabi-Yau threefold geometry, which requires the checking of whether or not all singularities in the walls
between these phases are avoided by the Calabi-Yau hypersurface.

There are generally many MPCP desingularization configurations possible, each of which sets different topological restrictions on how $X$ can be deformed within $\cA$. These deformations are parameterized by K\"ahler moduli, and the different configurations serve to divide up the space of moduli into discontinuous chambers. 
When passing from one chamber to another, some singular points are blown up, while others are blown down, so that the singularities cannot be consistently resolved at the boundary. This exchange is called a flop. Each distinct chamber defines a unique resolved ambient space $\tilde{\cA}$, and there are, in general, many more of these than the original ambient toric varieties $\cA$.

It often happens, however, that the smooth Calabi-Yau hypersurface $X$ does not intersect any of the singular points involved in a flop between chambers of the moduli space. If, in addition, the topology of $X$ is invariant under the flop, then the singularity can be neglected for our purposes, and the chambers can be effectively glued and associated with a single unique, smooth hypersurface $X$.

It is clear from the above discussion that obtaining an accurate count of unique, smooth Calabi-Yau geometries a priori is highly non-trivial. However, the calculations have been performed for geometries with Hodge number $h^{1,1}(X)\leq 6$~\cite{Altman:2014bfa}.

%%%%%%%%%%%%%%%%%%%%%%%%%%%%%%%%%%%%%%%%%%%%%%%%%%%%%%%%
%%%%%%%%%%%%%%%%%%%%%%%%%%%%%%%%%%%%%%%%%%%%%%%%%%%%%%%%
\subsection{Topology of Toric Divisors}
\label{subsec:topology}

We will assume the ambient space $\cA$ to be a resolved Gorenstein toric Fano variety with dimension $n=4$ whose anticanonical divisor $X=-K_{\cA}$ is a Calabi-Yau threefold hypersurface. 
Denote $x_i$ as the weighted homogeneous coordinates used to define $X$ inside the ambient space $\cA$. Then the divisor $D_{i}=\{x_i=0\}$ defines a 4-cycle on $X$ which is dual to a 2-cycle $\omega_i$.

In the context of toric geometry, a pair of ``Non-trivial Identical Divisors'' (NIDs) refers to a pair of divisors with distinct charges under the torus action, but whose intersections with the Calabi-Yau hypersurface have identical cohomology. In order to determine the Hodge number of an individual divisor on the Calabi-Yau threefold, we use the Koszul extension to the \texttt{cohomCalg} package \cite{Blumenhagen:2010pv, cohomCalg:Implementation} with the \texttt{HodgeDiamond} module to calculate these quantities. However, in many situations relevant to our purposes, this module cannot give an explicit result.

For an irreducible divisor $D$, the complex conjugation $h^{p,q}(D)=h^{q,p}(D)$ and Hodge star $h^{p,q}(D)=h^{2-p,2-q}(D)$ dualities constrain the independent Hodge numbers of $D$ down to only $h^{1,0}(D), h^{2,0}(D)$, and $h^{1,1}(D)$. As a result, when we encounter difficulties with \texttt{cohomCalg}, we first calculate the Euler number $\chi(D)=\int_{D}{c_{2}(D)}$ of the divisor on the hypersurface, and then determine $h^{1,0}(D)$ and $h^{2,0}(D)$ by calculating the trivial line bundle cohomology of the divisor $h^\bullet(D, \cO_{D}) = \{h^{0,0}(D), h^{1,0}(D), h^{2,0}(D)\}$ by chasing the Koszul sequence:
\bea
0\rightarrow {\cal O}_{\cA}(-X-D)\fto {\cal O}_{\cA}(-X) \oplus {\cal O}_{\cA}(-D) \fto {\cal O}_{\cA} \fto {\cal O}_{D/X} \rightarrow 0 \, .
\eea

Then, using the expression
\bea
\chi(D) = \sum\limits_{i=0}^{2} (-1)^i \,{\rm dim}\left(H^i_{\rm DR}(D)\right) = \sum\limits_{p+q=0}^{2}(-1)^{p+q} \,{\rm dim}\left(H^{q}(D,\Omega^{p})\right)\, ,
\eea
we can fix $h^{1,1}(D)$ and get the full Hodge diamond for any divisor.
The internal topology of these divisors plays an important role in string compactification and moduli stabilization. In our procedure for scanning divisor involutions, several divisors classifications are of particular phenomenological interest.\footnote{For the phenomenological relevance of these divisor types, see the discussion and references in~\cite{Gao:2013pra}.}
\\

\noindent {\it Completely rigid divisor:}  The Hodge numbers of these divisors are characterized by \\
$h^\bullet(D) = \{h^{0,0}(D), h^{0,1}(D), h^{0,2}(D), h^{1,1}(D)\}= \{1,0,0,h^{1,1}(D)\}$ 
such that  $h^{1,1}(D) \neq 0$. 
This class of divisors is further subdivided into either the del Pezzo surfaces $\{\IP^2 \equiv dP_0$, $dP_{n}$, with $n=1, \dots , 8\}$ (which may be shrinkable or non-shrinkable depending on the diagonalizability of its intersection tensor) with $n=h^{1,1}(D)-1$, and 
the ``non-shrinkable rigid divisors'' with $h^{1,1}(D)>9$.\\

\noindent {\it``Wilson'' divisor:} The Hodge numbers of these divisors are characterized by
$h^\bullet(D) = \{h^{0,0}(D), h^{0,1}(D), $ $h^{0,2}(D), h^{1,1}(D)\} =\{1, h^{1,0},0, h^{1,1}\}$ with $h^{1,0}(D),\, h^{1,1}(D) \neq 0$.
We will also further specify the ``Exact-Wilson'' divisor as $h^{\bullet}(D)=\{1,1,0,h^{1,1}\}$ with $h^{1,1} (D)\neq 0$. \\

\noindent {\it Deformation divisor:}  These divisors are characterized simply by $h^{2,0}(D) \neq 0$.
%
%Unlike the poly-instanton zero modes encoded in $h^{1,0}_{+}(D)$, one can lift these (extra) deformation zero modes by turning on background magnetic fluxes~\cite{Bianchi:2011qh}. Since these fluxes can rigidify some deformation divisors,  such circumstances facilitate the moduli stabilization process by introducing  more terms for superpotential contributions.
%Such mechanism has been utilized
%to build an explicit de Sitter \cite{Louis:2012nb}.
 %Here we focus on the following three kinds
 %of deformation divisors.
\begin{itemize}
\item
A K3 divisor is a deformation divisor with Hodge numbers $h^{\bullet}(D)=\{1,0,1,20\}$. 
%
%A K3-fibration is useful for obtaining an anisotropic shape of the Calabi-Yau compactification. This leads to some LVS models with what is effectively two large extra dimensions of micron size and a fundamental gravity scale around $\sim 1$ TeV \cite{Cicoli:2011it, Cicoli:2011yy}. The property of spaces which contain both $K3$ and Wilson surface are also studied in~\cite{Blumenhagen:2012Poly1,Lust:2013kt,Gao:2013pra}.
\item
Another deformation divisor that appears often in our scan is similar to a K3 divisor, but with an extra $h^{1,1}$ deformation degree of freedom, i.e. $h^{\bullet}(D)=\{1, 0, 1, 21\}$. We will refer to this as a type-1 special deformation divisor, $SD1$.
\item
Another deformation divisor that appears in our classification has Hodge numbers $h^{\bullet}(D)=\{1, 0, 2, 30\}$, which we refer to as $SD2$.
\end{itemize}

$SD1$ and $SD2$ appear to share similar properties, particularly in their volume forms. In our scan, we will first identify the Hodge numbers of all the toric divisors on the Calabi-Yau hypersurface $X$, which descend from the ambient space $\cA$. This calculation will be performed for each Calabi-Yau geometry in the database at \dburl, where the Hodge diamond  of each divisors in the defining Calabi-Yau manifold is presented.
%%%%%%%%%%%%%%%%%%%%%%%%%%%%%%%%%%%%%%%%%%%%%%%%%%%%%%%%
%%%%%%%%%%%%%%%%%%%%%%%%%%%%%%%%%%%%%%%%%%%%%%%%%%%%%%%%
\subsection{Holomorphic Divisor Exchange Involutions}
\label{subsec:involutions}

The map $\sigma: x_i \leftrightarrow x_j$ 
exchanging two homogeneous coordinates of the ambient toric variety $\cA$ can be pulled back to a holomorphic involution on the corresponding toric divisor cohomology classes $\sigma^{*}: D_i \leftrightarrow D_j$, which, on favorable manifolds, restricts in a straightforward way to the Calabi-Yau hypersurface $X$. We then define the even and odd parity eigendivisor classes $D_\pm \equiv D_i \pm D_j \in H^{1,1}_{\pm}(X/\sigma^{*})$. In general, a given geometry can allow multiple disjoint involutions $\sigma_1, \sigma_2,\dots,\sigma_{n}$. In this case, the full involution is given by $\sigma\equiv\sigma_{1}\circ\sigma_{2}\circ\cdots\circ\sigma_{n}$.

We consider two divisors $D_{i},D_{j}$ to be identical if $h^{\bullet}(D_{i})\cong h^{\bullet}(D_{j})$. They may nevertheless still be topologically distinct on $X$. This can readily be determined in favorable geometries by inspecting their toric $\mathbb{C}^{*}$ weights. If an identical pair of divisors also share the same weights, then an involution $\sigma$ exchanging them will act trivially on the hypersurface polynomial and will leave $h^{1,1}_-(X/\sigma^{*})=0$.

As mentioned earlier, from a phenomenological point of view, there are a variety of reasons to work on a Calabi-Yau manifold with non-trivial odd parity $h^{1,1}_{-}(X/\sigma^{*})>0$. Consequently, we consider only holomorphic divisor involutions between ``Non-trivial Identical Divisor pairs'' (NIDs) in this work.
After identifying the NIDs, we will check which involutions between these divisors are consistent with the topology of the Calabi-Yau manifold. Since the hypersurface is embedded in a desingularized ambient variety $\cA$, we will require our orientifold involution to be an automorphism of $\cA$, leaving invariant the exceptional divisors from resolved singularities. 
The information describing the desingularization is encoded in the Stanley-Reisner ideal $ \mathcal{I}_{SR}(\cA)$.
As a consequence, the involution should be a symmetry of $\mathcal{I}_{SR}(\cA)$.
We then further distinguish the ``proper'' involutions which are symmetries of the linear ideal  $\mathcal{I}_{lin}(\cA)$, encoding toric divisor redundancy. This ensures that the defining polynomial of the Calabi-Yau hypersurface generically remains homogeneous under the coordinate exchange without tuning any coefficients to zero. These involutions are then also symmetries of the graded Chow ring
\bea
\label{eq:chow}
A^{\bullet}(\cA)  \cong \frac{\IZ ( D_1, \cdots, D_k)}{\mathcal{I}_{lin}(\cA)+ \mathcal{I}_{SR}(\cA)}\, .
\eea
Due to the favorability condition on the Calabi-Yau threefold hypersurface we have
\bea
\label{eq:favor}
A^1(\cA) \cong H^{1,1}(\cA) \cong  {\rm Pic} (\cA) \cong {\rm Pic} (X) \cong H^{1,1}(X)  \cong A^1(X)\, ,
\eea
and thus the toric triple intersection tensor defined in the Chow ring $A^4(X)$ of Calabi-Yau threefolds by
\bea
\label{eq:triple}
d_{ijk}=\int\limits_{X}{D_{i}\wedge D_{j}\wedge D_{k}\equiv D_{i}\cdot D_{j}\cdot D_{k}\cdot X}\hspace{5mm}\text{and}\hspace{5mm}X=-K_{\cA}=\sum\limits_{i=1}^{k}{D_{i}}
\eea
will also remain invariant. If the involution involves non-rigid, deformation divisors, this condition will ensure that they can be exchanged in a consistent way.

Let us consider a simple example with $h^{1,1}(X) = 3$. The triple intersection tensor can be written in the basis of divisor class $\{J_1, J_2, J_3\}\in H^{1,1}(X;\mathbb{Z})$ as
\bea
\kappa_{ijk}=\int\limits_{X}{J_{i}\wedge J_{j}\wedge J_{k}\equiv J_{i}\cdot J_{j}\cdot J_{k}\cdot X}\, .
\eea
Suppose we have a proper involution $\sigma^{*}: J_2 \leftrightarrow J_3$. Then, under the basis change
\bea
\{J_1, J_2, J_3\}\mapsto\{J_0,J_+,J_-\}\hspace{5mm}\text{where}\hspace{5mm}J_\pm =J_2 \pm J_3\, ,
\eea
the intersection numbers with an odd number of minus indices ($\kappa_{00-}, \kappa_{++-}, \kappa_{0+-}, \kappa_{---}$) should vanish identically. In this basis, we can write down the K\"ahler form $J=t^{0}J_{0}+t^{+}J_{+}+t^{-}J_{-}$. For a consistent orientifold, we must have both
$\sigma^{*}J=J$ and  $\sigma^{*}\Omega_{3}=-\Omega_{3}$, 
 where $\Omega_{3}$ is the unique holomorphic (3,0)-form on $X$. The constraint on the K\"ahler form given by eq.(\ref{eq:orientifold}) can be used to to show that the K\"ahler cone condition ($\int_{\mathcal{C}^{i}}{J}>0$) will be consistently satisfied on the orientifold. The holomorphic (3,0)-form $\Omega_{3}$ can be constructed in terms of the projective coordinates using the techniques in Section 5.6 of \cite{Denef:2008wq},
%As mentioned in Section \ref{subsec:involutions}, we determine the parity following the formula for the volume form given in \cite{Denef:2008wq}:
\bea
\label{eq:3form}
\Omega = \frac{1}{2\pi i}\oint_{P=0}\frac{\omega \cdot \Pi_{a} V^{a}}{P}\, .
\eea
Here $P$ is the  hypersurface polynomial and $\omega = R(x) dx^{1} \wedge \cdots \wedge dx^{k}$, with $R(x)$ a homogeneous rational function of the $x_{i}$ that makes $\omega / {P}$ have zero weight. Since the hypersurface polynomial always has weights equal to the sum of the toric divisor weights, for a toric CY we can take $R(x) \equiv 1$. Finally, the $V^{a}$ are the holomorphic vector fields generating the gauge symmetries, determined in terms of the weights $Q^{a}_{i}$:
\bea
\label{eq:V}
V^{a}=\Sigma_{i} Q^{a}_{i}x_{i}\frac{\partial}{\partial x_{i}}\, .
\eea
Since we will be restricting the hypersurface polynomial $P$ to be invariant under $\sigma$, the numerator of the integrand, which we will here call $\mathcal{Q}$, determines the parity.   Its behavior under $\sigma$ can be calculated in a straightforward manner.

%The resulting orientifold projection that breaks $\mathcal{N}=2$ SUGRA down to $\mathcal{N}=1$ is given by $\Omega\sigma (-1)^{F_{L}}$, where $\Omega$ is the world-sheet parity transformation and $F_{L}$ is the left-moving space-time fermion number.

The procedure, then, is to first determine the  Hodge numbers for all individual divisors in a given triangulation, scan the database of Calabi-Yau threefolds extracted from the Kreuzer-Skarke list up to $h^{1,1}(X)=6$ and pick out the desingularized configurations (i.e. triangulations of reflexive 4D polytopes) that support an exchange of NIDs. We then identify all the ``proper'' involutions that allow for consistent orientifold geometries. Our results are available via the database search engine at \dburl, and the statistics of this scan are summarized in Section \ref{sec:orientifolddiscuss}. In the tables found in that section, we further classify the different involutions into exchanges that include completely rigid surfaces, Wilson surfaces, and some deformation surfaces. 

Before continuing, let us take the opportunity to distinguish between two types of involution:
\begin{itemize}
\item {\it Triangulation-wise proper involution}:   The involutions present at the triangulation level - that is, within a single chamber of the K\"ahler cone of a given geometry. 
\item {\it  Geometry-wise proper involution}:  The involutions which are globally consistent across all disjoint phases of the K\"ahler cone for each unique Calabi-Yau geometry.  Each of the geometry-wise proper involutions may correspond to several triangulation-wise involutions which can span an entire CY geometry.
\end{itemize}

%
%These methods can and will be systemically generalized to higher $h^{1,1}(X)\geq 7$, and database will be updated accordingly.

%\newpage
%%%%%%%%%%%%%%%%%%%%%%%%%%%%%%%%%%%%%%%%%%%%%%%%%%%%%%%%
%%%%%%%%%%%%%%%%%%%%%%%%%%%%%%%%%%%%%%%%%%%%%%%%%%%%%%%%
\subsection{Classifications of the Fixed Orientifold Planes}
\label{subsec:fixedloci}

The next question to ask is whether there exist any point-wise fixed loci for a given involution on the Calabi-Yau threefold. If there is a fixed codimension-$n$ subvariety defined by the simultaneous vanishing of $n$ polynomials, then $X$ will define an orientifold with an O$m$ plane, where $m=3+2(3-n)$. For a Calabi-Yau threefold, we must have $0\leq n\leq 3$, and so we can only have O9, O7, O5, or O3 planes. If we have none of these, and the hypersurface determined by the $\sigma$-invariant Calabi-Yau polynomial is smooth, then the involution is a free $\mathbb{Z}_{2}$ action on $X$.

We now present an algorithm for identifying and classifying fixed-point sets under an orientifold of the type described in the previous section. We first define the MPCP-desingularized ambient 4D toric variety
\bea\label{eq:defamb}
\cA=\frac{\mathbb{C}^{k}\smallsetminus Z}{\left(\mathbb{C}^{*}\right)^{k-4}\times G}\, ,
\eea
where $Z$ is the locus of points in $\mathbb{C}^{k}$ ruled out by the Stanley-Reisner ideal $\mathcal{I}_{SR}(\cA)$, and $G$ is the fundamental group (i.e. trivial in most cases\footnote{In the total favorable Kreuzer-Skarke list with $h^{1,1}(X) \leq 6$, there are only 14 of them which have non-trivial fundamental group.  In this paper, we only consider the cases with trivial fundamental group. }). The geometry on this toric variety can be described by the projective coordinates $\{x_{1},...,x_{k}\}$ and their toric $\mathbb{C}^{*}$ equivalence classes
\bea
(x_{1},...,x_{k})\sim (\lambda^{\mathbf{W}_{i1}}x_{1},...,\lambda^{\mathbf{W}_{ik}}x_{k})\, ,
\eea
which define a projective weight matrix $\mathbf{W}$. 
\\ \\
Throughout this section, we use the following notation:
\begin{itemize}
\item $k$ is the number of toric divisors
\item $p$ is the number of (disjoint) coordinate exchanges (in Section~\ref{sss:cyp})
\item $k' = |\mathcal{G}|$ is the number of definite parity polynomial generators (in Section~\ref{sss:fixpoint})
\item $\tilde{\mathbf{W}}$ is the weight matrix of the definite parity generators (in Section~\ref{sss:fixpoint})
\item $r$ is the rank of $\tilde{\mathbf{W}}$ (in Section~\ref{sss:fixpoint})
\end{itemize}
%
%In addition, the image of the symplectic moment map $\mu: \cA\rightarrow\mathbb{R}^{4}$ defines a convex Newton polytope by Atiyah and Guillemin-Sternberg. In cases gleaned from the Kreuzer-Skarke list, as in our study, these are reflexive lattice polytopes $\Delta=\mu(\cA)$ with polar dual $\Delta^{*}$.

%%%%%%%%%%%%%%%%%%%%%%%%%%%%%%%%%%%%%%%%%%%%%%%%%%%%%%%%
\subsubsection{Invariant Calabi-Yau hypersurface polynomial}\label{subsec:CY}
\label{sss:cyp}

Consider the smooth Calabi-Yau anticanonical hypersurface $X=-K_{\cA}$ defined by the vanishing of a homogeneous polynomial $\{P=0\}$. The polynomial $P$ can be expressed in terms of the known vertices $m\in M, n\in N$ of the Newton and dual polytopes, respectively
\bea
P = \sum_{m\in\Delta}{a_{m}M_{m}}=0,\hspace{5mm}\text{where}\hspace{5mm}M_{m}=\prod_{i=1}^{k}{x_{i}^{\langle m,n_{i}\rangle +1}} \, .
\eea
An involution of the type that we are considering will consist of a set of $p$ fully disjoint coordinate exchanges, $\sigma_{s}: x_{i_{s}} \leftrightarrow x_{j_{s}},\;\; s=1, \dotsc, p$ and $\sigma=\sigma_{1}\circ\cdots\circ\sigma_{p}$. In order for the Calabi-Yau hypersurface to be invariant under $\sigma$, we must restrict to the subset of moduli space in which the defining polynomial is invariant. To check this, we define the set of monomials $\mathcal{M}=\{M_{m}\vert m\in\Delta\}$. Then for $M_{m},M_{m'}\in\mathcal{M}$, we identify three cases:
\begin{enumerate}
\item
$\sigma(M_{m})=M_{m}\hspace{5mm}\Rightarrow\hspace{5mm}a_{m}$ is generic,
\item
$\sigma(M_{m})=M_{m'},\;\;m\neq m'\hspace{5mm}\Rightarrow\hspace{5mm}a_{m}=a_{m'}$,
\item
$\sigma(M_{m})\not\in\mathcal{M}\hspace{5mm}\Rightarrow\hspace{5mm}a_{m}=0$.
\end{enumerate}

Clearly, imposing these restrictions requires some tuning in the complex structure moduli space, but the end result is that $P\mapsto P_{symm}$ such that $\sigma(P_{symm})=P_{symm}$, in addition to $\sigma^{*}J=J$. It is important to note that this tuning may introduce singularities into the CY hypersurface. The MPCP desingularization of the ambient space $\cA$ only guarantees resolution of singularities with codimension larger than two, and by Bertini's theorem we can expect that the Calabi-Yau polynomial avoids them. However, now that we have required some coefficients to vanish, it is not necessarily the case that the hypersurface will avoid these singularities.

We can now search for the set of points fixed under $\sigma$. We first locate the fixed-point set in the ambient space, and only later restrict this set to the Calabi-Yau hypersurface.\footnote{This restriction can be done in a straightforward manner since we only consider favorable manifolds.}

%%%%%%%%%%%%%%%%%%%%%%%%%%%%%%%%%%%%%%%%%%%%%%%%%%%%%%%%
\subsubsection{Minimal generating set of homogeneous polynomials with definite-parity}

Any subvariety $V$ can be described as an intersection of hypersurfaces, each generated by a single homogenous polynomial. When $V$ is point-wise fixed under an involution $\sigma$, it must be contained within the eigenspace of $\sigma$. This, however, does not imply that each individual hypersurface is in the eigenspace of $\sigma$. Consider, for example, the following case:
\bea
&V=\{x\in\mathcal{A}\mid f(x)=g(x)=0\}\text{, with }\left\{
\begin{array}{ll}
	\sigma\circ f=g \\
	\sigma\circ g=f \\
\end{array} 
\right.\, .
\eea
Clearly, in this scenario the subvariety (i.e. intersection of hypersurfaces) remains fixed under the involution while each individual hypersurface does not\footnote{Note, however, that while these hypersurfaces are not individually point-wise fixed, they do individually satisfy the involution condition with $\sigma^{2}\circ f=f$ and $\sigma^{2}\circ g=g$.}, and therefore does not have a definitely parity under $\sigma$. In order to reduce the search space, and therefore the computational complexity of our scan, we restrict to only fixed point loci and corresponding orientifold planes whose hyperplane generators have definite parity under the involution $\sigma$. We leave the identification of the remaining orientifold planes to future work.

With this in mind, we look for the minimal generating set $\mathcal{G}$ of homogeneous polynomials $y(x_{1},...,x_{k})$ that are (anti-)invariant under $\sigma$. Each of these generators then has definite parity $\sigma(y)=\pm y$ for each $y\in\mathcal{G}$.

Clearly, the monomials in $\mathcal{M}$ satisfying case (1) above are included in $\mathcal{G}$. In addition, we define the subset $\mathcal{G}_{0}\subset\{x_{1},...,x_{k}\}$ of the projective coordinates that are left unexchanged by $\sigma$.  Then, for ease of notation, we can define the orthogonal decomposition
\bea
\mathcal{G}=\mathcal{G}_{0}\cup\mathcal{G}_{+}\cup\mathcal{G}_{-} \, .
\eea
In order to determine these generating polynomials, we note that because $\sigma^{2}=1$, any (anti-)invariant polynomial can be written in terms of monomials $N_{i}$ as
\begin{align}\label{eq:genpoly}
Q_{0,+}&=\sum_{i}{c_{i}N_{i}}+\sum_{j}{d_{j}\left(N_{j}+\sigma(N_{j})\right)}\notag\\
Q_{-}&=\sum_{j}{d_{j}\left(N_{j}-\sigma(N_{j})\right)},\hspace{3mm}\text{where}\hspace{3mm}\sigma(N_{i})=N_{i}\hspace{3mm}\text{and}\hspace{3mm}\sigma(N_{j})\neq N_{j}\, .
\end{align}
Thus, $\mathcal{G}_{0}\cup\mathcal{G}_{+}$ will only contain monomial and binomial generators, and $\mathcal{G}_{-}$ only binomial generators.

The unexchanged coordinates in $\mathcal{G}_{0}$ are known {\em a priori} from our choice of involution, so we restrict our attention to finding the non-trivial even and odd parity generators in $\mathcal{G}_{+}$ and $\mathcal{G}_{-}$. To do this, we must consider not only $\sigma$, but the $2^{n}-1$ non-trivial ``sub-involutions'' $\rho\subseteq\sigma$ given by the non-empty subsets of $\{\sigma_{1},...,\sigma_{n}\}$ of size $1\leq m\leq n$.

For $m=1$, $\rho\equiv\sigma_{s}: x_{i_{s}} \leftrightarrow x_{j_{s}}$, and it is clear that $\mathcal{G}_{+}=\{x_{i_{s}}x_{j_{s}}\}$ and $\mathcal{G}_{-}=\emptyset$. Since we are considering only exchanges of NIDs, we do not need to consider binomials (i.e. $x_{i_{s}}\pm x_{j_{s}}$) as these are not homogeneous.

For $m>1$ sub-involutions, any invariant monomials are just products of the generators from $m=1$ sub-involutions. Therefore, we will now turn our attention to binomial generators. In particular, we look for those of the form
\bea
y_{\pm}(\mathbf{a})=  x_{i_{1}}^{a_{1}}x_{i_{2}}^{a_{2}}\dotsm x_{i_{m}}^{a_{m}} \pm x_{j_{1}}^{a_{1}}x_{j_{2}}^{a_{2}}\dotsm x_{j_{m}}^{a_{m}}\, ,
\eea
where  $\mathbf{a} = (a_{1}, a_{2}, \dotsc, a_{m}) \in \mathbb{Z}^{m}$ are such as to ensure that the binomial is homogeneous. If $\mathbf{a}$ contains negative entries, we simply multiply through by the necessary monomial factor to clear any denominators and obtain a proper binomial. The condition for homogeneity, in terms of the columns $\mathbf{w}_{i_{s}}$ and $\mathbf{w}_{j_{s}}$ of the weight matrix $\mathbf{W}$ is given by
\bea
a_{1}\mathbf{w}_{i_{1}} + a_{2}\mathbf{w}_{i_{2}} + \dotsb a_{m}\mathbf{w}_{i_{m}} = a_{1}\mathbf{w}_{j_{1}} + a_{2}\mathbf{w}_{j_{2}} + \dotsb + a_{m}\mathbf{w}_{j_{m}}\, ,
\eea
\noindent which we rewrite as
\bea
a_{1}(\mathbf{w}_{i_{1}}-\mathbf{w}_{j_{1}}) + a_{2}(\mathbf{w}_{i_{2}}-\mathbf{w}_{j_{2}}) + \dotsb + a_{m}(\mathbf{w}_{i_{m}}-\mathbf{w}_{j_{m}})=0\, .
\eea
Let $\mathbf{D}$ be the matrix whose columns are the difference vectors $\mathbf{d}_{s} = \mathbf{w}_{i_{s}} - \mathbf{w}_{j_{s}}$. The above equality implies that $\mathbf{a}$ lies in $\ker \mathbf{D} \cap \mathbb{Z}^{m}$. The generators of $\ker \mathbf{D} \cap \mathbb{Z}^{m}$ as a $\mathbb{Z}$-module give the exponents of binomial generators in $\mathcal{G}_{\pm}$, which we can see via the following argument:

Let $\mathbf{a}, \mathbf{b} \in \ker \mathbf{D} \cap \mathbb{Z}^{m}$, with associated binomials $y_{\pm}(\mathbf{a}) = A \pm B, y_{\pm}(\mathbf{b}) = C \pm D$, where $A,B,C,D$ are monomials in the $x_{i}$ with $\sigma(A) = B, \sigma(C) = D$. Consider $y_{+}(\mathbf{a} + \mathbf{b})$ at some fixed point of $\sigma$. We need only consider the region where no $x_{i}$ in these expressions vanishes (if $x_{i} = 0$ for some $i$, this point is already contained within the vanishing set of $x_{i}x_{\sigma(i)} \in \mathcal{G}_{+}$ and thus would not be a new fixed point). In this region, the vanishing of $y_{+}(\mathbf{a} + \mathbf{b})$ is equivalent to the vanishing of $AC + BD$. As $y_{+}(\mathbf{a}), y_{-}(\mathbf{a})$ have the same weights with opposite parity, at least one must vanish at every fixed point. Thus at our fixed point we have $A = \pm B$. As the same is true for $y_{+}(\mathbf{b}), y_{-}(\mathbf{b})$, we also have $C = \pm D$.

If $A = B, C = -D$ or $A = -B, C = D$, then $AC = -BD$ and $AC + BD$ vanishes trivially. Otherwise we have $AC = BD$, and $AC + BD = 2AC$. This vanishes iff $A = B = 0$ or $C = D = 0$, which as we noted earlier lies in the vanishing set of $x_{i}x_{\sigma(i)}$ for some $x_{i}$. Hence any fixed points in the vanishing set of $y_{+}(\mathbf{a} + \mathbf{b})$ are contained within the vanishing sets of $y_{\pm}(\mathbf{a})$, $y_{\pm}(\mathbf{b})$, or the monomial elements of $\mathcal{G}_{+}$, and we need not consider it further. A similar argument applies for $y_{-}(\mathbf{a} + \mathbf{b})$.

We repeat this procedure for each sub-involution $\rho\subseteq\sigma$. The sets $\mathcal{G}_{\pm}$ are given by the union of the generators found for each of the sub-involutions.

%%%%%%%%%%%%%%%%%%%%%%%%%%%%%%%%%%%%%%%%%%%%%%%%%%%%%%%%
\subsubsection{Naive fixed point loci}
\label{sss:fixpoint}

We first perform a version of the Segre embedding,\footnote{A similar Segre embedding is used in the F-theory lifts from Type IIB orientifold models \cite{Collinucci:2009uh}.} transforming the projective coordinates into the (anti-)invariant generators $\{x_{1},...,x_{k}\}\mapsto\{y_{1},...,y_{k'}\}\equiv\mathcal{G}$. The cardinality $k'=\left\lvert\mathcal{G}\right\rvert$ may be less than, equal to, or greater than $k$. We construct the weight matrix $\tilde{\mathbf{W}}$ for these generators by taking appropriate linear combinations of the original weight matrix columns $\mathbf{w}_{i}$.

In order for a codimension-1 subvariety $D\subset X$ to be point-wise fixed under the involution $\sigma^{*}$, the corresponding coordinate exchange must force its defining polynomial to vanish, i.e. $\sigma : y\mapsto -y$, so that $D=\{y=0\}$ is fixed. This implies that the defining polynomial $y$ of every point-wise fixed codimension-1 subvariety must be generated by odd-parity generators in $\mathcal{G}_{-}$. In the case where $\mathcal{G}_{-}$ is empty, there is no non-trivial fixed subvariety. Then, naively, the only locus in the ambient space (if it exists) fixed under $\sigma$ is the set
\[\{x_{i_{1}} = x_{i_{2}} = \dotsb = x_{i_{n}} = x_{j_{1}} = x_{j_{2}} = \dotsb = x_{j_{n}} = 0\},\]
\noindent which is trivially unaffected by $\sigma$. However, when $\mathcal{G}_{-}$ is non-empty, this trivial locus is immediately a subspace of any generator $y\in\mathcal{G}_{-}$, and is therefore redundant.

To determine the point-wise fixed loci for codimension larger than one, we must check whether the involution forces a subset of generators $\mathcal{F}\subseteq\mathcal{G}$ to vanish simultaneously. The number of checks required can be reduced by noting that if a locus is not point-wise fixed, then no locus containing it will be either. For this reason, we consider only the subsets $\mathcal{F}$ where $\mathcal{F}\cap\mathcal{G}_{-}\neq\emptyset$.

It would seem, then, that the only point-wise fixed locus would be $G:=\bigcap_{y\in\mathcal{G}_{-}}{\{y=0\}}$ while all other fixed loci $F:=\bigcap_{y\in\mathcal{F}}{\{y=0\}}$ correspond to the proper subsets $\mathcal{F}\subsetneq\mathcal{G}_{-}$, and are therefore redundant. However, it is important to note that the torus $\mathbb{C}^{*}$ actions provide $r=\text{rank}(\tilde{\mathbf{W}})$ additional degrees of freedom for the generators to avoid being forced to zero. That is, there may be a toric equivalence class that neutralizes the odd parity of some set of generators, while adding odd parity to another set. In each subset of generators $\mathcal{F}$, we check for this by solving the system of equations
\bea
\lambda_{1}^{\tilde{W}_{1i}} \lambda_{2}^{\tilde{W}_{2i}} \dotsb \lambda_{r}^{\tilde{W}_{ri}} = \sigma(y_{i})/y_{i},\quad i=1,...,k' \, .
\eea
By the construction of the generator $y_{i}$, the right-hand side is equal to $\pm 1$. The set is point-wise fixed if this equation is solvable in the $\lambda_{i}$.

Since the right-hand side of each of these equations has unit magnitude, we only need to look for solutions with $\lambda_{i}$ on the complex unit circle. As any element of the unit circle can be written $\lambda_{i} = e^{i \pi u_{i}}$ with $0 \le u_{i} < 2$, each of the equations above can be rewritten as a linear congruence in the $x$ as
\bea
\tilde{W}_{1i}u_{1} + \dotsb + \tilde{W}_{ri}u_{r}\equiv\left\{\begin{array}{rl}
0\pmod{2},&y_{i}\in\mathcal{G}_{0}\cup\mathcal{G}_{+}\\
1\pmod{2},&y_{i}\in\mathcal{G}_{-}
\end{array}\right. \, .
\eea
We can rewrite this as a system of linear Diophantine equations as
\bea
\tilde{W}_{1i}u_{1} + \dotsb + \tilde{W}_{ri}u_{r}-2q_{i}\equiv\left\{\begin{array}{rl}
0,&y_{i}\in\mathcal{G}_{0}\cup\mathcal{G}_{+}\\
1,&y_{i}\in\mathcal{G}_{-}
\end{array}\right.
\eea
for $q_{i}\in\mathbb{Z}$. The fact that $0 \le u_{i} < 2$ means that $0 \le q_{i} < \tilde{W}_{1i} + \dotsb + \tilde{W}_{ri}$, and thus there are only a finite number of vectors $\mathbf{q}:=(q_{1},...,q_{k'})\in\mathbb{Z}^{k'}$ for which the solvability of this linear system needs to be checked. This can easily be done using standard matrix techniques. If the system has a solution $\mathbf{u}:=(u_{1},...,u_{r})\in\mathbb{Q}^{r}$ for any of the allowed $\mathbf{q}$ vectors, then the set is point-wise fixed under $\sigma$.

One may notice that the generators we have defined above are not all entirely independent. For example, consider an involution $\sigma: x_{1} \leftrightarrow x_{2}, x_{3} \leftrightarrow x_{4}$ such that we get generators $y_{1}=x_{1}x_{2}, y_{2}=x_{3}x_{4}, y_{3}= x_{1}x_{3}+x_{2}x_{4}, y_{4}=x_{1}x_{3}-x_{2}x_{4}$. Then the generators are related via the consistency condition $y_{3}^{2} = y_{4}^{2} + 4y_{1}y_{2}$. This manifests in a constraint on the vanishing of the generators. For example, if $y_{1} = y_{3} = 0$, then $y_{4}$ is necessarily equal to zero. When scanning the subsets $\mathcal{F}$ of vanishing generators, we implement a consistency check to filter out any spurious sets that violate these restrictions.

Before moving forward, it is important to note that it is indeed sufficient to consider only the zero loci of the minimal generators, and not more general combinations. If we have $y_{1},y_{2}\in\mathcal{G}$ with zero loci $D_{1}=\{y_{1}=0\}$ and $D_{2}=\{y_{2}=0\}$, the vanishing of the product is simply the union of the two subvarieties $\{y_{1}y_{2}=0\}=D_{1}\cup D_{2}$. Thus, $D_{1}\cup D_{2}$ is only fixed when $D_{1}$ and $D_{2}$ are already fixed independently. We can also consider the vanishing of the sum (or difference) $\{y_{1}+y_{2}=0\}$, such that both $y_{1}$ and $y_{2}$ have the same $\mathbb{C}^{*}$ weights and the sum has definite parity. If either one vanishes, then both do, and the resulting subvariety lies on $D_{1}\cap D_{2}$, so that clearly both must be fixed independently for their intersection to be fixed. Then, finally, we assume that neither generator vanishes. Because $\sigma$ acts as a homomorphism on the generators, the definite parity condition ensures that $y_{1}$, $y_{2}$, and $y_{1}+y_{2}$ must all have the same parity via
\begin{equation}
 \sigma(y_{1}+y_{2})=\sigma(y_{1})+\sigma(y_{2})=\left\{\begin{array}{rl}
 y_{1}+y_{2},&y_{1},y_{2}\in\mathcal{G}_{0}\cup\mathcal{G}_{+}\\
 -(y_{1}+y_{2}),&y_{1},y_{2}\in\mathcal{G}_{-}
\end{array}\right. \, .
\end{equation}
We saw above that determining whether or not a set is point-wise fixed under $\sigma$ is dependent only on its weights and parity, so $y_{1}+y_{2}$ defines a fixed locus only if $D_{1}$ and $D_{2}$ are fixed as well. The same argument applies to differences $y_{1}-y_{2}$.

%%%%%%%%%%%%%%%%%%%%%%%%%%%%%%%%%%%%%%%%%%%%%%%%%%%%%%%%
\subsubsection{Stanley-Reisner ideal and Calabi-Yau transversality}

All of our calculations have thus far depended only on the toric weights, but eq.(\ref{eq:defamb}) tells us that the ambient space $\cA$ has a set of points $Z$ ruled out by the Stanley-Reisner (SR) ideal $\mathcal{I}_{SR}$. Now that we have found all possible point-wise fixed loci for each subset of generators $\mathcal{F}$, we must check that none of these lie in $Z$. $\mathcal{I}_{SR}$ is defined as the square-free ideal of non-intersections of toric divisors. For example, if $D_{1}D_{2}\in \mathcal{I}_{SR}$, then $d_{12ij}=D_{1}\cdot D_{2}\cdot D_{i} \cdot D_{j}=0$ for all $i,j=1,...,k$. However, the concept works equally well with the projective coordinates. One can say that if $x_{1}x_{2}$ is in the SR ideal, then the simultaneously vanishing set $\{x_{1}=x_{2}=0\}$ will not exist in $\cA$. As our fixed sets will often consist of the zero loci of binomials, compatibility with the SR ideal cannot be generally read off in this way. We check compatibility with the Stanley-Reisner ideal by expressing the ideal and the generators as Boolean expressions. For each generator $y\in\mathcal{G}$, we define a Boolean variable $Y$ whose value is \texttt{True} when $y=0$.

To make this more clear, we give a few examples. The monomial $y_{1}y_{2}$, which is zero when $y_{1}=0$ or $y_{2}=0$, is assigned the Boolean expression $Y_{1}\lor Y_{2}$. We build the expressions for binomials from those of their constituent parts. For example, consider $y_{1}y_{2}-y_{3}y_{4}$. This expression is zero if both $y_{1}y_{2}$ and $y_{3}y_{4}$ are zero, or if neither of them are (which means that none of the individual $y_i=0$). We thus get the Boolean expression
\begin{equation}
y_{1}y_{2}-y_{3}y_{4}\mapsto \left[(Y_{1}\lor Y_{2})\land(Y_{3}\lor Y_{4})\right]\lor \left[(\neg Y_{1}) \land (\neg Y_{2}) \land (\neg Y_{3}) \land (\neg Y_{4})\right]\, .
\end{equation}
By this iterative method, we can create the Boolean expression for any polynomial. Although tedious, this process is easily automated. Using this method, we create the expressions for the SR ideal and the polynomial generators. If these Boolean expressions together form a contradiction, the fixed-point set does not intersect the Calabi-Yau and is thus discarded. Unfortunately, this Boolean approach will not catch every case. In the above example, if $y_{1}y_{2}=y_{3}y_{4}\neq 0$, this method is insufficient and we need something more robust. However, this first pass will rule out many spurious cases.

At least one coordinate in each element of the Stanley-Reisner ideal must be non-zero at each point in $\cA$. By finding the minimal generating sets of coordinates that, when set non-zero, satisfy this condition, we can split $\cA$ up into disjoint regions $U_{i}$. In each of these sectors, we implement this non-zero condition by setting the coordinates equal to unity. By working in these sectors, we ensure that any remaining fixed sets that contradict the SR ideal will be filtered out.

We now check whether each set can be restricted to the Calabi-Yau hypersurface. For a given fixed set, we compute in each sector $U_{i}$ the dimension of the ideal generated by the Calabi-Yau polynomial $P_{symm}$ and the fixed set generators $\mathcal{F}\equiv\{y_{1},...,y_{p}\}$
\begin{equation}
\label{eq:fixed}
\cI^{fixed}_{ip}=\langle U_{i},P_{symm},y_{1},...,y_{p}\rangle \, .
\end{equation}
If the dimension $\text{dim }\cI^{fixed}_{ip}<0$ for all $U_{i}$, then there is no Groebner basis, and we conclude that this set does not intersect the CY hypersurface.

For each set that is not discarded, we repeat this calculation for the ideal with one fixed set generator $\text{dim }\cI^{fixed}_{i1}$, and then two $\text{dim }\cI^{fixed}_{i2}$, etc. until $\text{dim }\cI^{fixed}_{i\ell}=\text{dim }\cI^{fixed}_{ip}$ when adding more generators to the ideal no longer changes the dimension for any region $U_{i}$\footnote{Note that while this procedure will yield the correct dimension of the fixed point locus, it may occasionally overlook the existence of non-transversally intersecting generators. This may cause some points to be falsely identified as fixed, but does not affect the overall dimension of the set. Since, however, this will not lead to a misidentification of orientifold planes, a more thorough treatment will be left to future work.} Then, the intersection $\{y_{1}=\cdots =y_{\ell}=0\}$ of these generators gives the final point-wise fixed locus, with redundancies eliminated. Since each generator is codimension-1 in $X$, the length $\ell$ of this set of generators is the codimension in $X$ of the fixed point locus. An O3 plane corresponds to a codimension-3 point-wise fixed subvariey, an O5 plane has codimension-2, an O7 plane has codimension-1, and an O9 plane has codimension-0 (which means the entire CY is fixed under $\sigma$).

%%%%%%%%%%%%%%%%%%%%%%%%%%%%%%%%%%%%%%%%%%%%%%%%%%%%%%%%
\subsubsection{Smoothness}

Finally, we check whether the invariant Calabi-Yau hypersurface defined by $P_{symm}$ is smooth.  This is  important to determine whether an involution is a free action. We do this by checking if there is any solution to the condition $P_{symm}=dP_{symm}=0$ that is not ruled out by the Stanley-Reisner ideal. In practice, this is done by setting up the ideals
\begin{equation}
\label{eq:smooth}
\cI^{smooth}_{i}=\langle U_{i},\; P_{symm},\; \frac{\partial P_{symm}}{\partial x_{1}},\; ...,\; \frac{\partial P_{symm}}{\partial x_{k}}\rangle
\end{equation}
\noindent for each region $U_{i}$ allowed by the Stanley-Reisner ideal, and computing the dimension. If $\text{dim }\cI^{smooth}_{i}<0$ for all $U_{i}$, then the invariant Calabi-Yau hypersurface is smooth.
If no O-planes exist and the invariant Calabi-Yau hypersurface is smooth, then the involution defines a $\mathbb{Z}_{2}$ free action on $X$. 

\subsubsection{Tadpole cancelation and string vacua}
Since we are interested in the orientifold Type IIB string vacua, it is reasonable to consider the possible tadpole cancelation in the $O3/O7$ system.
In this paper we are not concerned with the physics on the $D7$-branes and for the concrete orientifolds we will always cancel the $D7$-brane tadpole by simply placing eight $D7$-branes on top of the $O7$-plane. Clearly, for concrete model building this simple assumption has to be relaxed.
The $O7$-planes and the $D7$-branes induce a $D3$-brane tadpole condition which simplifies to
\eq{
\label{eq:tadpole}
      N_{D3} + \frac{N_{\text{flux}}}{2}+ N_{\rm gauge}= \frac{N_{O3}}{4}+\frac{\chi(D_{O7})}{4}\, \equiv - \, Q_{D3}^{loc}.
}
with $N_{\text{flux}}=\frac{1}{(2\pi)^4 \alpha^{'2}}\int H_3\wedge F_3 $, $N_{\text{gauge}}=-\sum_{a} \frac{1}{8\pi^2} \int_{D_a} \text{tr}{\cal F}_a^2$, and $N_{D3}$, $N_{O3}$ the number of $D3$-branes, $O3$-planes respectively. The $D3$-tadpole cancelation condition requires the total D3-brane charge $Q_{D3}^{loc}$ of the seven-brane stacks and $O3$-planes to be an integer.
It will serve as a  consistency check that this number is indeed an integer. If the involution passes this naive tadpole cancellation check, we will denote our geometry as a \lq\lq  naive orientifold  Type IIB string vacua".  Of course, when an involution results in a geometry with only $O7$-planes, it can automatically be classified as a naive string vacuum. 
The classification of these orientifold involutions is summarized in Tables~\ref{tab:exchangegeom}-\ref{tab:involutiongeom}.

%
%In Section \ref{sec:example}, we demonstrate this algorithm on a few explicit examples.

%%%%%%%%%%%%%%%%%%%%%%%%%%%%%%%%%%%%%%%%%%%%%%%%%%%%%%%%
%%%%%%%%%%%%%%%%%%%%%%%%%%%%%%%%%%%%%%%%%%%%%%%%%%%%%%%%
\subsection{Hodge Number Splitting}
\label{subsec:splitting}

The orientifold involution $\sigma$ exchanges two projective coordinates on $X$.  
The holomorphicity condition requires that the pullback $\sigma^{*}$ maps $(p,q)$-forms on $X$ to $(p,q)$-forms. This is also true at the level of cohomology as the Dolbeault operator $\bar\partial$ commutes with the pullback $\sigma^*$. This implies that in the orientifold limit, 
the dimensions of Hodge numbers split as:
\bea
\label{eq:hodgesplit}
H^{p,q}(X/\sigma^{*})=H^{p,q}_{+}(X/\sigma^{*}) \, \oplus H^{p,q}_{-}(X/\sigma^{*})\, .
\eea

Because we work only with favorable geometries, the calculations are simplified by the fact that the toric classes of ambient space $\cA$ always restrict in a straightforward way to the Calabi-Yau hypersurface, and the divisor classes are the same, as shown in eq.(\ref{eq:favor}). In order to determine the $h^{1,1}_\pm (X/\sigma^*)$ splitting,  we can always expand the K\"ahler form in terms of these divisor classes.

To illustrate, we choose a toy example with $h^{1,1}(X)=3$, admitting a proper orientifold involution $\sigma^{*}: D_{2}\leftrightarrow D_{3}$. Suppose the divisor classes $\{D_1, D_2, D_3\}$ form a basis for $H^{1,1}(X;\mathbb{Z})$. Then, the K\"ahler form can be expanded as the linear combination
\begin{equation}\label{eq:kform}
J= t_{1}J_{1}+t_{2}J_{2}+t_{3}J_{3} = t_{1}D_{1}+t_{2}D_{2}+t_{3}D_{3}\, ,
\end{equation}
with $t_{1},t_{2},t_{3}\in\mathbb{Z}$. But the K\"ahler form must obey the constraint of even parity under the orientifold involution, and must therefore only have components in $H^{1,1}_+(X)$, so that
\bea\label{eq:kformswap}
J=\sigma^{*}J =t_{1}D_{1}+t_{2}D_{3}+t_{3}D_{2} =t_{1}J_{1}+t_{3}J_{2}+t_{2}J_{3}\, .
\eea
Then, comparing eq.(\ref{eq:kform}) and eq.(\ref{eq:kformswap}), we note that we must have $t_{2}=t_{3}=t_{+}$, for some $t_{+}\in\mathbb{Z}$. Defining the even and odd parity eigendivisors $D_{\pm}=D_{2}\pm D_{3}$, we can write
\begin{equation}
J=t_{1}D_{1}+t_{+}D_{+}\, .
\end{equation}

Thus, on the orientifold $X/\sigma^{*}$, there are only two independent directions in the K\"ahler moduli space, $D_{1}$ and $D_{+}$, while $D_{-}\in H^{1,1}_{-}(X/\sigma^{*})$ does not appear in the K\"ahler form. We therefore find the equivariant cohomology $h^{1,1}_{+}(X/\sigma^{*})=2 $ and $h^{1,1}_{-}(X/\sigma^{*})=1$. This can only be done in a consistent way when $\sigma$ is a proper involution\footnote{In fact, in the case of a proper involution, it is extremely rare that we can write down an involution exchanging only a single pair of NIDs. However, for clarity, we use this simple toy example only to illustrate the underlying concept.} respecting the linear ideal $\mathcal{I}_{lin}$, which addresses the redundancies among the toric divisor classes. Since the involution exchanges pairs of NIDs, one can always expand the K\"ahler form in the orientifold-invariant basis, depending on some pairs of the divisors involved in the involution.

 %Then, the following formula applies:
%
% \begin{align}\label{eq:euler}
 %\chi(X/\sigma^{*})&=\frac{\chi(X)}{2}=h^{1,1}(X)-h^{2,1}(X)\notag\\
 %&=2\left(h^{1,1}_{+}-h^{2,1}_{+}\right) \, .
% \end{align}
 %
%We then find $h^{2,1}_+= h^{1,1}_+ - \frac{\chi(X)}{4}$. 

  In addition to the even and odd parity splitting of $h^{1,1}(X/\sigma^{*})$,  in principle we can apply the Lefschetz fixed point theorem (for a brief introduction, see \cite{Shanahan} and the Appendix of \cite{Blumenhagen:2010ja}) to determine the $h^{2,1}_-(X/\sigma^*) $ in the $\IZ_2$-orbifold limit. 
In general the $\IZ_2$ involution $\sigma$ induces a fixed-point set $ {\cal F}$.  Due to the hodge number splitting eq.(\ref{eq:hodgesplit}), we can 
define the Leftschetz number of $\sigma^*$ as $ L(\sigma^*, X) $:
%Defining an induced action on the De Rham cohomology of $M$, $H^*(M)$, such that $H^i = H^i_+\oplus H^i_-$, then we have
\bea
  L(\sigma^*, X) \equiv \sum_i (-)^i (b^i_+-b^i_-) = \chi({\cal F})\,,
\eea
where  $b_\pm^i$ are the split Betti numbers and the the right-hand side is  the Euler number of the fixed locus ${\cal F}$. 
There is a very useful theorem to calculate the Euler number of the $\IZ_2$-orbifold space:
\bea
\label{eq:chi}
    \chi(X/{\sigma^*}) =  \frac{1}{2}\,\big(L(\sigma^*, X) + \chi(X)\big) \, = \sum_i  (-)^i (b^i_+),
\eea
which is the average of the Lefschetz number and the Euler number of $X$. 
For the divisor exchange involution $\sigma^*$, we have further:\footnote{For a reflection $\sigma$, the formula changes to $h^{2,1}_- = h^{1,1}_- + \frac{L(\sigma^*, X)  - \chi (X) }{4}  - 1$. The difference is due to the fact  that the equivariant cohomology fora  reflection is $h^{\bullet}(X, \cO) = \{1_+, 0, 0, 1_-\}$ while for a divisor exchange involution it is $h^{\bullet}(X, \cO) = \{1_+, 0, 0, 1_+\}$.}
\bea
\label{eq:h21split}
h^{2,1}_-  (X/\sigma^*) = h^{1,1}_-  (X/\sigma^*) + \frac{L(\sigma^*, X)  - \chi (X) }{4} ,
\eea
where
\begin{eqnarray}
\label{eq:lefschtz}
 L(\sigma^*, X) = \chi(\cal F)  \supset \begin{cases}
                    \, \, \chi(O7) = \int_{O7} \,c_2(O7)\, ,  \\
                    \, \, \chi(O5) = \int_{O5} \,c_1(O5)\, , \\
                   \,\,   \chi(O3) = \int_{O3} \,c_0(O3) \,=  N_{O3}\,  .
\end{cases}
\end{eqnarray}
It is interesting to see that for an $O3/O7$-system with eight $D7$ on top of the $O3$-planes, the Lefschetz number divided by 4 is exactly the $Q_{D3}^{loc}$ in eq.(\ref{eq:tadpole}) for tadpole cancellation.  
%a few $h^{2,1}_-$ get from eq.(\ref{eq:h21split}) may be negative or fractional even when the tadpole cancelation work due to the fact that the Euler number of the original Calabi-Yau manifold can't be divided by 4.  Of course, once people resolve the singularity to get a smooth geometry, the new $h^{2,1}_-$ of the resolved geometry should be non-negative  integer. This can also be seen in the CICY case \cite{Carta:2020ohw} and we left it for a future work.  

If the involution is a free action, $L(\sigma^*, X) = 0$ and eq.(\ref{eq:chi}) reduces to the standard form in which the Euler characteristic of a free action quotient group obeys $\chi(X/\sigma^{*})=\chi(X)/\left\lvert\sigma^{*}\right\rvert$, where $|\sigma^{*}| = 2$ is the order of the group action defined by $\sigma^{*}\cong\mathbb{Z}_{2}$.
These results are enumerated in Table \ref{tab:involutiongeom}.

However, the full Hodge number such as $h^{2,1}_{\pm}$ is only well-defined in the smooth case.  In the orbifold limit, $h^{2,1}_{\pm}$ may contain some ambiguity - for example, it may be negative or fractional even when the tadpole cancellation is satisfied. So one can consider resolving the possible conifold singularities in the orbifold limit to get  a smooth geometry and therefore a well-defined  $h^{2,1}_{\pm}$.  On the other hand, the $h^{1,1}_-$ we obtain are robust and not affected by blowing up the singularity. Since most of the string model-building in which people are interested is in the orbifold limit, in this paper we only present the robust $h^{1,1}_-(X/\sigma^*)$ results and leave determining $h^{2,1}_{\pm}(X/\sigma^*)$ in the smooth case for a future work.

%%%%%%%%%%%%%%%%%%%%%%%%%%%%%%%%%%%%%%%%%%%%%%%%%%%%%%%%
%%%%%%%%%%%%%%%%%%%%%%%%%%%%%%%%%%%%%%%%%%%%%%%%%%%%%%%%
%%%%%%%%%%%%%%%%%%%%%%%%%%%%%%%%%%%%%%%%%%%%%%%%%%%%%%%%
\section{Illustrative Examples of the Algorithm}
\label{sec:example}

In this section, we demonstrate three explicit examples of finding and classifying the point-wise fixed sets of a Calabi-Yau orientifold, following the method described in the previous section. The example in Section \ref{subsec:nontriv} has nonempty $\mathcal{G}_{+}, \mathcal{G}_{-}$ and demonstrates our method in full. Additionally, the involution allows us to compute the split Hodge numbers on the orientifold. The example in Section \ref{subsec:free1} demonstrates a potential free action which has fixed points in the ambient space, but not on the Calabi-Yau hypersurface itself. Finally, the example in Section \ref{subsec:free3} demonstrates the only proper, smooth, geometry-wide free action found by our scan.

%%%%%%%%%%%%%%%%%%%%%%%%%%%%%%%%%%%%%%%%%%%%%%%%%%%%%%%%
%%%%%%%%%%%%%%%%%%%%%%%%%%%%%%%%%%%%%%%%%%%%%%%%%%%%%%%%
\subsection{Proper Involution with O3 and O7 planes}
\label{subsec:nontriv}

%%%%%%%%%%%%%%%%%%%%%%%%%%%%%%%%%%%%%%%%%%%%%%%%%%%%%%%%
Our first example is from the database of Calabi-Yau threefolds at \dburl with Hodge numbers $h^{1,1}(X)=4,\;\; h^{2,1}(X)=64$. It can be identified by its index in the database: 
\begin{equation*}
\begin{array}{|c|c|c|}
\hline
\text{Polytope ID}&\text{Geometry ID}&\text{Triangulation ID}\\
\hline
566&1&1\\
\hline
\end{array}
\end{equation*}
\noindent Although we will examine this example within the context of this particular triangulation, this involution is valid in each chamber of this geometry's K\"ahler cone. This example defines an MPCP desingularized ambient toric variety with weight matrix $\mathbf{W}$ given by
\bea
\label{eq:toricdata1}
\begin{array}{|c|c|c|c|c|c|c|c|}
\hline
x_{1} & x_{2} & x_{3} & x_{4} & x_{5} & x_{6} & x_{7} & x_{8}\\\hline
0 & 0 & 0 & 1 & 0 & 1 & 0 & 0 \\\hline
0 & 0 & 1 & 0 & 0 & 0 & 1 & 0 \\\hline
0 & 1 & 0 & 0 & 1 & 0 & 0 & 1 \\\hline
1 & 0 & 0 & 1 & 0 & 0 & 1 & 1 \\\hline
\end{array}
\eea
\noindent and Stanley-Reisner (SR) ideal
\bea
\cI_{SR}=\langle x_{1}x_{8},\; x_{3}x_{7},\; x_{4}x_{6},\; x_{1}x_{4}x_{7},\; x_{2}x_{3}x_{5},\; x_{2}x_{5}x_{6},\; x_{2}x_{5}x_{8}\, \rangle.
\eea
The Hodge numbers of the corrsponding individual toric divisors $D_i \equiv \{x_i = 0\} $ are
\bea\label{eq:divhodge1}
h^{\bullet}(D_{1})&=\{1,0,0,9\}\notag\\
h^{\bullet}(D_{2})=h^{\bullet}(D_{4})=h^{\bullet}(D_{5})=h^{\bullet}(D_{7})&=\{1,0,1,21\}\notag\\
h^{\bullet}(D_{3})=h^{\bullet}(D_{6})&=\{1,0,0,12\}\\
h^{\bullet}(D_{8})&=\{1,0,2,30\}\notag \, .
\eea

For this example, there is only one proper involution exchanging NIDs, which leaves the SR ideal together with Linear ideal invariant, and thereby  intersection numbers invariant. This proper NID involution is formed by exchanging two pair of coordinates:
\bea
\label{eq:nontrivinvol}
\sigma: \,\,x_{3} \leftrightarrow x_{6}, \quad x_{4} \leftrightarrow x_{7}\, .
\eea
From eq.(\ref{eq:divhodge1}) we see that this is an exchange of two dP$_{9}$ divisors and two exact Wilson divisors.

As is necessary for a consistent orientifold, the volume form eq.(\ref{eq:3form}) has a definite parity under $\sigma$. For completeness, we demonstrate our calculation of the parity. For this example, the vector fields eq.(\ref{eq:V}) can be read off from the toric data eq.(\ref{eq:toricdata1}), and are:
\begin{align}
\notag
V^{1} &= x_{4} \frac{\partial}{\partial x_{4}} + x_{6} \frac{\partial}{\partial x_{6}} \\ \notag
V^{2} &= x_{3} \frac{\partial}{\partial x_{3}} + x_{7} \frac{\partial}{\partial x_{7}}  \\
V^{3} &= x_{2} \frac{\partial}{\partial x_{2}} + x_{5} \frac{\partial}{\partial x_{5}} + x_{8} \frac{\partial}{\partial x_{8}} \\ \notag
V^{4} &= x_{1} \frac{\partial}{\partial x_{1}} + x_{4} \frac{\partial}{\partial x_{4}} + x_{7} \frac{\partial}{\partial x_{7}} + x_{8} \frac{\partial}{\partial x_{8}}\, .
\end{align}
Performing the contractions with $\omega$ is straightforward, and yields
{\footnotesize
\begin{align}
\notag
\mathcal{Q}   = & \,\,x_{5} x_{6} x_{7} x_{8} dx_{1} \wedge dx_{2} \wedge dx_{3} \wedge dx_{4} + x_{4} x_{6} x_{7} x_{8} dx_{1} \wedge dx_{2} \wedge dx_{3} \wedge dx_{5} - x_{4} x_{5} x_{7} x_{8} dx_{1} \wedge dx_{2} \wedge dx_{3} \wedge dx_{6} \nonumber\\
& - x_{4} x_{5} x_{6} x_{7} dx_{1} \wedge dx_{2} \wedge dx_{3} \wedge dx_{8} + x_{3} x_{6} x_{7} x_{8} dx_{1} \wedge dx_{2} \wedge dx_{4} \wedge dx_{5} + x_{3} x_{5} x_{6} x_{8} dx_{1} \wedge dx_{2} \wedge dx_{4} \wedge dx_{7} \nonumber\\
& - x_{3} x_{5} x_{6} x_{7} dx_{1} \wedge dx_{2} \wedge dx_{4} \wedge dx_{8} + x_{3} x_{4} x_{7} x_{8} dx_{1} \wedge dx_{2} \wedge dx_{5} \wedge dx_{6} + x_{3} x_{4} x_{6} x_{8} dx_{1} \wedge dx_{2} \wedge dx_{5} \wedge dx_{7} \nonumber\\
& - x_{3} x_{4} x_{5} x_{8} dx_{1} \wedge dx_{2} \wedge dx_{6} \wedge dx_{7} + x_{3} x_{4} x_{5} x_{7} dx_{1} \wedge dx_{2} \wedge dx_{6} \wedge dx_{8} + x_{3} x_{4} x_{5} x_{6} dx_{1} \wedge dx_{2} \wedge dx_{7} \wedge dx_{8} \nonumber\\
& - x_{2} x_{6} x_{7} x_{8} dx_{1} \wedge dx_{3} \wedge dx_{4} \wedge dx_{5} - x_{2} x_{4} x_{7} x_{8} dx_{1} \wedge dx_{3} \wedge dx_{5} \wedge dx_{6} - x_{2} x_{4} x_{6} x_{7} dx_{1} \wedge dx_{3} \wedge dx_{5} \wedge dx_{8} \nonumber\\
& + x_{2} x_{3} x_{6} x_{8} dx_{1} \wedge dx_{4} \wedge dx_{5} \wedge dx_{7} - x_{2} x_{3} x_{6} x_{7} dx_{1} \wedge dx_{4} \wedge dx_{5} \wedge dx_{8} + x_{2} x_{3} x_{4} x_{8} dx_{1} \wedge dx_{5} \wedge dx_{6} \wedge dx_{7} \nonumber\\
& - x_{2} x_{3} x_{4} x_{7} dx_{1} \wedge dx_{5} \wedge dx_{6} \wedge dx_{8} - x_{2} x_{3} x_{4} x_{6} dx_{1} \wedge dx_{5} \wedge dx_{7} \wedge dx_{8} - x_{1} x_{6} x_{7} x_{8} dx_{2} \wedge dx_{3} \wedge dx_{4} \wedge dx_{5} \nonumber\\
& + x_{1} x_{5} x_{6} x_{7} dx_{2} \wedge dx_{3} \wedge dx_{4} \wedge dx_{8} - x_{1} x_{4} x_{7} x_{8} dx_{2} \wedge dx_{3} \wedge dx_{5} \wedge dx_{6} - x_{1} x_{4} x_{5} x_{7} dx_{2} \wedge dx_{3} \wedge dx_{6} \wedge dx_{8} \nonumber\\
& + x_{1} x_{3} x_{6} x_{8} dx_{2} \wedge dx_{4} \wedge dx_{5} \wedge dx_{7} + x_{1} x_{3} x_{5} x_{6} dx_{2} \wedge dx_{4} \wedge dx_{7} \wedge dx_{8} + x_{1} x_{3} x_{4} x_{8} dx_{2} \wedge dx_{5} \wedge dx_{6} \wedge dx_{7} \nonumber\\
& - x_{1} x_{3} x_{4} x_{5} dx_{2} \wedge dx_{6} \wedge dx_{7} \wedge dx_{8} - x_{1} x_{2} x_{6} x_{7} dx_{3} \wedge dx_{4} \wedge dx_{5} \wedge dx_{8} - x_{1} x_{2} x_{4} x_{7} dx_{3} \wedge dx_{5} \wedge dx_{6} \wedge dx_{8} \nonumber\\
& + x_{1} x_{2} x_{3} x_{6} dx_{4} \wedge dx_{5} \wedge dx_{7} \wedge dx_{8} + x_{1} x_{2} x_{3} x_{4} dx_{5} \wedge dx_{6} \wedge dx_{7} \wedge dx_{8} \nonumber\, .
\end{align}
}
With this explicit expression for $\mathcal{Q}$ in terms of the $x_{i}$, we can find its image under $\sigma$ by exchanging $\{x_{3} \leftrightarrow x_{6},\,x_{4} \leftrightarrow x_{7}; \,\,dx_{2} \leftrightarrow dx_{4},\, dx_{7} \leftrightarrow dx_{8}\}$. As an example, the first term in $\mathcal{Q}$ is $x_{5} x_{6} x_{7} x_{8} dx_{1} \wedge dx_{2} \wedge dx_{3} \wedge dx_{4}$. After applying $\sigma$, this term becomes $x_{3} x_{4} x_{5} x_{8} dx_{1} \wedge dx_{2} \wedge dx_{6} \wedge dx_{7}$. This is the negative of the first term in the fourth line of the expression for $\mathcal{Q}$ above. These calculations, as well as the preceding contractions, are tedious to perform but easily automated. We find that the resulting form is exactly the negative of $\mathcal{Q}$, and thus the volume form has odd parity under $\sigma$, i.e. $\sigma^* \Omega_3 = - \, \Omega_3$. So we would expect if there are any fixed points under the involution, it should be $O3$ or $O7$-planes, or both.

\subsubsection*{Orientifold Planes}

Since the projective coordinates $x_{1}$, $x_{2}$, $x_{5}$, and $x_{8}$ are not affected by the involution, they are included in our list of (anti-)invariant polynomial generators
\bea
\mathcal{G}_{0}=\{x_{1},x_{2},x_{5},x_{8}\}\, .
\eea
If we define $\sigma_{1}:x_{3} \leftrightarrow x_{6}$ and $\sigma_{2}:x_{4} \leftrightarrow x_{7}$, such that $\sigma=\sigma_{1}\circ\sigma_{2}$, then we have the three sub-involutions: $\sigma_{1}$, $\sigma_{2}$, and $\sigma$. We now explore these cases:
\begin{enumerate}
\item $\mathbf{\sigma_{1}:x_{3} \leftrightarrow x_{6}}$\\
Because we only consider NIDs, $x_{3}$ and $x_{6}$ have different weights and cannot be combined into a homogenous binomial. Thus, we are left only with the invariant monomial $x_{3}x_{6}$.\\
$\Rightarrow\hspace{5mm}\mathcal{G}_{+}=\{x_{3}x_{6}\},\;\;\mathcal{G}_{-}=\emptyset $

\item $\mathbf{\sigma_{2}:x_{4} \leftrightarrow x_{7}}$\\
For the same reason, we are left only with the invariant monomial $x_{4}x_{7}$.\\
$\Rightarrow\hspace{5mm}\mathcal{G}_{+}=\{x_{7}x_{8}\},\;\;\mathcal{G}_{-}=\emptyset $

\item $\mathbf{\sigma: x_{3} \leftrightarrow x_{6},\;\; x_{4} \leftrightarrow x_{7}}$\\
Any invariant monomials in this case are just products of the ones we found above. However, we must now consider binomial generators of the form
\begin{equation}
x_{3}^{m}x_{4}^{n} \pm x_{6}^{m}x_{7}^{n}
\end{equation}
for $m,n\in\mathbb{Z}$. The homogeneity of this binomial is determined by the following condition on the weights
\begin{equation}
m(\mathbf{W}_{i3}-\mathbf{W}_{i4})+n(\mathbf{W}_{i6}-\mathbf{W}_{i7})=\mathbf{0}\, .
\end{equation}
Thus, the vector $(m,n)\in\mathbb{Z}^{2}$ lies in the kernel of the matrix of difference vectors
\begin{equation}
\begin{array}{|c|c|}
\hline
\mathbf{W}_{i2}-\mathbf{W}_{i4} & \mathbf{W}_{i7}-\mathbf{W}_{i8}\\\hline
\hline
-1 & 1\\\hline
1 & -1\\\hline
0 & 0\\\hline
0 & 0\\\hline
\end{array}
\end{equation}
The kernel is generated by the vector $(m,n)=(1,1)$, so that our binomial generators\footnote{Note that if the kernel was generated by $(m,n)=(1,-1)$, then the binomial generators would be given by $x_{3}x_{4}^{-1} \pm x_{6}x_{7}^{-1}$. If we multiply through by $x_{4}x_{7}$, we get $x_{3}x_{7} \pm x_{4}x_{6}$. This just implies that we assumed the wrong positions of $x_{4}$ and $x_{7}$, so they must be switched in the binomial.} are given by $x_{3}x_{4} \pm x_{6}x_{7}$. This implies that $\mathcal{G}_{+}=\{x_{3}x_{4}+x_{6}x_{7}\}$ and $\mathcal{G}_{-}=\{x_{3}x_{4}-x_{6}x_{7}\}$.
\end{enumerate}
Note that all the (anti-)invariant polynomial generations in $\mathcal{G}=\mathcal{G}_{0}\cup\mathcal{G}_{+}\cup\mathcal{G}_{-}$ are given by
\begin{equation}\label{eq:segre}
\begin{gathered}
y_{1}=x_{1},\;\;\; y_{2}=x_{2},\;\;\; y_{3}=x_{5},\;\;\; y_{4}=x_{8},\;\;\; y_{5}=x_{3}x_{6},\\
%\\
y_{6}=x_{4}x_{7},\;\;\; y_{7}=x_{3}x_{4}+x_{6}x_{7},\;\;\; y_{8}=x_{3}x_{4}-x_{6}x_{7}\, .
\end{gathered}
\end{equation}

This coordinate transformation defines the Segre embedding with consistency condition $y_{7}^{2}=y_{8}^{2}+4y_{5}y_{6}$ and new weight matrix
\begin{equation}
\label{eq:toricdata2}
\begin{array}{|c|c|c|c|c|c|c|c|l}
\cline{1-8}
y_{1} & y_{2} & y_{3} & y_{4} & y_{5} & y_{6} & y_{7} & y_{8} & \\\cline{1-8}
0 & 0 & 0 & 0 & 1 & 1 & 1 & 1 & (1)\\\cline{1-8}
0 & 0 & 0 & 0 & 1 & 1 & 1 & 1 & (2)\\\cline{1-8}
0 & 1 & 1 & 1 & 0 & 0 & 0 & 0 & (3)\\\cline{1-8}
1 & 0 & 0 & 1 & 0 & 2 & 1 & 1 & (4)\\\cline{1-8}
\end{array}
\end{equation}
However, this matrix only has rank 3, and is redundant. We see that row $(3)=2\times (2)-3\times (1)$, so we eliminate this row
\bea
\label{eq:toricdata3}
\begin{array}{|c|c|c|c|c|c|c|c|l}
\cline{1-8}
y_{1} & y_{2} & y_{3} & y_{4} & y_{5} & y_{6} & y_{7} & y_{8} & \\\cline{1-8}
0 & 0 & 0 & 0 & 1 & 1 & 1 & 1 & \lambda_1\\\cline{1-8}
0 & 1 & 1 & 1 & 0 & 0 & 0 & 0 & \lambda_2\\\cline{1-8}
1 & 0 & 0 & 1 & 0 & 2 & 1 & 1 & \lambda_3\\\cline{1-8} 
\end{array}
\eea

The involution $\sigma$ can be rewritten simply in these coordinates as $y_{8}\mapsto -y_{8}$. Thus, we see that $F_{1}=\{y_{8}=0\}$ is a point-wise fixed, codimension-1 subvariety. This defines an O7 plane on the orientifold of $X$. By inspecting eq.(\ref{eq:nontrivinvol}), we may, in addition, naively assume the fixed point set $F_{0}=\{x_{3}=x_{4}=x_{6}=x_{7}=0\}$, which is trivially unaffected by the involution. However, $F_{0}$ is a subset of $F_{1}$, and is therefore redundant. The redundancy of this trivial fixed point set is a general feature of any involution which admits odd parity binomial generators in $\mathcal{G}_{-}$.

By taking advantage of the toric degrees of freedom, however, there may be other non-trivial fixed sets in addition to $F_{1}$. We therefore check whether any subset $\mathcal{F}$ of the generators can neutralize the odd parity of $y_{8}$, becoming fixed themselves in the process. This scan can be simplified by noting that if a set of points is not fixed, then neither is any set containing it. Thus, if the simultaneous vanishing of a set of generators is not fixed, then neither is the vanishing of any subset. We therefore begin our scan with the largest set of generators and work our way down. The largest set we can choose has four generators, since their simultaneous vanishing defines a set of isolated points on the ambient space $\cA$.

Consider the subset $\{y_{1},y_{2},y_{3},y_{7}\}\equiv\mathcal{F}_{2}\subset\mathcal{G}$. In order for the locus $F_{2}=\{y_{1}=y_{2}=y_{3}=y_{7}=0\}$ to be fixed, we must use the three independent $\mathbb{C}^{*}$ actions to neutralize the odd parity of $y_{8}$ while leaving everything else invariant (as $y_{8}$ is the only non-zero generator with negative parity). This constraint is defined by the toric equivalence class
\begin{align}
(y_{4},y_{5},y_{6},-y_{8})&\sim (\lambda_{2}\lambda_{3} y_{4},\lambda_{1} y_{5},\lambda_{1}\lambda_{3}^{2}y_{6},\lambda_{1}\lambda_{3}y_{8})\notag\\
&=(y_{4},y_{5},y_{6},y_{8})
\end{align}
where $\lambda_{1},\lambda_{2},\lambda_{3}\in\mathbb{C}^{*}$. This can be written more simply as the system of multiplicative equations
\begin{align}
\lambda_{2}\lambda_{3} &=1&\lambda_{1} &=1\notag\\
\lambda_{1}\lambda_{3}^{2} &=1&\lambda_{1}\lambda_{3} &=-1\, .
\end{align}
Simple inspection reveals that this system can clearly be solved with solution $(\lambda_{1},\lambda_{2},\lambda_{3})=(1,-1,-1)$, but we will continue with our algorithm, as the computer algorithm, following our analysis in a systematic manner, would not be able to arrive at this conclusion at this juncture.

By defining $\lambda_{1}=e^{i\pi u}$, $\lambda_{2}=e^{i\pi v}$, $\lambda_{3}=e^{i\pi w}$ as explicit values on the complex unit circle with $u,v,w\in\mathbb{Q}$, and noting that $0\leq u,v,w <2$ is the primitive domain of unique values, we can replace the multiplicative equations with linear congruences
\begin{align}
v+w &\equiv 0\pmod{2}&u&\equiv 0\pmod{2}\notag\\
u+2w&\equiv 0\pmod{2}&u+w&\equiv 1\pmod{2} \, .
\end{align}
Finally, by adding in the auxiliary variables $q_{1},...,q_{4}\in\mathbb{Z}$, we can write this as a linear system of Diophantine equalities
\begin{align}\label{eq:dioph}
v+w-2q_{1} &=0&u-2q_{2}&=0\notag\\
u+2w-2q_{7}&=0&u+w-2q_{8}&=1 \, .
\end{align}
Since $0\leq u,v,w <2$, we notice that the vector $(q_{1},...,q_{4})\in\mathbb{Z}^{4}$ is in the lattice $\Lambda\subset\mathbb{Z}^{4}$ defined by
\begin{equation}
\begin{gathered}
0\leq q_{1},q_{4}<2\hspace{8mm}0\leq q_{2}<1\hspace{8mm}0\leq q_{3}<3
\end{gathered}\, .
\end{equation}
This gives us a finite range to scan over, and the search is relatively fast. For each point $(q_{1},...,q_{4})\in\Lambda$, we search for a solution to eq.(\ref{eq:dioph}). If any solution is found for any lattice point, then the set of generators $\mathcal{F}$ has a point-wise fixed vanishing locus. As we saw earlier, this particular system can indeed be solved, and so $F_{2}=\{y_{1}=y_{2}=y_{3}=y_{7}=0\}$ defines a point-wise fixed set. We do not find any additional point-wise fixed sets that are not subsets of $F_{1}$ and $F_{2}$.

%%%%%%%%%%%%%%%%%%%%%%%%%%%%%%%%%%%%

We must now check that the fixed sets intersect the Calabi-Yau hypersurface transversally, so that they are not redundant. The orientifold-symmetric Calabi-Yau polynomial has the form
\begin{align}
P_{symm}&=a_{1}x_{1}^{4}x_{2}^{3}x_{3}^{2}x_{6}^{2}+a_{2}x_{1}^{4}x_{2}^{2}x_{3}^{2}x_{5}x_{6}^{2}+a_{3}x_{1}^{4}x_{2}x_{3}^{2}x_{5}^{2}x_{6}^{2}+a_{4}x_{1}^{4}x_{3}^{2}x_{5}^{3}x_{6}^{2} +a_{5}x_{1}^{3}x_{2}^{3}x_{3}^{2}x_{4}x_{6}\nonumber\\
&+a_{5}x_{1}^{3}x_{2}^{3}x_{3}x_{6}^{2}x_{7}+a_{6}x_{1}^{3}x_{2}^{2}x_{3}^{2}x_{4}x_{5}x_{6} +a_{6}x_{1}^{3}x_{2}^{2}x_{3}x_{5}x_{6}^{2}x_{7}+a_{7}x_{1}^{3}x_{2}x_{3}^{2}x_{4}x_{5}^{2}x_{6}\nonumber\\
&+a_{7}x_{1}^{3}x_{2}x_{3}x_{5}^{2}x_{6}^{2}x_{7} +a_{8}x_{1}^{3}x_{3}^{2}x_{4}x_{5}^{3}x_{6}+a_{8}x_{1}^{3}x_{3}x_{5}^{3}x_{6}^{2}x_{7}+a_{9}x_{1}^{3}x_{2}^{2}x_{3}^{2}x_{6}^{2}x_{8}\nonumber\\
&+a_{10}x_{1}^{3}x_{2}x_{3}^{2}x_{5}x_{6}^{2}x_{8}+a_{11}x_{1}^{3}x_{3}^{2}x_{5}^{2}x_{6}^{2}x_{8}+a_{12}x_{1}^{2}x_{2}^{3}x_{3}^{2}x_{4}^{2} +a_{12}x_{1}^{2}x_{2}^{3}x_{6}^{2}x_{7}^{2}\nonumber\\
&+a_{13}x_{1}^{2}x_{2}^{2}x_{3}^{2}x_{4}^{2}x_{5}+a_{13}x_{1}^{2}x_{2}^{2}x_{5}x_{6}^{2}x_{7}^{2} +a_{14}x_{1}^{2}x_{2}x_{3}^{2}x_{4}^{2}x_{5}^{2}+a_{14}x_{1}^{2}x_{2}x_{5}^{2}x_{6}^{2}x_{7}^{2}\nonumber\\
&+a_{15}x_{1}^{2}x_{3}^{2}x_{4}^{2}x_{5}^{3}+a_{15}x_{1}^{2}x_{5}^{3}x_{6}^{2}x_{7}^{2} +a_{16}x_{1}^{2}x_{2}^{3}x_{3}x_{4}x_{6}x_{7}+a_{17}x_{1}^{2}x_{2}^{2}x_{3}x_{4}x_{5}x_{6}x_{7}\nonumber\\
&+a_{18}x_{1}^{2}x_{2}x_{3}x_{4}x_{5}^{2}x_{6}x_{7} +a_{19}x_{1}^{2}x_{3}x_{4}x_{5}^{3}x_{6}x_{7}+a_{20}x_{1}^{2}x_{2}^{2}x_{3}^{2}x_{4}x_{6}x_{8}+a_{20}x_{1}^{2}x_{2}^{2}x_{3}x_{6}^{2}x_{7}x_{8}\nonumber\\
& +a_{21}x_{1}^{2}x_{2}x_{3}^{2}x_{4}x_{5}x_{6}x_{8}+a_{21}x_{1}^{2}x_{2}x_{3}x_{5}x_{6}^{2}x_{7}x_{8}+a_{22}x_{1}^{2}x_{3}^{2}x_{4}x_{5}^{2}x_{6}x_{8} +a_{22}x_{1}^{2}x_{3}x_{5}^{2}x_{6}^{2}x_{7}x_{8}\nonumber\\
&+a_{23}x_{1}^{2}x_{2}x_{3}^{2}x_{6}^{2}x_{8}^{2}+a_{24}x_{1}^{2}x_{3}^{2}x_{5}x_{6}^{2}x_{8}^{2} +a_{25}x_{1}x_{2}^{3}x_{3}x_{4}^{2}x_{7}+a_{25}x_{1}x_{2}^{3}x_{4}x_{6}x_{7}^{2}\nonumber\\
&+a_{26}x_{1}x_{2}^{2}x_{3}x_{4}^{2}x_{5}x_{7} +a_{26}x_{1}x_{2}^{2}x_{4}x_{5}x_{6}x_{7}^{2}+a_{27}x_{1}x_{2}x_{3}x_{4}^{2}x_{5}^{2}x_{7}+a_{27}x_{1}x_{2}x_{4}x_{5}^{2}x_{6}x_{7}^{2}\nonumber\\
&+a_{28}x_{1}x_{3}x_{4}^{2}x_{5}^{3}x_{7}+a_{28}x_{1}x_{4}x_{5}^{3}x_{6}x_{7}^{2}+a_{29}x_{1}x_{2}^{2}x_{3}^{2}x_{4}^{2}x_{8}+a_{29}x_{1}x_{2}^{2}x_{6}^{2}x_{7}^{2}x_{8}\nonumber\\
&+a_{30}x_{1}x_{2}x_{3}^{2}x_{4}^{2}x_{5}x_{8}+a_{30}x_{1}x_{2}x_{5}x_{6}^{2}x_{7}^{2}x_{8}+a_{31}x_{1}x_{3}^{2}x_{4}^{2}x_{5}^{2}x_{8} +a_{31}x_{1}x_{5}^{2}x_{6}^{2}x_{7}^{2}x_{8}\nonumber\\
&+a_{32}x_{1}x_{2}^{2}x_{3}x_{4}x_{6}x_{7}x_{8}+a_{33}x_{1}x_{3}x_{4}x_{5}^{2}x_{6}x_{7}x_{8} +a_{34}x_{1}x_{2}x_{3}^{2}x_{4}x_{6}x_{8}^{2}+a_{34}x_{1}x_{2}x_{3}x_{6}^{2}x_{7}x_{8}^{2}\nonumber\\
&+a_{35}x_{1}x_{3}^{2}x_{4}x_{5}x_{6}x_{8}^{2} +a_{35}x_{1}x_{3}x_{5}x_{6}^{2}x_{7}x_{8}^{2}+a_{36}x_{1}x_{3}^{2}x_{6}^{2}x_{8}^{3}+a_{37}x_{2}^{3}x_{4}^{2}x_{7}^{2}+a_{38}x_{2}^{2}x_{4}^{2}x_{5}x_{7}^{2}\nonumber\\
&+a_{39}x_{2}x_{4}^{2}x_{5}^{2}x_{7}^{2}+a_{40}x_{4}^{2}x_{5}^{3}x_{7}^{2}+a_{41}x_{2}^{2}x_{3}x_{4}^{2}x_{7}x_{8}+a_{41}x_{2}^{2}x_{4}x_{6}x_{7}^{2}x_{8} +a_{42}x_{2}x_{3}x_{4}^{2}x_{5}x_{7}x_{8}\nonumber\\
&+a_{42}x_{2}x_{4}x_{5}x_{6}x_{7}^{2}x_{8}+a_{43}x_{3}x_{4}^{2}x_{5}^{2}x_{7}x_{8}+a_{43}x_{4}x_{5}^{2}x_{6}x_{7}^{2}x_{8} +a_{44}x_{2}x_{3}^{2}x_{4}^{2}x_{8}^{2}+a_{44}x_{2}x_{6}^{2}x_{7}^{2}x_{8}^{2}\nonumber\\
&+a_{45}x_{3}^{2}x_{4}^{2}x_{5}x_{8}^{2}+a_{45}x_{5}x_{6}^{2}x_{7}^{2}x_{8}^{2} +a_{46}x_{2}x_{3}x_{4}x_{6}x_{7}x_{8}^{2}+a_{47}x_{3}x_{4}x_{5}x_{6}x_{7}x_{8}^{2}\nonumber\\
&+a_{48}x_{3}^{2}x_{4}x_{6}x_{8}^{3}+a_{48}x_{3}x_{6}^{2}x_{7}x_{8}^{3}\, .
\end{align}

The fixed point set $F_{2}=\{y_{1}=y_{2}=y_{3}={y_{7}}=0\}$ can be written in terms of the original projective coordinates $\{x_{1}=x_{2}=x_{5}=0\}\cap\{x_{3}x_{4}=-x_{6}x_{7}\}$. If we make these substitutions in $P_{symm}$, it reduces to
\begin{align}
P_{symm}&=a_{48}(x_{3}^{2}x_{4}x_{6}x_{8}^{3} + x_{3}x_{6}^{2}x_{7}x_{8}^{3})\notag\\
&=a_{48}x_{3}x_{6}{ x_8^3 y_{7}}\, .
\end{align}

As we are considering the subset where $x_{1}=x_{2}=x_{5}=0$, and the Stanley-Reisner ideal forbids $x_{2}=x_{3}=x_{5}=0$, then $x_{3}$ cannot vanish. Similarly, { $x_{2}=x_{5}=x_{6}=0$ is forbidden and so $x_6$ cannot vanish.}  Finally, $x_{2}=x_{5}=x_{8}=0$ is forbidden, so $x_{8}$ cannot vanish. Thus, the only way for $P_{symm}$ to vanish is to have { $y_{7}=0$}.

Hence, $P_{symm}=0$ implies { $y_{7}=0$} due to the SR constraints, and $y_{8}$ is redundant when restricting the fixed set to $X$. The reduced set is then $F'_{2}=\{y_{1}=y_{2}=y_{3}=0\}$.

In practice, we combine the transversality and SR ideal checks by performing Groebner basis calculations to check the dimension of the ideal $\cI^{fixed}_{ij}$ as in eq.(\ref{eq:fixed}) for $U_{i}$ a region allowed by the Stanley-Reisner ideal. 
\begin{align}
\cI^{fixed}_{ij} =\langle U_{i},\; P_{symm},\; F_{j}\rangle
\end{align}
%\noindent
 where given the $\cI_{SR}$, the allowed regions in this case are
\begin{align}
U_{1}&=&\langle x_{1}-t_{1},x_{2}-t_{2},x_{3}-t_{3},x_{4}-t_{4}\rangle & &U_{2}&=&\langle x_{1}-t_{1},x_{2}-t_{2},x_{3}-t_{3},x_{6}-t_{6}\rangle \notag\\
U_{3}&=&\langle x_{1}-t_{1},x_{2}-t_{2},x_{4}-t_{4},x_{7}-t_{7}\rangle & &U_{4}&=&\langle x_{1}-t_{1},x_{2}-t_{2},x_{6}-t_{6},x_{7}-t_{7}\rangle \notag\\
U_{5}&=&\langle x_{1}-t_{1},x_{3}-t_{3},x_{4}-t_{4},x_{5}-t_{5}\rangle & &U_{6}&=&\langle x_{1}-t_{1},x_{3}-t_{3},x_{5}-t_{5},x_{6}-t_{6}\rangle \notag\\
U_{7}&=&\langle x_{1}-t_{1},x_{3}-t_{3},x_{6}-t_{6},x_{8}-t_{8}\rangle & &U_{8}&=&\langle x_{1}-t_{1},x_{4}-t_{4},x_{5}-t_{5},x_{7}-t_{7}\rangle \notag\\
U_{9}&=&\langle x_{1}-t_{1},x_{5}-t_{5},x_{6}-t_{6},x_{7}-t_{7}\rangle & &U_{10}&=&\langle x_{2}-t_{2},x_{3}-t_{3},x_{4}-t_{4},x_{8}-t_{8}\rangle \notag\\
U_{11}&=&\langle x_{2}-t_{2},x_{4}-t_{4},x_{7}-t_{7},x_{8}-t_{8}\rangle & &U_{12}&=&\langle x_{2}-t_{2},x_{6}-t_{6},x_{7}-t_{7},x_{8}-t_{8}\rangle \notag\\
U_{13}&=&\langle x_{3}-t_{3},x_{4}-t_{4},x_{5}-t_{5},x_{8}-t_{8}\rangle & &U_{14}&=&\langle x_{3}-t_{3},x_{4}-t_{4},x_{6}-t_{6},x_{8}-t_{8}\rangle \notag\\
U_{15}&=&\langle x_{3}-t_{3},x_{6}-t_{6},x_{7}-t_{7},x_{8}-t_{8}\rangle & &U_{16}&=&\langle x_{4}-t_{4},x_{5}-t_{5},x_{7}-t_{7},x_{8}-t_{8}\rangle \notag\\
U_{17}&=&\langle x_{5}-t_{5},x_{6}-t_{6},x_{7}-t_{7},x_{8}-t_{8}\rangle\, ,
\end{align}
for auxiliary variables $t_{1},...,t_{8}\in\mathbb{C}$. In order to simplify the calculation, however, we set $t_{1}=...=t_{8}=1$. If the dimension $\text{dim }\cI >0$, and removing any generator from $F_{i}$ changes the dimension, then we know that $F_{i}$ intersects $X$ transversally and is allowed by the SR ideal.

Finally, we can tell from the number of intersecting codimension-1 subvarieties in each of the fixed sets $F_{1}$ and $F'_{2}$ that they have complex codimension 1 and 3 in $X$ respectively. This implies that $F_{1}$ is an O7 plane, while $F'_{2}$ is an O3 plane locus.

{%\color{red}
 These fixed point sets intersect the respective $\sigma$-invariant hypersurface so that we get a number of $O7$ and $O3$-planes. The  homology class of  the single $O7$
plane is
\bea
\label{eq:o7}
D(O7_{F_1}) = D_3 + D_4,
\eea 
with Euler characteristic $\chi(D(O7)) = 39$.
In order to determine the number of $O3$-planes, we can use the intersection form to
compute the triple intersection numbers as
\bea
\label{eq:o3}
O3_{F'_2}: D_1 D_2 D_5 =  1
\eea
so that in total there is one $O3$-plane.  Using eq.(\ref{eq:tadpole}) the contribution to the D3-brane tadpole is
\bea
 N_{D3} + \frac{N_{\text{flux}}}{2}+ N_{\rm gauge}= \frac{N_{O3}}{4}+\frac{\chi(D_{O7})}{4} = \frac{1 + 39}{4} = 10 \,.
\eea
Thus $Q_{D3}^{loc} = -10 $ and this is a \lq\lq naive orientifold Type~IIB string vacua".
}

%%%%%%%%%%%%%%%%%%%%%%%%%%%%%%%%%%%%%%%%%%%%%%%%%%%%%%%%
\subsubsection*{Hodge Splitting}

We now turn to the procedure for computing the splitting of the Hodge numbers on the Calabi-Yau orientifold. The linear ideal, which fixes toric divisor redundancies, is given by
\bea
\begin{array}{cccccccccccccccccc}
\cI_{lin}=&\langle &-D_{1}&-&D_{2}&-&D_{3}&-&D_{4}&+&0&+&D_{6}&+&D_{7}&+&D_{8},&\\
&+&0&+&0&+&D_{3}&+&D_{4}&+&0&-&D_{6}&-&D_{7}&&0,&\\
&-&D_{1}&&0&-&D_{3}&-&D_{4}&-&D_{5}&+&D_{6}&+&D_{7}&+&D_{8},&\\
&+&0&+&0&+&0&+&D_{4}&+&D_{5}&-&D_{6}&+&0&-&D_{8}&\rangle\, ,
\end{array}
\eea
\noindent and a basis in $H^{1,1}(X;\mathbb{Z})$ is given by $J_{1}=D_{4},\; J_{2}=D_{5},\; J_{3}=D_{6},\; J_{4}=D_{8}$. %

By the definition of the holomorphic involution eq.(\ref{eq:orientifold}),  the K\"ahler form is even and must therefore belong to $H^{1,1}_+(X/\sigma^{*})$ under the involution
\bea
\sigma^*: \,\, D_3 \leftrightarrow D_6, \quad D_4 \leftrightarrow D_7.
\eea
In a case with favorable geometry, as in this example, the calculation is simplified by the fact that the toric divisor classes of the ambient space $\cA$ always restrict in a straightforward way to the Calabi-Yau threefold hypersurface via eq.(\ref{eq:favor}).
We can therefore expand the K\"ahler form in terms of these divisor classes
\bea\label{eq:kformex1}
J=t_{1}J_{1}+t_{2}J_{2}+t_{3}J_{3}+t_{4}J_{4}=t_{1}D_{5}+t_{2}D_{6}+t_{3}D_{7}+t_{4}D_{8}
\eea
with $t_{1},t_{2},t_{3},t_{4}\in\mathbb{Z}$. But the K\"ahler form must obey the constraint of even parity under the orientifold involution, and must therefore only have components in $H^{1,1}_+(X)$, so that
\begin{align}\label{eq:kformswapex1}
J=\sigma^{*}J%=t_{1}\sigma^{*}D_{4}+t_{2}\sigma^{*}D_{5}+t_{3}\sigma^{*}D_{6}+t_{4}\sigma^{*}D_{8} 
=t_{1}D_{5}+t_{2}D_{3}+t_{3}D_{4}+t_{4}D_{8}
=t_{1}J_{1}+t_{2}D_{3}+t_{3}D_{4}+t_{4}J_{4}\, .
\end{align}
In order to relate eq.(\ref{eq:kformex1}) and eq.(\ref{eq:kformswapex1}), we must be able to write $D_{3}$ and $D_{4}$ (restricted to the CY hypersurface $X=\sum\limits_{i=1}^{8}{D_{i}}$) in terms of our chosen basis. This can be done by reducing $D_{3}$ and $D_{4}$ by the linear ideal $\cI_{lin}$ (also restricted to the CY hypersurface). We find that on $X$, $D_{3}$ and $D_{4}$ are uniquely given by\footnote{We do this calculation using the symbolic algebraic geometry software packages \texttt{Singular} \cite{DGPS} and \texttt{Sage} \cite{sage}.}
\bea
D_{3}=J_{1}+J_{3}-J_{4}\hspace{7mm}\text{and}\hspace{7mm}D_{4}=-J_{1}+J_{2}+J_{4}\, .
\eea

Plugging these into eq.(\ref{eq:kformswapex1}), we deduce
\bea\label{eq:kformnewex1}
J=(t_{1}+t_{2}-t_{3})J_{1} + t_{3}J_{2} + t_{2}J_{3} + (-t_{2}+t_{3}+t_{4})J_{4}\, .
\eea
Now, comparing eq.(\ref{eq:kformex1}) and eq.(\ref{eq:kformnewex1}), we obtain the following system of linear equations
\begin{align}
t_{1}+t_{2}-t_{3}=t_{1}, && 
t_{3}=t_{2}, &&\notag\\
t_{2}=t_{3},&&
-t_{2}+t_{3}+t_{4}=t_{4}\, ,&&
\end{align}
for which the only independent solution is $t_{2}=t_{3}$. Thus, we see that only 3 directions in the K\"ahler moduli space are independent, and so $h^{1,1}_{+}(X/\sigma^{*})=3$, and by extension $h^{1,1}_{-}(X/\sigma^{*})=1$. This will be the case for any choice of integral basis on the K\"ahler moduli space. In fact, choosing even and odd parity eigendivisors $D_{\pm,1}=D_{3}\pm D_{6},\;\; D_{\pm,2}=D_{4}\pm D_{7}$, the K\"ahler form can be written
\begin{align}
J=(t_{1}-t_{2})J_{1} + (t_{4}+t_{2})J_{4}+t_{2}(D_{+,1}-D_{+,2})
=(t_{1}-t_{+})J_{1} + (t_{4}+t_{+})J_{4}+t_{+}(D_{+})\, ,
\end{align}
where the latter equality makes the redefinitions $t_{+}=t_{2}$ and $D_{+}=D_{+,1}-D_{+,2}=(D_{3}+D_{6})-(D_{4}+D_{7})$. The even parity of the K\"ahler form is now manifest.  In this example, the involution $\sigma^{*}$ induces an O7 plane and an O3 plane, and so is not a free action and we have the Hodge number splitting\footnote{ We can apply Lefschetz fixed point theorem to compute $h^{2,1}_-$ as eq.(\ref{eq:h21split}) in the orbifold limit.  Then we have $h^{2,1}_- = h^{1,1}_- + \frac{ \chi_{O7}+ N_{O3}-\chi(X)}{4}$. Since $\chi_{O7} = 39$ and  $N_{O3} =1$,  we  get $h^{2,1}_- = 41$. }
\bea
h^{1,1}_+ (X/\sigma^{*}) = 3, \,\,\quad h^{1,1}_-  (X/\sigma^{*}) = 1\, .
\eea

%{\red Finally we check that the locus $\{P_{symm}=0\}$ is smooth by computing the dimension $\text{dim }\cI^{smooth}_{i}$ in eq.(\ref{eq:smooth}) for each disjoint region $U_{i}$  ($i = 1, \dots, 15$) allowed by the Stanley-Reisner ideal. We find that the maximum dimension is $-1$, so that $\{P_{symm}=0\}$ is smooth. This allows us to state that $H^{1,1}_{\pm}(X/\sigma^{*})$ are, in fact, the split Hodge numbers on the orientifold.}

%%%%%%%%%%%%%%%%%%%%%%%%%%%%%%%%%%%%%%%%%%%%%%%%%%%%%%%%
%%%%%%%%%%%%%%%%%%%%%%%%%%%%%%%%%%%%%%%%%%%%%%%%%%%%%%%%
\subsection{Improper Involution: Single Coordinate Exchange}
\label{subsec:free1}

%%%%%%%%%%%%%%%%%%%%%%%%%%%%%%%%%%%%%%%%%%%%%%%%%%%%%%%%
%\subsubsection{Orientifold Planes}

In this section, we demonstrate an explicit example of a Calabi-Yau orientifold for which an involution with a single coordinate exchange exists with an empty fixed-point locus, which makes the involution a would-be free-action.  However, this single coordinate exchange will violate the $\cI_{lin}$ and fails to keep the defining hypersurface polynomial homogenous. As discussed around eq.(\ref{eq:chow}-\ref{eq:triple}) this may fail to leave the the intersection numbers invariant. This turns out to be a general feature of single coordinate exchange involutions at low $h^{1,1}(X)$, and we will see why that is.  In fact, we do not begin to see proper fixed-point free involutions until $h^{1,1}(X)=6$ with upwards of four disjoint coordinate exchanges.  

We have chosen an example from the database of Calabi-Yau threefolds with Hodge numbers $h^{1,1}(X)=4,\;\; h^{2,1}(X)=82$. It can be identified by its index in the database:
\begin{equation*}
\begin{array}{|c|c|c|}
\hline
\text{Polytope ID}&\text{Geometry ID}&\text{Triangulation ID}\\
\hline
917&2&1\\
\hline
\end{array}
\end{equation*}
\noindent This example defines an MPCP desingularized ambient toric variety with weight matrix $\mathbf{W}$ given by
\bea
\label{eq:toricdata4}
\begin{array}{|c|c|c|c|c|c|c|c|c|}
\hline
x_{1} & x_{2} & x_{3} & x_{4} & x_{5} & x_{6} & x_{7} & x_{8}\\\hline
0 & 0 & 0 & 1 & 0 & 1 & 3 & 1\\\hline
0 & 0 & 1 & 1 & 1 & 0 & 3 & 0\\\hline
0 & 1 & 0 & 0 & 1 & 0 & 2 & 0\\\hline
1 & 1 & 0 & 1 & 0 & 2 & 5 & 0\\\hline
\end{array}
\eea
\noindent and Stanley-Reisner ideal
\bea
\cI_{SR}=\langle x_{1}x_{3},\; x_{1}x_{6},\; x_{3}x_{4},\; x_{6}x_{8},\; x_{1}x_{2}x_{5},\; x_{2}x_{5}x_{7},\; x_{4}x_{7}x_{8}\rangle\, .
\eea

For this example there is only one involution exchanging NIDs that leaves the SR ideal invariant. However, it is not ``proper''; it fails to leave the   the intersection numbers invariant.  In the interest of keeping this example simple, we will carry on with this choice of ``improper'' involution. This involution is given by
\bea
\sigma: x_{2} \leftrightarrow x_{5}\, .
\eea
Since the projective coordinates $x_{1}$, $x_{3}$, $x_{4}$, $x_{6}$, $x_{7}$, and $x_{8}$ are not affected by the involution, they are included in our list of (anti-)invariant polynomial generators
\bea
\mathcal{G}_{0}=\{x_{1},x_{3},x_{4},x_{6},x_{7},x_{8}\}\, .
\eea
In this case, there is only one sub-involution, $\sigma$ itself. Because we only consider NIDs, $x_{2}$ and $x_{5}$ have different weights and cannot be combined into a homogenous binomial. Thus, we are left only with the invariant monomial $x_{2}x_{5}$. This implies that the even and odd parity generator sets are $\mathcal{G}_{+}=\{x_{2}x_{5}\}$ and $\mathcal{G}_{-}=\emptyset $.

Now, all the (anti-)invariant polynomial generations in $\mathcal{G}=\mathcal{G}_{0}\cup\mathcal{G}_{+}\cup\mathcal{G}_{-}$ are given by
\begin{equation*}
\begin{gathered}
y_{1}=x_{1},\;\;\; y_{2}=x_{3},\;\;\; y_{3}=x_{4},\;\;\; y_{4}=x_{6},\;\;\; y_{5}=x_{7},\;\;\; y_{6}=x_{8},\;\;\; y_{7}=x_{2}x_{5}\, .
\end{gathered}
\end{equation*}
There are no generators with odd parity under the involution which are manifestly fixed. It is for this reason that single coordinate exchange involutions are fixed-point free. Naively, however, we may still have the fixed set $F_{0}=\{x_{2}=x_{5}=0\}$. Note that this set is allowed by the Stanley-Reisner ideal and does, in fact, exist in the ambient space. However, the orientifold-symmetric Calabi-Yau polynomial has the form
\begin{align*}
P_{symm}&=a_{1}x_{7}^{2}+ a_{1}x_{1}^{2}x_{2}^{2}x_{4}^{4}x_{5}^{2}x_{6}x_{8}+a_{3}x_{1}^{4}x_{2}^{2}x_{3}^{2}x_{4}^{2}x_{5}^{2}x_{6}x_{8}^{3}+a_{4}x_{1}^{2}x_{2}^{2}x_{3}^{2}x_{4}^{2}x_{5}^{2}x_{6}^{2}x_{8}^{2}+a_{5}x_{1}^{3}x_{2}^{2}x_{3}x_{4}^{3}x_{5}^{2}x_{6}x_{8}^{2}\\
&+a_{6}x_{1}^{5}x_{2}^{2}x_{3}^{3}x_{4}x_{5}^{2}x_{6}x_{8}^{4}+a_{7}x_{1}^{6}x_{2}^{2}x_{3}^{2}x_{4}^{2}x_{5}^{2}x_{8}^{4}+a_{8}x_{2}^{2}x_{3}^{4}x_{5}^{2}x_{6}^{4}x_{8}^{2}+a_{9}x_{2}x_{4}^{2}x_{5}x_{6}x_{7}\\
&+a_{10}x_{1}^{2}x_{2}x_{4}^{2}x_{5}x_{7}x_{8}+a_{11}x_{1}^{3}x_{2}x_{3}x_{4}x_{5}x_{7}x_{8}^{2}+a_{12}x_{1}^{5}x_{2}^{2}x_{3}x_{4}^{3}x_{5}^{2}x_{8}^{3}+a_{13}x_{1}^{8}x_{2}^{2}x_{3}^{4}x_{5}^{2}x_{8}^{6}\\
&+a_{14}x_{1}^{6}x_{2}^{2}x_{3}^{4}x_{5}^{2}x_{6}x_{8}^{5}+a_{15}x_{1}^{4}x_{2}^{2}x_{4}^{4}x_{5}^{2}x_{8}^{2}+a_{16}x_{1}^{4}x_{2}x_{3}^{2}x_{5}x_{7}x_{8}^{3}+a_{17}x_{1}^{7}x_{2}^{2}x_{3}^{3}x_{4}x_{5}^{2}x_{8}^{5}\\
&+a_{18}x_{1}x_{2}x_{3}x_{4}x_{5}x_{6}x_{7}x_{8}+a_{19}x_{1}x_{2}^{2}x_{3}x_{4}^{3}x_{5}^{2}x_{6}^{2}x_{8}+a_{20}x_{1}x_{2}^{2}x_{3}^{3}x_{4}x_{5}^{2}x_{6}^{3}x_{8}^{2}+a_{21}x_{1}^{3}x_{2}^{2}x_{3}^{3}x_{4}x_{5}^{2}x_{6}^{2}x_{8}^{3}\\
&+a_{22}x_{2}x_{3}^{2}x_{5}x_{6}^{2}x_{7}x_{8}+a_{23}x_{1}^{2}x_{2}x_{3}^{2}x_{5}x_{6}x_{7}x_{8}^{2}+a_{24}x_{1}^{4}x_{2}^{2}x_{3}^{4}x_{5}^{2}x_{6}^{2}x_{8}^{4}+a_{25}x_{2}^{2}x_{3}^{2}x_{4}^{2}x_{5}^{2}x_{6}^{3}x_{8}\\
&+a_{26}x_{1}^{2}x_{2}^{2}x_{3}^{4}x_{5}^{2}x_{6}^{3}x_{8}^{3}+a_{27}x_{2}^{2}x_{4}^{4}x_{5}^{2}x_{6}^{2}\, ,
\end{align*}

\noindent where $a_{i}\in\mathbb{C}$ are arbitrary coefficients. On the ambient space fixed set, where $\{x_{2}=x_{5}=0\}$, this reduces simply to
\begin{align}
P_{symm}&=a_{1}x_{7}^{2}\, .
\end{align}
Thus, the vanishing with the Calabi-Yau polynomial on this set is equivalent to the vanishing of $x_{7}$. However, as  $x_{2}x_{5}x_{7}$ is an element of the Stanley-Reisner ideal, $x_{7}$ cannot vanish on this set. Hence, the region where the Calabi-Yau hypersurface intersects the fixed-point locus of the ambient space is ruled out, giving us what is potentially a free action on the orientifold, provided that it is also non-singular by checking the dimension of ideal $\cI^{smooth}_{i}$ eq.(\ref{eq:smooth}) for each disjoint region $U_{i}$ allowed by the Stanley-Reisner ideal.

%%%%%%%%%%%%%%%%%%%%%%%%%%%%%%%%%%%%%%%%%%%%%%%%%%%%%%%%
%\subsubsection{Hodge Splitting}

We now turn to the procedure for computing the splitting of the Hodge numbers on the Calabi-Yau orientifold. The linear ideal, which fixes toric divisor redundancies, is given by
\bea
\begin{array}{cccccccccccccccccc}
\cI_{lin}=&\langle & -D_{1}&-&D_{2}&-&D_{3}&-&D_{4}&-&D_{5}&-&D_{6}&+&D_{7}&-&D_{8},&\\
&&0&+&0&+&0&+&D_{4}&+&2D_{5}&+&2D_{6}&-&D_{7}&+&0,&\\
&&-D_{1}&-&D_{2}&+&D_{3}&-&D_{4}&+&3D_{5}&+&4D_{6}&-&D_{7}&+&0,&\\
&&0&+&D_{2}&+&0&+&D_{4}&-&D_{5}&-&D_{6}&+&0&+&0&\rangle\, .
\end{array}
\eea
As we saw before, $x_{2}$ and $x_{5}$ have different weights and cannot be combined into a homogenous binomial.  So  we only need to consider the coordinate exchange  $x_{2} \leftrightarrow x_{5}$ in the defining polynomial of the Calabi-Yau hypersurface. The involution fails to keep the defining polynomial homogeneous  without tuning any coefficients to zero.  For example the monomial $x_4 x_5^2 x_6^2 x_7 $ in the original defining polynomial with degree $||{6,6,4,10}||$ will be changed to monomial  $x_4 x_2^2 x_6^2 x_7 $ with degree  $||{6,2,4,14}||$ after the involution. This violates the linear ideal $\cI_{lin}$ and it is also the reason why the triple intersection number changes under such an involution.

\subsection{Free Action Involution}
\label{subsec:free3}

%%%%%%%%%%%%%%%%%%%%%%%%%%%%%%%%%%%%%%%%%%%%%%%%%%%%%%%%

%%%%%%%%%%%%%%%%%%%%%%%%%%%%%%%%%%%%%%%%%%%%%%%%%%%%%%%%

In this section, we demonstrate an explicit example of a Calabi-Yau orientifold for which the involution is a free action. Additionally, this is the sole example we found in which the resulting orientifold is smooth. This example has Hodge numbers $h^{1,1}(X)=6,\;\; h^{2,1}(X)=46$. It can be identified by its index in the database
\begin{equation*}
\begin{array}{|c|c|c|}
\hline
\text{Polytope ID}&\text{Geometry ID}&\text{Triangulation ID}\\
\hline
7916&1&1\\
\hline
\end{array}
\end{equation*}
\noindent As this geometry consists of only one triangulation, this involution spans the entire geometry. This example defines an MPCP desingularized ambient toric variety with weight matrix $\mathbf{W}$ given by
\bea
\label{eq:toricdata6}
\begin{array}{|c|c|c|c|c|c|c|c|c|c|}
\hline
x_{1} & x_{2} & x_{3} & x_{4} & x_{5} & x_{6} & x_{7} & x_{8} & x_{9} & x_{10}\\\hline
0 & 0 & 0 & 0 & 1 & 1 & 0 & 0 & 0 & 0\\\hline
0 & 0 & 0 & 1 & 0 & 0 & 1 & 0 & 0 & 0\\\hline
0 & 0 & 1 & 0 & 0 & 0 & 0 & 1 & 0 & 0\\\hline
0 & 1 & 0 & 0 & 0 & 0 & 0 & 0 & 1 & 0\\\hline
1 & 0 & 0 & 0 & 0 & 0 & 0 & 0 & 0 & 1\\\hline
1 & 0 & 0 & 1 & 1 & 0 & 0 & 1 & 1 & 0\\\hline
\end{array}
\eea
\noindent with the Stanley-Reisner ideal
\begin{align}		
	\cI_{SR}=&\langle x_{1}x_{10},\; x_{2}x_{9},\; x_{3}x_{8},\; x_{4}x_{7},\; x_{5}x_{6},\; x_{1}x_{4}x_{5},\; x_{1}x_{4}x_{8},\; x_{1}x_{4}x_{9},\; x_{1}x_{5}x_{8},\;\notag\\
	&x_{1}x_{5}x_{9},\; x_{1}x_{8}x_{9},\; x_{2}x_{3}x_{6},\;x_{2}x_{3}x_{7},\; x_{2}x_{3}x_{10},\; x_{2}x_{6}x_{7},\; x_{2}x_{6}x_{10},\; x_{2}x_{7}x_{10},\;\\
	&x_{3}x_{6}x_{7},\; x_{3}x_{6}x_{10},\; x_{3}x_{7}x_{10},\; x_{4}x_{5}x_{8},\; x_{4}x_{5}x_{9},\; x_{4}x_{8}x_{9},\; x_{5}x_{8}x_{9},\; x_{6}x_{7}x_{10} \rangle\notag,
\end{align}
and all toric divisors have the same Hodge numbers, $h^{\bullet}(D_{i}) = \{1,0,0,11\}$.
In fact, this example is the only one which leads to a free action involving five coordinate exchanges, and is given by
\bea
\label{eq:freeinvolution}
	\sigma: \,\, x_{1} \leftrightarrow x_{10}, x_{2} \leftrightarrow x_{9}, x_{3} \leftrightarrow x_{8}, x_{4} \leftrightarrow x_{7}, x_{5} \leftrightarrow x_{6}.
\eea
\subsubsection*{Orientifold Planes}

This involution eq.(\ref{eq:freeinvolution}) affects all ten projective coordinates, and thus $\mathcal{G}_{0} = \emptyset$. Computing the (anti-)invariant polynomials, we find the following:
\begin{align}
\mathcal{G}_{+} = &\{ x_{1}x_{10}, x_{2}x_{9}, x_{3}x_{8}, x_{4}x_{7}, x_{5}x_{6}, x_{1}x_{2} + x_{9}x_{10}, x_{1}x_{3} + x_{8}x_{10},\notag\\
&x_{1}x_{6} + x_{5}x_{10}, x_{1}x_{7} + x_{4}x_{10}, x_{2}x_{8} + x_{3}x_{9}, x_{2}x_{4} + x_{7}x_{9}, x_{2}x_{5} + x_{6}x_{9},\\
& x_{3}x_{4} + x_{7}x_{8}, x_{3}x_{5} + x_{6}x_{8}, x_{4}x_{6} + x_{5}x_{7} \},\notag
\end{align}
\begin{align}
\mathcal{G}_{-} = &\{ x_{1}x_{2} - x_{9}x_{10}, x_{1}x_{3} - x_{8}x_{10}, x_{1}x_{6} - x_{5}x_{10}, x_{1}x_{7} - x_{4}x_{10}, \notag\\
&x_{2}x_{8} - x_{3}x_{9}, x_{2}x_{4} - x_{7}x_{9}, x_{2}x_{5} - x_{6}x_{9}, x_{3}x_{4} - x_{7}x_{8}, x_{3}x_{5} - x_{6}x_{8}\\
&, x_{4}x_{6} - x_{5}x_{7} \}.\notag
\end{align}

We note that all of the monomial elements of $\mathcal{G}_{+}$ are also members of the Stanley-Reisner ideal. As all of these monomials have the form $x_{i}x_{\sigma(i)}$, they thus cannot vanish in the ambient space. Also, recall that for our pairs of binomial generators, at least one must vanish at any fixed point as the expressions have the same weights but opposite parity. (For example, for this involution at least one of $x_{1}x_{2} \pm x_{9}x_{10}$ must vanish at any fixed point).

This means that there are $2^{10}$ possible fixed sets, corresponding to a sign choice in each of the 10 pairs of binomial generators. Scanning over these sets, there are many choices whose weights allow them to be fixed. However, all of these choices lead to sets which fail to intersect the ambient space due to the coordinate restrictions imposed by the Stanley-Reisner ideal.
Since this involution is indeed a free action, there is no fixed locus and therefore no O-plane.  This manifold can also be considered as one of the \lq\lq naive orientifold Type IIB string vacua".

%%%%%%%%%%%%%%%%%%%%%%%%%%%%%%%%%%%%%%%%%%%%%%%%%%%%%%%%
\subsubsection*{Hodge Splitting}

We now turn to the procedure for computing the splitting of the Hodge numbers on the Calabi-Yau orientifold. The linear ideal, which fixes toric divisor redundancies, is given by
\bea
{\footnotesize
\begin{array}{cccccccccccccccccccccc}
\cI_{lin}=&\langle & -D_{1}&-&D_{2}&-&D_{3}&-&D_{4}&+&0&+&0&+&D_{7}&+&D_{8}&+&D_{9}&+&D_{10},&\\
&&-D_{1}&-&D_{2}&-&D_{3}&+&0&-&D_{5}&+&D_{6}&+&0&+&D_{8}&+&D_{9}&+&D_{10},&\\
&&-D_{1}&-&D_{2}&+&0&-&D_{4}&+&D_{5}&-&D_{6}&+&D_{7}&+&0&+&D_{9}&+&D_{10},&\\
&&-D_{1}&+&0&-&D_{3}&-&D_{4}&+&D_{5}&-&D_{6}&+&D_{7}&+&D_{8}&+&0&+&D_{10}&\rangle\, ,
\end{array}
}
\eea
and a basis in $H^{1,1}(X;\mathbb{Z})$ is given by $J_{1} = D_{5}, J_{2} = D_{6}, J_{3} = D_{7}, J_{4} = D_{8}, J_{5} = D_{9}, J_{6} = D_{10}$. 
In this example the involution acts on the divisor classes as 
\bea
	\sigma^{*} : D_{1} \leftrightarrow D_{10}, D_{2} \leftrightarrow D_{9}, D_{3} \leftrightarrow D_{8}, D_{4} \leftrightarrow D_{7}, D_{5} \leftrightarrow D_{6},
\eea
thus all five of the exchanges in this example are of non-shrinkable rigid divisors.
This case is favorable, and we can thus expand the K\"ahler form in terms of these divisor classes
	$J = t_{1}J_{1} + t_{2}J_{2} + t_{3}J_{3} + t_{4}J_{4} + t_{5}J_{5} + t_{6}J_{6}$,
with $t_{1}, ... t_{6} \in \mathbb{Z}$. The constraint that the K\"ahler form must only have components in $H^{1,1}_{+}(X)$ implies
\begin{align}
	\label{eq:kformswapex4}
	J&=\sigma^{*}J	=t_{1}D_{6}+t_{2}D_{5}+t_{3}D_{4}+t_{4}D_{3}+t_{5}D_{2}+t_{6}D_{1}
	%&=t_{1}J_{2}+t_{2}J_{1}+t_{3}D_{4}+t_{4}D_{3}+t_{5}D_{2}+t_{6}D_{1} \, .
\end{align}

As in previous examples, we rewrite $D_{1}, D_{2}, D_{3},$ and $D_{4}$ in terms of our chosen basis using the linear ideal. Performing the algebra, and plugging the relations 
%\begin{align}
%	J_{1} - J_{2} + J_{6} = D_{1}, &&
%	-J_{1} + J_{2} + J_{5} = D_{2}, &&\nonumber\\
%	-J_{1} + J_{2} + J_{4} = D_{3},  &&
%	J_{1} - J_{2} + J_{3} = D_{4}. &&
%\end{align}
 into eq.(\ref{eq:kformswapex4}), we obtain
\bea
	\label{eq:kformnewex4}
	J = (t_{2} + t_{3} - t_{4} - t_{5} + t_{6})J_{1} + (t_{1} - t_{3} + t_{4} + t_{5} - t_{6})J_{2} + t_{3}J_{3} + t_{4}J_{4} + t_{5}J_{5} + t_{6}J_{6}.
\eea
Note that the last four terms are unchanged from the original expansion of the K\"ahler form and we get
\begin{align}
t_{2} + t_{3} - t_{4} - t_{5} + t_{6} = t_{1}, \quad\quad
t_{1} - t_{3} + t_{4} + t_{5} - t_{6} = t_{2}.
\end{align}
Solving for $t_{1}$ in the latter equation shows that these are in fact the same, and hence we can write either $t_{1}$ or $t_{2}$ in terms of the other K\"ahler moduli. Hence $h^{1,1}_{+}(X/\sigma^{*}) = 5$ and $h^{1,1}_{-}(X/\sigma^{*}) = 1$. Choosing the independent even-parity eigendivisors $D_{1+,} = D_{5}+D_{6}, D_{+,2}=D_{4}+D_{7}, D_{+,3}=D_{3}+D_{8}, D_{+,4}=D_{2}+D_{9}, D_{+,5}=D_{1}+D_{10}$, we can write the K\"ahler form in the manifestly even-parity form
\bea
	J = t_{+,1}D_{+,1} + t_{+,2}D_{+,2} + t_{+,3}D_{+,3} + t_{+,4}D_{+,4} + t_{+,5}D_{+,5},
\eea
with coefficients $t_{+,i}$ given by
\bea
t_{+,1} = t_{1} + \frac{1}{2}(-t_{3} + t_{4} + t_{5} - t_{6}), \quad t_{+,2}=\frac{1}{2}t_{3},\nonumber\\ t_{+,3}=\frac{1}{2}t_{4}, \quad t_{+,4}=\frac{1}{2}t_{5}, \quad t_{+,5}=\frac{1}{2}t_{6}.
\eea

Finally we check whether the locus $\{P_{symm}=0\}$ is smooth by %
computing the dimension $\text{dim }\cI^{smooth}_{i}$ as eq.(\ref{eq:smooth}) for each disjoint region $U_{i}$ allowed by the Stanley-Reisner ideal. We find that the maximum dimension is $-1$, so that $\{P_{symm}=0\}$ is smooth. The involution considered in eq.(\ref{eq:freeinvolution}) is indeed a free action.
 Furthermore, the Hodge number splits under the Lefschetz fixed point theorem:
\bea
h^{2,1}_- (X/\sigma^{*}) = h^{1,1}_- (X/\sigma^{*})  + \frac{L(\sigma, X) -\chi(X)}{4} = h^{1,1}_- (X/\sigma^{*})  - \frac{\chi(X)}{4} =  21\, ,
\eea
and then the Hodge number of this smooth Calabi-Yau threefold splits as:
\bea
h^{1,1}_+ (X/\sigma^{*})  = 5, \,\, h^{1,1}_- (X/\sigma^{*})  = 1; \,\, h^{2,1}_+ (X/\sigma^{*})  = 25, \,\, h^{2,1}_- (X/\sigma^{*})  = 21\, .
\eea
Since the manifold is smooth, there is no ambiguity in defining $h^{2,1}_- (X/\sigma^{*}) $ and eq.(\ref{eq:h21split}) gives the true Hodge number splitting.

%\newpage
%%%%%%%%%%%%%%%%%%%%%%%%%%%%%%%%%%%%%%%%%%%%%%%%%%%%%%%%
%%%%%%%%%%%%%%%%%%%%%%%%%%%%%%%%%%%%%%%%%%%%%%%%%%%%%%%%
%%%%%%%%%%%%%%%%%%%%%%%%%%%%%%%%%%%%%%%%%%%%%%%%%%%%%%%%
\section{Scanning Results}
\label{sec:orientifolddiscuss}

In a systematic scan within the database \cite{Altman:2014bfa} up to $h^{1,1} = 6$, we analyzed 22,974 favorable polytopes, from which we obtain  646,903 MPCP triangulations.  As discussed in Section \ref{subsec:favorable}, some  subset of the triangulations of a dual polytope encode identical topological information, with the primary difference being the content of the K\"ahler cone. In such cases, we must glue them together into a larger K\"ahler cone corresponding to
a single Calabi-Yau geometry. This gluing results in 100,368 distinct favorable Calabi-Yau geometries. However, due to computational restraints, we were unable to examine every triangulation. The percentages of geometries scanned are shown in Table.\ref{tab:scan}.

\begin{table}[h!]
{\footnotesize
  \centering
    \begin{tabular}{|r|r||c|c|c|c|c|c||c|}
    \hline
    \multicolumn{2}{|c||}{\parbox[c][2em][c]{5cm}{\centering$\mathbf{h^{1,1}(X)}$}} & \textbf{1} & \textbf{2} & \textbf{3} & \textbf{4} & \textbf{5} & \textbf{6}  & \bf{Total} \\\hline
    \multicolumn{2}{|r||}{\parbox[c][3em][c]{5cm}{\centering\textbf{\# of Favorable Polytopes}}} & 5     & 36    & 243   & 1185  & 4897  & 16608 & 22974\\    \hline
    \multicolumn{2}{|r||}{\parbox[c][3em][c]{5cm}{\centering\textbf{\# of Favorable Triangulations}}} & 5     & 48    & 525   & 5330  & 56714 & 584281 & 646903 \\\hline
           \multicolumn{2}{|r||}{\parbox[c][3em][c]{5cm}{\centering\textbf{\# of Favorable Geometries}}} & 5     & 39    & 305   & 2000  & 13494 & 84525 & 100368  \\\hline
    \hline
     % \multicolumn{2}{|r||}{\parbox[c][3em][c]{5cm}{\centering\textbf{\% of Favorable Geometry Scanned}}} & 80  &  100 &  &  &  &  & \\\hline
     \multicolumn{2}{|r||}{\parbox[c][3em][c]{5cm}{\centering\textbf{\% of Favorable Triangulations Scanned}}} & 80 & 100 & 99.8 & 99.66 & 99.41 & 99.01 & 99.01 \\
     \hline
    \end{tabular}
      \caption{The favorable polytopes,  triangulations, geometries for $h^{1,1}(X) \leq 6$. }
      \label{tab:scan}
      }
    \end{table}

\subsection{Classification of Proper Involutions}

According to the definition of orientifold projection eq.(\ref{eq:orientifold}), each of the proper involutions will lead to an orientifold Calabi-Yau threefold.  As a result, we will classify various properties of orientifold Calabi-Yau threefolds in the $\IZ_2$ orbifold limit according to different kinds of proper involutions.

%which includes an analysis of particular classes of exchanged divisors. 
We  first consider the so-called \lq\lq triangulation-wise" involutions. In Section \ref{subsec:involutions}, we outline the procedure for obtaining the desired Non-trivial Identical Divisor (NID) involutions.
In total, we find 107,171 such involutions present at the triangulation level which exist within a single chamber of the K\"ahler cone of a given geometry. Of these, after considering the favorable triangulations and trivial fundamental group, 28,463 are ``proper'' in the sense that they preserve the intersection structure of $X$ and allow for consistent orientifold geometries as described in Section~\ref{subsec:involutions}.  We also find that 8,449 favorable CY geometries admit a consistent involution within at least one chamber of their K\"ahler cones. 
These 28,463 triangulation-wise proper involutions are distributed in 25,375 different triangulations. 

After considering the gluing of K\"ahler cones  corresponding to a single Calabi-Yau geometry, we refer to those involutions which 
span all disjoint phases of the K\"ahler cone for a unique Calabi-Yau geometry as \lq\lq geometry-wise" proper NID involutions. 
 We find a total of 5,660 geometry-wise proper involutions, each of which may correspond to several triangulation-wise involutions. We find that there are only 1,401 favorable polytopes and 4,482 favorable geometries which contain a geometry-wise proper NID involution, which account for $6.1\%$ and $4.47\%$ of scanned polytopes and geometries respectively. These results are summarized in Table.\ref{tab:stat}.  

\begin{table}[h!]
{\footnotesize
  \centering
    \begin{tabular}{|r|r||P{0.7cm}|P{0.7cm}|P{1cm}|P{1cm}|P{1cm}|P{1cm}||c|}
    \hline
    \multicolumn{2}{|c||}{\parbox[c][2em][c]{5cm}{\centering$\mathbf{h^{1,1}(X)}$}} & \textbf{1} & \textbf{2} & \textbf{3} & \textbf{4} & \textbf{5} & \textbf{6}  & \bf{Total} \\\hline

     \multicolumn{9}{|c|}{\parbox[c][2em][c]{15cm}{\centering\textbf{Triangulation-wise  proper NID exchange  involutions}}} \\\hline 
       \multicolumn{2}{|r||}{\parbox[c][3em][c]{5cm}{\centering\textbf{\# of Polytopes \\ contains  Involutions}}} & 0  & 1 & 25  & 166  & 712  & 2172  & 3076 \\\hline
       \multicolumn{2}{|r||}{\parbox[c][3em][c]{5cm}{\centering\textbf{\# of Geometries contains Involutions}}} & 0 & 1 & 26 &  273 & 1559  & 6590  & 8449  \\
   \hline
 \multicolumn{2}{|r||}{\parbox[c][3em][c]{5cm}{\centering\textbf{\# of Triangulations contains  Involutions}}} & 0 & 1 & 31 &  405 & 3372  & 21566  & 25375  \\\hline
 \multicolumn{2}{|r||}{\parbox[c][3em][c]{5cm}{\centering\textbf{\# of   Involutions}}} & 0 & 6 & 51 & 516 & 4085 & 23805 & 28463  \\
\hline
    \multicolumn{9}{|c|}{\parbox[c][2em][c]{15cm}{\centering\textbf{Geometry-wise   proper NID exchange  involutions}}} \\\hline
   \multicolumn{2}{|r||}{\parbox[c][3em][c]{5cm}{\centering\textbf{\# of Polytope \\  contains Involutions}}} & 0 & 1 & 16 & 96 & 330 & 958 & 1401 \\\hline
   \multicolumn{2}{|r||}{\parbox[c][3em][c]{5cm}{\centering\textbf{\# of Geometries contains Involutions}}} & 0 & 1 & 17 &  183 & 911  & 3370  & 4482 \\
   \hline
 \multicolumn{2}{|r||}{\parbox[c][3em][c]{5cm}{\centering\textbf{\# of  Involutions}}} & 0 & 6 & 28 & 259 & 1219 & 4148 & 5660 \\
 \hline
  \multicolumn{2}{|r||}{\parbox[c][3em][c]{5cm}{\centering\textbf{\% of  Polytope \\ contains Involutions }}} & 0 & 2.78 & 6.58 & 8.10 & 6.74 & 5.77 & 6.10 \\
\hline
 \multicolumn{2}{|r||}{\parbox[c][3em][c]{5cm}{\centering\textbf{\% of  Geometries contains Involutions }}} & 0 & 2.56 & 5.57 & 9.15 & 6.75 & 3.99 & 4.47\\
	\hline	
    \end{tabular}
      \caption{Statistic counting on the  triangulation/geometry-wide Non-trivial Identical Divisors exchange involutions in favorable polytopes,  triangulations and geometries.}
      \label{tab:stat}
      }
    \end{table}

Each of the geometry-wise proper NID involutions may exchange several pairs of topologically distingushed divisors. Thus we  enumerate the number of different pairs of Non-trivial Identical Divisors for each of the involutions as shown in Table \ref{tab:exchangegeom}.  We note that it is very rare for an involution to exchange both del Pezzo, K3, and exact-Wilson divisors simultaneously for small $h^{1,1}(X)$.

As an example, consider $h^{1,1}(X) = 2$. There are 36 polytopes which contains 48 MPCP triangulations. After gluing the K\"ahler cone of several triangulations with the same topology, we end up with 39 distinguished Calabi-Yau manifolds (Table~\ref{tab:scan}). Among them only one polytope, which contains a single triangulation, (and thereby corresponds to a single geometry) contains a proper involution (which is both a triangulation-wise and geometry-wise involution). In fact, there are 6 different kinds of involutions acting on the geometry which result in 6 different orientifold Calabi-Yau manifolds (Table~\ref{tab:stat}). Each of these 6 involutions exchanges three pairs of Special Deformation (SD2) divisors, which results in a total of 18 exchanged pairs of SD2 surfaces (Table~\ref{tab:exchangegeom}).  

\begin{table}[ht!]
{\footnotesize
  \centering
  \hspace*{-.5cm}
    \begin{tabular}{|r||c|c|c|c|c|c||c|}
	\hline
	\multicolumn{8}{|c|}{\parbox[c][2em][c]{15cm}{\centering\textbf{Number of pairs of Non-trivial Identical Divisors (NID) under involutions}}} \\
	\hline
	\parbox[c][2em][c]{6cm}{\centering$\mathbf{h^{1,1}(X)}$} & \textbf{1} & \textbf{2} & \textbf{3} & \textbf{4} & \textbf{5} & \textbf{6} & {\bf Total} \\\hline
	\multicolumn{8}{|c|}{\parbox[c][2em][c]{15cm}{\centering\textbf{Triangulation-wise proper Involutions}}} \\
	\hline
	\parbox[c][2em][c]{6cm}{\centering\textbf{\# of  Involutions}} & 0 & 6 & 51 & 516 & 4085 & 23805 & 28463 \\\hline
	\parbox[c][2em][c]{6cm}{\centering\textbf{del Pezzo surface }$\mathbf{dP_{n}}$, $\mathbf{n\leq 8}$} & 0 & 0 & 12 & 238 & 2233 & 14507&17090 \\
	\hline
	\parbox[c][2em][c]{6cm}{\centering\textbf{Rigid surface} $\mathbf{dP_{n}}$, $\mathbf{n > 8}$} & 0 & 0 & 14 & 512 & 5659 & 32481 & 38666 \\
	\hline
	\parbox[c][2em][c]{6cm}{\centering\textbf{(exact-)Wilson}\textbf{ surface}} & 0 (0) & 0 (0) & 5 (0) & 40 (5) & 177 (80) & 744 (411)&  966 (496) \\
	\hline
	\parbox[c][2em][c]{6cm}{\centering\textbf{K3 surface}} & 0 & 0 & 65 & 300 & 619 & 1976&  2960 \\
	\hline
	\parbox[c][2em][c]{6cm}{\centering\textbf{SD1 surface}} & 0 & 0 & 9 & 47 & 418 & 2190& 2664 \\
	\hline
	\parbox[c][2em][c]{6cm}{\centering\textbf{SD2 surface}}& 0 & 18 & 8 & 33 & 109 & 459 & 627 \\
	\hline
	\parbox[c][2em][c]{6cm}{\centering\textbf{del Pezzo and K3}} & 0 & 0 & 0 & 9 & 98 & 572 & 679\\
	\hline
	\parbox[c][2em][c]{6cm}{\centering\textbf{del Pezzo and (Exact-)Wilson}} & 0 (0) & 0 (0) & 1 (0) & 28 (0) & 95 (9) & 667 (286) & 791 (295) \\
	\hline
	\parbox[c][2em][c]{6cm}{\centering\textbf{K3 and (Exact-)Wilson}} & 0 (0) & 0 (0) & 8 (0) & 12 (4) & 43 (7) & 101 (9) & 156 (20)  \\
	\hline
	\parbox[c][2em][c]{6cm}{\centering\textbf{del Pezzo, K3 and (Exact-)Wilson}} & 0 (0) & 0 (0) & 0 (0) & 0 (0) & 28 (0) & 87 (2) & 115 (2) \\
	\hline
%	\hline
%	\parbox[c][3em][c]{5cm}{\centering\textbf{\% of Favorable}\\\textbf{Triangulations Scanned}} & 80 & 100 & 99.8 & 99.66 & 99.41 & 99.01 \\
	\multicolumn{8}{|c|}{\parbox[c][2em][c]{15cm}{\centering\textbf{Geometry-wise proper Involutions}}} \\
	\hline
	\parbox[c][2em][c]{6cm}{\centering\textbf{\# of  Involutions}} & 0 & 6 & 28 & 259 & 1219 & 4148 & 5660 \\\hline
	\parbox[c][2em][c]{6cm}{\centering\textbf{del Pezzo surface } $\mathbf{dP_{n}}$, $\mathbf{n\leq 8}$} & 0 & 0 & 8 & 107 & 634 & 2660& 3409\\
	\hline
	\parbox[c][2em][c]{6cm}{\centering\textbf{Rigid surface} $\mathbf{dP_{n}}$, $\mathbf{n > 8}$} & 0 & 0 & 8 & 259 & 1973 & 6198 & 8438\\
	\hline
	\parbox[c][2em][c]{6cm}{\centering\textbf{(Exact-)Wilson surface}} & 0 (0) & 0 (0) & 5 (0) & 28 (2) & 48 (4) & 136 (75)& 217 (81) \\
	\hline
	\parbox[c][2em][c]{6cm}{\centering\textbf{K3 surface}} & 0 & 0 & 28 & 215 & 219 & 527& 989  \\
	\hline
	\parbox[c][2em][c]{6cm}{\centering\textbf{SD1 surface}} & 0 & 0 & 8 & 23 & 102 & 216& 349  \\
	\hline
	\parbox[c][2em][c]{6cm}{\centering\textbf{SD2 surface}} & 0 & 18 & 6 & 18 & 39 & 84 &  165\\
	\hline
	\parbox[c][2em][c]{6cm}{\centering\textbf{del Pezzo and K3}} & 0 & 0 & 0 & 0 & 26 & 156 &  182\\
	\hline
	\parbox[c][2em][c]{6cm}{\centering\textbf{del Pezzo and (Exact-)Wilson}} & 0 (0) & 0 (0) & 1 (0) & 19 (0) & 40 (1) & 109 (40) &169 (41) \\
	\hline
	\parbox[c][2em][c]{6cm}{\centering\textbf{K3 and (Exact-)Wilson}} & 0 (0) & 0 (0) & 8 (0) & 12 (4) & 13 (4) & 23 (3) & 56 (11)  \\
	\hline
	\parbox[c][2em][c]{6cm}{\centering\textbf{del Pezzo, K3 and (Exact-)Wilson}} & 0 (0) & 0 (0) & 0 (0) & 0 (0) & 4 (0) & 16 (2) & 20 (2)\\
	\hline
    \end{tabular}
    \caption{Number of  pairs of NIDs exchanged under triangulation/geometry-wise proper involutions. }
  \label{tab:exchangegeom}
  }
\end{table}%

%{\blue [Q: 5702 is not the same as 5692 in the Table. Where is the missing 10 coming from?]}

\subsection{Classification of O-planes}

The fixed points of the involution correspond to orientifold planes, which acquire charges that must be cancelled by appropriate configurations of D3 and D7 branes in order to avoid anomalies in our theory. We scan for these orientifold planes in Section \ref{subsec:fixedloci} by seeking the fixed points in the ambient toric manifold and restricting down to components transversal with the involution-invariant part of the Calabi-Yau hypersurface. Note than in all but one case, the fixed sets allowed by a geometry-wise proper involution across an entire consistent CY geometry are either individual O3, O5, or O7 planes, or combinations of O3 and O7 planes. In one case, discussed in Section \ref{subsec:free3}, we find a freely acting $\mathbb{Z}_{2}$ involution. In every case, the parity of the volume form under $\sigma$ is in agreement with the orientifold planes found by our algorithm, i.e,  $\sigma^* \Omega = - \Omega$ for O3, O7, and O3/O7 cases, and  $\sigma^* \Omega = \Omega$ for O5 cases. The results of this scan can be found in Table~\ref{tab:classificationgeomproper}.\footnote{For $h^{1,1}(X)=6$, due to the calculation time limit, we did not obtain the fixed loci for 33 triangulation-wise involutions, which account for $0.14\%$ of the 23805 triangulation-wise proper involutions in $h^{1,1} (X) = 6$.  So to classify O-planes and naive Type IIB string vacua (Tables~\ref{tab:classificationgeomproper}, \ref{tab:vacua}), we only take into account the 23772 triangulation-wise involutions for $h^{1,1}(X) = 6$.}  It shows that 23,961 out of the total 28,430  (84.3\% of)  triangulation-wise proper involutions and 4,108 out of the total 5,660 (72.6\%) geometry-wise proper involutions will end up with an orientifold Calabi-Yau threefold with an $O3/O7$-plane system.  

As a consistency check, we note that there are no simultaneous $O3$ and $O5$-planes existing under a single involution, as well as no $O7$ and $O5$-planes coexisting. Of course, it then follows that there are no instances of an orientifold Calabi-Yau threefold containing $O3$, $O5$ and $O7$-plane under a single involution.

We note that our results (in particular, only finding one free action) are in agreement with \cite{Braun:2017juz}. In that work, a search for freely-acting discrete symmetries of a more general variety was performed for $h^{1,1}(X)\leq 3$, with free actions being found for five toric Calabi-Yaus. Of these five, two do not admit any $\mathbb{Z}_{2}$ symmetries, while the $\mathbb{Z}_{2}$ symmetries of the other three consist entirely of coordinate reflections of the form $x_{i} \leftrightarrow -x_{i}$. In particular, none of these cases support an NID involution. This is consistent with our results, which found no NID involutions below $h^{1,1}(X) = 6$.

As an example, consider $h^{1,1}(X) = 3$. There are 25 polytopes with 31 triangulations containing 51 triangulation-wise proper involutions. Among these involutions, 9, 20 and 31 will result in O3, O5 and O7-plane loci respectively. Among them, there are 9 involutions which contain both O3- and O7- planes simultaneously. These 51 triangulation-wise involutions reduce to 28 geometry-wise proper involutions when requiring the involution span all disjoint phases of the K\"ahler cone for a unique  Calabi-Yau geometry. In these 28 involutions, 4 contain O3 loci, 16 contain O5 loci, 12 contain O7 loci, while 4 contain a combination of O3 and O7-planes (Table \ref{tab:classificationgeomproper}).

\begin{table}[ht!]
{\footnotesize
  \centering
   \begin{tabular}{|P{4.5cm}||P{1cm}|P{1cm}|P{1cm}|P{1cm}|P{1cm}|P{1cm}||P{1.5cm}|}
%	\hline
%	\parbox[c][3em][c]{6cm}{\centering\textbf{\% of Favorable Triangulations Scanned}} & 100 & 100 & 99.8 & 99.66 & 99.41 & 99.01 \\
%	\hline
%	\parbox[c][3em][c]{4.5cm}{\centering\textbf{\% of NID Involutions Scanned for Fixed Loci}} & 100 & 100 & 100 & 100 & 100 & 99.55 ? & \\
	\hline
	\multicolumn{8}{|c|}{\parbox[c][2em][c]{15cm}{\centering\textbf{Classification of O-plane fixed point locus}}} \\
	\hline
	\parbox[c][2em][c]{4.5cm}{\centering$\mathbf{h^{1,1}(X)}$} & \textbf{1} & \textbf{2} & \textbf{3} & \textbf{4} & \textbf{5} & \textbf{6} & {\bf Total} \\\hline
	\multicolumn{8}{|c|}{\parbox[c][2em][c]{15cm}{\centering\textbf{Triangulation-wise proper Involutions}}} \\\hline
	\parbox[c][2em][c]{4.5cm}{\centering\textbf{\# of  Involutions}} & 0 & 6 & 51 & 516 & 4085 & {23772} & {28430} \\\hline
	%\parbox[c][4em][c]{4.5cm}{\centering\textbf{\% of Fixed Point Free  Involutions Scanned for Smoothness}} & 100 & 100 & 100 & 100 & 99.58 & 79.24  &\\\hline
			\parbox[c][2em][c]{3.5cm}{\centering\textbf{O3}} & 0 & 0 & 9 & 253 & 2640  & 18193  & 21083 \\
	\hline
	\parbox[c][2em][c]{3.5cm}{\centering\textbf{O5}} & 0 & 6 & 20 & 157&  1006& 3279  & 4468\\
	\hline
	\parbox[c][2em][c]{3.5cm}{\centering\textbf{O7}} & 0 & 0 & 31 & 328 & 3005 & 20137  & 23501 \\
		\hline
	\parbox[c][2em][c]{3.5cm}{\centering\textbf{O3 and O7}} & 0 & 0 & 9 & 222 & 2566 & 17826 & 20623 \\
		\hline
	\parbox[c][2em][c]{3.5cm}{\centering\textbf{Free Action}} & 0 & 0 & 0 & 0 & 0 & 1 & 1\\	
	\hline
	\multicolumn{8}{|c|}{\parbox[c][2em][c]{15cm}{\centering\textbf{Geometry-wise proper Involutions}}} \\\hline
	\parbox[c][2em][c]{4.5cm}{\centering\textbf{\# of  Involutions}} & 0 & 6 & 28 & 259 & 1219& 4148 & 5660 \\\hline
	%\parbox[c][4em][c]{4.5cm}{\centering\textbf{\% of Fixed Point Free  Involutions Scanned for Smoothness}} & 100 & 100 & 100 & 100 & 99.58 & 79.24  &\\\hline
			\parbox[c][2em][c]{3.5cm}{\centering\textbf{O3}} & 0 & 0 & 4& 82& 557 & 2611 & 3254 \\
	\hline
	\parbox[c][2em][c]{3.5cm}{\centering\textbf{O5}} & 0 & 6 & 16 & 106&  488& 929  & 1545\\
	\hline
	\parbox[c][2em][c]{3.5cm}{\centering\textbf{O7}} & 0 & 0 & 12 & 124 & 691 & 3082  & 3909 \\
		\hline
	\parbox[c][2em][c]{3.5cm}{\centering\textbf{O3 and O7}} & 0 & 0 & 4 & 53 & 523 & 2475 & 3055 \\
	\hline
	\parbox[c][2em][c]{3.5cm}{\centering\textbf{Free Action}} & 0 & 0 & 0 & 0 & 0 & 1 & 1\\	
%	\parbox[c][2em][c]{3.5cm}{\centering\textbf{\# of O3 }} &  &  &  &  &  & & \\
%	\hline
%		\parbox[c][2em][c]{3.5cm}{\centering\textbf{\# of O7 }} &   &   &  &  &  &  &\\
%	\hline
%		\parbox[c][3em][c]{4.5cm}{\centering\textbf{\# of Naive Type IIB \\String Vacua }} & 0  & 0  & 4/9  & 48/206 & 452/2341 & 1917/15294 & 3055/20623 \\
	 \hline
  \end{tabular}
  \caption{ Classification of  O-plane fixed point locus and free actions under the triangulation/geometry-wise proper involutions.  
  }
  \label{tab:classificationgeomproper}
  }
\end{table}

\subsection{String Landscape}
Under the assumption of placing eight D7-branes on top of the O7-plane to cancel the D7-tadpole, we also count the naive orientifold Type IIB string vacua with an $O3/O7$-system by considering cases that satisfy D3-tadpole cancellation, i.e, $Q_{D3}^{loc}$ in eq.(\ref{eq:tadpole}) is an integer. 
%We also require the split Hodge number be integer for consistent check.  
If under an involution there is only an $O7$-plane involved, we just count it as one naive Type IIB string vacuum.
%\footnote{Since the $D5$-tadpole canceled automatically by the involution, in principle we can count  all the configurations with $O5$ as naive string vacua also. Here we just consider $O3/O7$ system for example.}.  
It turns out that for most of the Calabi-Yau threefolds admitting a proper involution, they will end up with an $O3/O7$-system and obtain a naive orientifold Type IIB string vacua. The results are summarized in Table~\ref{tab:vacua}. It shows that 20,715  (72.9\%) of triangulation-wise proper involutions  and  3,334 (58.9\%) of geometry-wise proper involutions result in a naive Type IIB string vacuum. 

In the $O5/O9$-system, if we take account that the $D5$-tadpole is cancelled automatically by the involution, in principle we can count all the configurations with $O5$ as naive string vacua also. Then  25,183 (88.6\% of) triangulation-wise proper involutions and 4,879 (86.2\%) of  geometry-wise involutions will end up with a naive type IIB string vacuum. We did not count these vacua in the present work.

For example, for $h^{1,1}(X) = 3$ there are 51 triangulation-wise and 28 geometry-wise involutions, respectively. Under the triangulation-wise involutions, there are 9 orientifold geometries containing both $O3$ and $O7$-planes which satisfy the naive D3-tadpole cancellation. There are 22 geometries containing only $O7$-planes and, as explained before, we take all of them as naive string vacua. In total there are 31 naive string vacua among the triangulation-wise involutions. Similarly, among the geometry-wise involutions, there are four Calabi-Yau threefolds with $O3/O7$-planes and eight Calabi-Yau threefolds with only $O7$-planes which satisfy the D3-tadople cancellation condition, adding up to a total of 12 naive orientifold Type IIB string vacua (Table~\ref{tab:vacua}). 

\begin{table}[ht!]
{\footnotesize
  \centering
   \begin{tabular}{|P{4.5cm}||P{1cm}|P{1cm}|P{1cm}|P{1cm}|P{1cm}|P{1cm}||P{1.5cm}|}
      \hline
		\multicolumn{8}{|c|}{\parbox[c][2em][c]{15cm}{\centering\textbf{Naive Orientifold Type IIB String Vacua with $O3/O7$-system}}} \\
	\hline
		\parbox[c][2em][c]{4.5cm}{\centering$\mathbf{h^{1,1}(X)}$} & \textbf{1} & \textbf{2} & \textbf{3} & \textbf{4} & \textbf{5} & \textbf{6} & {\bf Total} \\\hline
	\multicolumn{8}{|c|}{\parbox[c][2em][c]{15cm}{\centering\textbf{Triangulation-wise proper Involutions}}} \\\hline
	\parbox[c][2em][c]{4.5cm}{\centering\textbf{\# of  Involutions}} & 0 & 6 & 51 & 516 & 4085 & {23772} & {28430} \\
	\hline
	\parbox[c][2em][c]{3.5cm}{\centering\textbf{Contains O3 \& O7}} & 0  & 0  & 9  & 206 & 2346 & 15234 & 17795 \\
	\hline
	\parbox[c][2em][c]{3.5cm}{\centering\textbf{Contains Only O3}} & 0 &  0 & 0 & 31 &  74&  355 & 460 \\
	\hline
	\parbox[c][2em][c]{3.5cm}{\centering\textbf{Contains Only O7}} & 0 & 0 & 22 & 102 & 386 & 1950  & 2460 \\
		\hline\hline
			\parbox[c][2em][c]{3.5cm}{\centering\textbf{Total String Vacua}} & 0 & 0 & 31 & 339  & 2806  & 17539   &  20715\\
		\hline
	\multicolumn{8}{|c|}{\parbox[c][2em][c]{15cm}{\centering\textbf{Geometry-wise proper Involutions}}} \\\hline
	\parbox[c][2em][c]{4.5cm}{\centering\textbf{\# of  Involutions}} & 0 & 6 & 28 & 259 & 1219 & 4148 & 5660 \\
	\hline
	\parbox[c][2em][c]{3.5cm}{\centering\textbf{Contains O3 \& O7}} & 0  & 0  & 4  & 48 & 455 & 1874 & 2381 \\
	\hline
	\parbox[c][2em][c]{3.5cm}{\centering\textbf{Contains Only O3}} & 0 &  0 & 0 & 29 &  34&  136 & 199 \\
	\hline
	\parbox[c][2em][c]{3.5cm}{\centering\textbf{Contains Only O7}} & 0 & 0 & 8 & 68 & 149 & 529  & 754 \\
		\hline\hline
			\parbox[c][2em][c]{3.5cm}{\centering\textbf{Total String Vacua}} & 0 & 0 & 12 & 145  & 638  & 2539   &  3334\\
		\hline

  \end{tabular}
  \caption{Classification of  naive orientifold Type IIB string vacua under the triangulation/geometry-wise proper  involutions.  
  }
  \label{tab:vacua}
  }
\end{table}%

\begin{figure}[ht!]
\centering
\includegraphics[width=7.5cm]{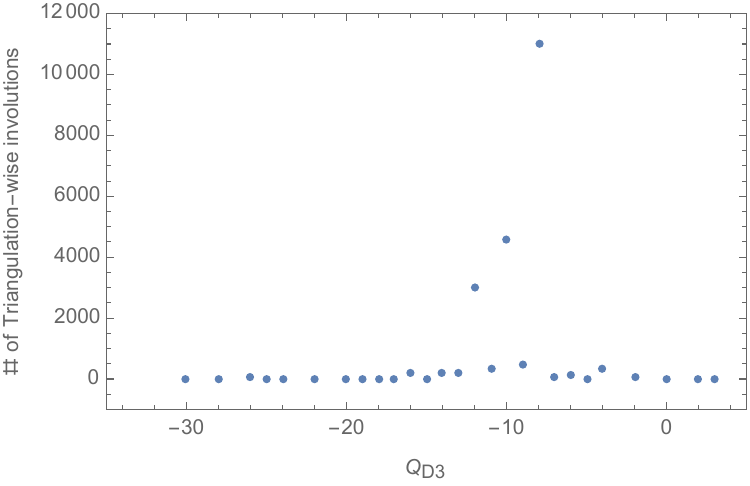}\quad
\includegraphics[width=7.5cm]{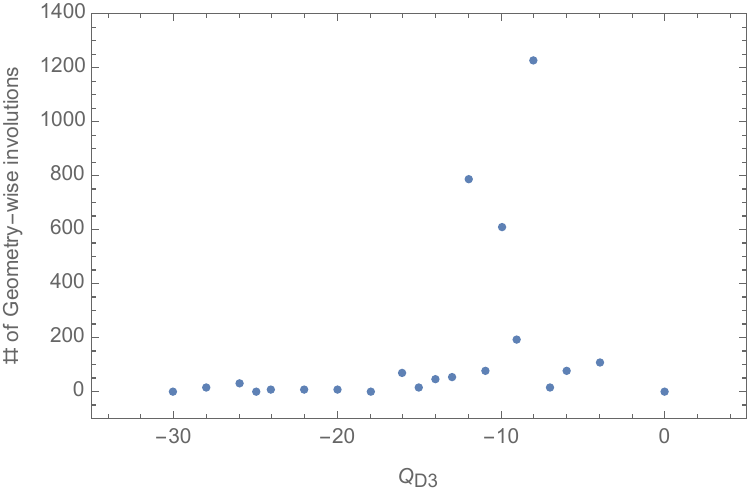}
\caption{Distribution of $Q_{D3}^{loc}$ under triangulation/geometry-wise proper involutions for naive orientifold Type IIB string vacua for $h^{1,1} (X) \leq 6$.}
\label{figure}
\end{figure}

For those 20,715 triangulation-wise and 3,334 geometry-wise naive orientifold Tyep IIB string vacua with an $O3/O7$-system, the distribution of $Q_{D3}^{loc}$  is shown in Fig.~\ref{figure}.  It shows that most of the involutions end up with an orientifold Calabi-Yau threefold with $Q_{D3}^{loc} $ around $-8$ in our scan. Again, the geometry-wise involutions put a strong constraint on the geometry and reduce dramatically the number of possible geometries. Here we again see that $Q_{D3}^{loc}$ is generally around $-8$. For triangulation-wise involutions, the smallest and largest $Q_{D3}^{loc}$ are $-30$ and $3$ respectively, while for geometry-wise involutions, the range of integer $Q_{D3}^{loc}$ shrinks to $ [-30, 0] $. 

\subsection{Hodge Number Splitting}

Finally, in Section \ref{subsec:splitting}, we discuss the decomposition of the K\"ahler moduli space into odd and even parity equivariant cohomology $H^{1,1}(X/\sigma^{*})=H^{1,1}_{+}(X/\sigma^{*})\oplus H^{1,1}_{-}(X/\sigma^{*})$. The constraint that the K\"ahler form must be invariant $\sigma^{*}J=J$ ensures that we can always find the dimension of the even parity space, and then by deduction, the dimension of the odd party space $h^{1,1}_{-}(X/\sigma^{*})$, which, as has been discussed, must be non-trivial in our case. The results of this K\"ahler moduli space splitting can be found in Table \ref{tab:involutiongeom}.  By utilizing the Lefschetz fixed point theorem  eq.(\ref{eq:h21split}) we can further determine the $h^{2,1}_{\pm}(X/\sigma^*)$ splitting in the orbifold limit.  The value of $h^{2,1}_{\pm}(X/\sigma^*)$ may get changed by a possible conifold resolution while $h^{1,1}_-(X/\sigma^*)$ is robust.  In this paper, we only present the robust $h^{1,1}_-(X/\sigma^*)$ results and leave the results of  $h^{2,1}_{\pm}(X/\sigma^*)$ after blowing up the singularities for a future work.
The entire procedure is explicitly performed in Section \ref{sec:example}.  
As an example, for $h^{1,1}(X) = 3$ there are 51 triangulation-wise and 28 geometry-wise involutions, respectively. Under these involutions, the Hodge numbers of the orientifold Calabi-Yau threefolds all split with $h^{1,1}_-(X/\sigma^*) = 1$.  

\begin{table}[h!]
{\footnotesize
  \centering
    \begin{tabular}{|r|r||P{1cm}|P{1cm}|P{1cm}|P{1cm}|P{1cm}|P{1cm}||P{1.5cm}|}
	\hline
\multicolumn{9}{|c|}{\parbox[c][2em][c]{14cm}{\centering\textbf{Hodge number splitting}}} \\
	\hline
	\multicolumn{2}{|c||}{\parbox[c][2em][c]{3.5cm}{\centering$\mathbf{h^{1,1}(X)}$}} & \textbf{1} & \textbf{2} & \textbf{3} & \textbf{4} & \textbf{5} & \textbf{6} & {\bf Total} \\\hline
%	\multicolumn{2}{|r||}{\parbox[c][2em][c]{3.5cm}{\centering\textbf{\# of  Involutions}}} & 0 & 12 & 32 & 291 & 1219 & 4148 & \\\hline
%	\hline
	\multicolumn{9}{|c|}{\parbox[c][2em][c]{14cm}{\centering\textbf{Triangulation-wide proper Involutions}}} \\
	\hline
%{\parbox[c][2em][|r||]{3.5cm}{\centering\textbf{\# of  Involutions}}} & 0 & 6/6 & 28/51 & 259/516 & 1219/4085 & 4148/23805 & 5660/28463 \\\hline
\multicolumn{2}{|r||}{\parbox[c][2em][c]{3.5cm}{\centering\textbf{\# of  Involutions}}} & 0 & 6  & 51 & 516 & 4085 & {23805} & {28463} \\
\hline
	\multirow{6}[12]{*}{\parbox[c][3em][c]{2.5cm}{\centering\textbf{\# of} $\mathbf{h^{1,1}_{-}}$}} & \parbox[c][2em][c]{0.5cm}{\centering\textbf{1}} & -- & 6 & 51 & 477 & 3682 & 20985 & 25201\\
	\cline{2-9} & \parbox[c][2em][c]{0.5cm}{\centering\textbf{2}} & -- & -- & 0 & 39 & 483 & 2618 & 3140 \\
	\cline{2-9} & \parbox[c][2em][c]{0.5cm}{\centering\textbf{3}} & -- & -- & -- & 0 & 0 & 202 & 202 \\
	\cline{2-9} & \parbox[c][2em][c]{0.5cm}{\centering\textbf{4}} & -- & -- & -- & -- & 0 & 0 & 0 \\
	\cline{2-9} & \parbox[c][2em][c]{0.5cm}{\centering\textbf{5}} & -- & -- & -- & -- & -- & 0 & 0 \\
	\hline
	\multicolumn{9}{|c|}{\parbox[c][2em][c]{14cm}{\centering\textbf{Geometry-wide proper Involutions}}} \\\hline
%{\parbox[c][2em][|r||]{3.5cm}{\centering\textbf{\# of  Involutions}}} & 0 & 6/6 & 28/51 & 259/516 & 1219/4085 & 4148/23805 & 5660/28463 \\\hline
\multicolumn{2}{|r||}{\parbox[c][2em][c]{3.5cm}{\centering\textbf{\# of  Involutions}}} & 0 & 6  & 28 & 259 & 1219 & 4148& 5660\\
\hline
	\multirow{6}[12]{*}{\parbox[c][3em][c]{2.5cm}{\centering\textbf{\# of} $\mathbf{h^{1,1}_{-}}$}} & \parbox[c][2em][c]{0.5cm}{\centering\textbf{1}} & -- & 6 & 28 & 277 & 1048 & 3413 & 4772\\
	\cline{2-9} & \parbox[c][2em][c]{0.5cm}{\centering\textbf{2}} & -- & -- & 0 & 32 & 171 & 661 & 864 \\
	\cline{2-9} & \parbox[c][2em][c]{0.5cm}{\centering\textbf{3}} & -- & -- & -- & 0 & 0 & 74 & 74 \\
	\cline{2-9} & \parbox[c][2em][c]{0.5cm}{\centering\textbf{4}} & -- & -- & -- & -- & 0 & 0 & 0 \\
	\cline{2-9} & \parbox[c][2em][c]{0.5cm}{\centering\textbf{5}} & -- & -- & -- & -- & -- & 0 & 0 \\
	\hline

  \end{tabular}
  \caption{ Classification of $h^{1,1}(X/\sigma^{*})$ splitting under the triangulation/geometry-wise proper  involutions.}
  \label{tab:involutiongeom}
  }
\end{table}%

%{ Here we also present the distribution of the splitting $h^{2,1}_-(X/\sigma^*)$ in the orbifold limit for naive string vacua with $O3/O7$-system for $h^{1,1} (X) \leq 6$  as shown in Fig. \ref{figure2}, where the geometries data are taken from Table.\ref{tab:vacua}.  From both Table. \ref{tab:involutiongeom} and Fig. \ref{figure2}, it shows that both $h^{1,1}_-(X/\sigma^*)$ and $h^{2,1}_-(X/\sigma^*)$ prefer a small splitting and the distribution fall off at a large number.  It would be interesting to see the distribution for much larger database beyond $h^{1,1} (X) > 7$. Since the number of Calabi-Yau geometries growing exponentially with $h^{1,1}(X)$ increasing, the distribution of Hodge number splitting will be dominated by large $h^{1,1}(X)$.

%\begin{figure}[ht!]
%\centering
%\includegraphics[width=7.5cm]{triang_h21.pdf}\quad
%\includegraphics[width=7.5cm]{geo_h21.pdf}
%\caption{Distribution of $h^{2,1}_-(X/\sigma^*)$ for triangulation/geometry-wise naive orientifold Type IIB string vacua with $O3/O7$-planes in the orbifold limit for $h^{1,1} (X) \leq 6$.}
%\label{figure2}
%\end{figure}
%}

%%%%%%%%%%%%%%%%%%%%%%%%%%%%%%%%%%%%%%%%%%%%%%%%%%%%%%%%
%%%%%%%%%%%%%%%%%%%%%%%%%%%%%%%%%%%%%%%%%%%%%%%%%%%%%%%%
%%%%%%%%%%%%%%%%%%%%%%%%%%%%%%%%%%%%%%%%%%%%%%%%%%%%%%%%

\section{Conclusions and Outlook}
\label{sec:conc}

In this paper, we extend and improve on a previous study \cite{Gao:2013pra} of the database \cite{Altman:2014bfa} (\dburl) of Calabi-Yau threefolds with $h^{1,1}(X) \leq 6$, by asking for the existence of a holomorphic $\mathbb{Z}_{2}$ orientifold involution $\sigma$.  First, we determined the topology of each divisor in defining the Calabi-Yau threefold by calculating its Hodge diamond. By requiring that the pullback of $\sigma$ to cohomology classes exchange only toric divisors with identical surface topology, but separate cohomology on the Calabi-Yau, we ensure that the orientifold has non-trivial odd equivariant cohomology $h^{1,1}_{-}(X/\sigma^{*})\neq 0$. We also classified the different kinds of Non-trivial Identical Divisors for each of the involutions and showed that consistency of this involution across the full K\"ahler cone is very restrictive. We further determined all possible fixed-point loci, i.e, the locations of $O3$, $O5$ and $O7$-planes, for each of the proper involutions. By checking the D3 tadpole cancellation condition, a class of naive Type IIB string vacua with $O3/O7$-system was obtained. We found that under the proper involutions one ends up with a majority of $O3/O7$-planes systems, most of which further admit a naive Type IIB string vacuum. Moreover, one free action was identified. We further calculated the Hodge number splitting to even/odd cohomology in the orbifold limit.  This dataset of orientifold Calabi-Yau threefolds, with all possible proper divisor exchange involutions, the classification and counts of orientifold planes under the involution, together with the non-trivial Hodge number splitting in the orbifold limit, represent a rich phenomenological starting  point for the construction of concrete string models for both particle physics and cosmology.

 In this paper we considered only involutions of type $\sigma^{*}J=J$, which result in Type IIB string vacua. A natural extension is to consider anti-holomorphic involutions $\sigma^{*}J=-\,J$ to classify the Type IIA vacua. We leave examination of these involutions to a future work. 
 As discussed in the Introduction, although reflections ${x_i \leftrightarrow -x_i}$ will in general not generate non-trivial $h^{1,1}_-(X)$ in a favorable geomtry, it might still be very interesting to classify their point-wise fixed loci and free actions on the Calabi-Yau hypersurface under the reflection. 
 %However, the fix locus on the Calabi-Yau hypersurface under these  reflections are still of interest for phenomenology reason, we leave  the discussion in a future work. 
 This will lead to a primary classification of the Type II string vacua landscape of the Kreuzer-Skarke database in our upcoming work \cite{Altman}.

Additionally, it would be ideal to extend these analyses beyond our current computational limit into the region $h^{1,1}(X) \geq 7$ of the Kreuzer-Skarke database. There has recently been progress on the triangulation of polytopes with large $h^{1,1}(X)$ \cite{Long:2014fba, Demirtas:2018akl, Demirtas:2020dbm}, although limitations on these techniques still preclude examining all possible MPCP triangulations for a general given polytope. Due to the large size of the Kreuzer-Skarke database, it is therefore natural to expect that in addition to the formal progress, supervised machine learning techniques will be necessary to understand the landscape, an approach applied to counting MPCP triangulations in \cite{Altman:2018zlc}. We will explore applying machine learning techniques to orientifold Calabi-Yau threefolds under divisor exchange involutions in a coming work \cite{Gao:2021xbs}.

In this paper, we focused on orientifold Calabi-Yaus under an involution without resolving the singularity, i.e, in the $\IZ_2$ orbifold limit, which is mostly considered in the literature for string model building. For other string applications, we can consider the resolution of the conifold singularities as considered early in the CICY  case \cite{Candelas:1989js,Candelas:1989ug} and recently in the CICY landscape \cite{Carta:2020ohw, Carta:2021uwv}. This will yield sets of CY threefolds that can be reached via conifold transitions, and possibly result in some new threefolds beyond the Kreuzer-Skarke list. We leave the resolution of conifold singularity for a future work. 
  
Besides the Kreuzer-Skarke and CICY database, it was found that one can relax the condition that the configuration matrix or weighted matrix entries be non-negative to construct a new class of Calabi-Yau manifold, called \lq\lq generalized Complete Intersection Calabi-Yau" (gCICYs) \cite{Anderson:2015iia} and its toric variations \cite{Berglund:2016yqo, Berglund:2016nvh}.  Some new mathematical  aspects \cite{Candelas:2016fdy, Garbagnati:2017rtb, Jia:2018iza} and physical applications  \cite{Anderson:2015yzz, Larfors:2020weh} of these geometries have recently been studied. Examining involutive, or more general quotient, symmetries of these new manifolds as well could prove interesting.

\section*{Acknowledgments}

We would like to thank James Halverson, Ralph Blumenhagen, Andres Collinucci, James Gray, Fengjun Xu, Hao Zou for helpful discussions and correspondence. XG was supported in part by the Humboldt Resaerch Fellowship and NSFC under grant numbers 12005150. BDN and JC were supported by the National Science Foundation under grant PHY-1913328.

%%%%%%%%%%%%%%%%%%%%%%%%%%%%%%%%%%%%%%%%%%%%%%%%%%%%%%%%
%%%%%%%%%%%%%%%%%%%%%%%%%%%%%%%%%%%%%%%%%%%%%%%%%%%%%%%%
%%%%%%%%%%%%%%%%%%%%%%%%%%%%%%%%%%%%%%%%%%%%%%%%%%%%%%%%
%\clearpage
\appendix
\section{Pseudocode Description of Fixed-Point Algorithm}
\label{appendix:pseudocode}
In this appendix we provide a pseudocode description of the fixed point classification algorithm described in Section \ref{subsec:fixedloci}. The input consists of a proper NID exchange involution $\sigma$ for a CY hypersurface $X$ in a toric variety $\mathcal{A}$ described by a fine, regular, star triangulation of a reflexive 4d polytope $\Delta^{\circ}$. The output is the set of fixed loci together with their codimensionality (classification into O7, O5, O3). We focus only on the calculations specific to this procedure, ignoring well-known procedures like finding the kernel or rank of a matrix. In addition, we have favored algorithmic simplicity over optimization in our descriptions of these algorithms.

The first calculation we describe is finding monomials with definite parity under the involution. This process is described in the routine below. The input is $\sigma$, which for these calculations can be expressed as a list of pairs, with each pair being the indices of exchanged coordinates, along with the resolved weight matrix $W$. The output is the sets $\mathcal{G}_{0}, \mathcal{G}_{+}, \mathcal{G}_{-}$ of invariant monomials. \\

{\footnotesize
\LinesNumbered
\SetAlgoLined

\begin{algorithm}[H]

let $\mathcal{G}_{0}, \mathcal{G}_{+}, \mathcal{G}_{-}$ be empty lists\\
$k \gets \sigma.length$\\
$m \gets W.rows$\\
$n \gets W.columns$\\

\tcc{Construct $\mathcal{G}_{0}$}
$\mathcal{G}_{0} \leftarrow \{x_{0},...,x_{n}\}$\\
\For{$p \leftarrow 0$ \KwTo $k$}{
    \For{$q \leftarrow 0$ \KwTo $1$}{
        $i \gets \sigma[p][q]$\\
        \If{$x_{i} \in \mathcal{G}_{0}$}{
            remove $x_{i}$ from $\mathcal{G}_{0}$
        }
    }
}

\tcc{Add monomials to $\mathcal{G}_{+}$}
\For{$p \leftarrow 0$ \KwTo $k$}{
	$i \gets \sigma[p][0]$\\
	$j \gets \sigma[p][1]$\\
	append $x_{i}x_{j}$ to $\mathcal{G}_{+}$
}

\tcc{Construct the matrix of difference vectors}
let $D$ be a new matrix of size $m \times k$\\
\For{$p \leftarrow 0$ \KwTo $k$}{
	$i \gets \sigma[p][0]$\\
	$j \gets \sigma[p][1]$\\
	\For{$r \leftarrow 0$ \KwTo $m$}{
		$D_{pr} = W_{ri} - W_{rj}$\\
	}
}

\tcc{Add binomials to $\mathcal{G}_{+}, \mathcal{G}_{-}$}
let $M$ be a basis for the integer kernel of $D$, with vectors as columns\\
\If{$M$ is nonempty}{
	\For{$i \leftarrow 0$ \KwTo $M.columns$}{
		$T1 \gets 1$\\
		$T2 \gets 1$\\
		\For{$j \leftarrow 0$ \KwTo $k$}{
			$a = \sigma[j][0]$\\
			$b = \sigma[j][1]$\\
			$v = M_{ji}$\\
			\eIf{$v >= 0$}{
				$T1 = T1 * x_{a}^{|v|}$\\
				$T2 = T2 * x_{b}^{|v|}$\\
			}{
				$T1 = T1 * x_{b}^{|v|}$\\
				$T2 = T2 * x_{a}^{|v|}$\\
			}
		}
		append $T1 + T2$ to $\mathcal{G}_{+}$ if not already a member\\
		append $T1 - T2$ to $\mathcal{G}_{-}$ if not already a member\\
	}
}

\KwRet $\mathcal{G}_{0}, \mathcal{G}_{+}, \mathcal{G}_{-}$

\caption{$\textrm{INVARIANT\_GENERATORS}(\sigma, W)$}
\end{algorithm}
}

\vskip0.2cm
The next procedure describes the check to determine whether or not the toric weights allow a locus, defined by the vanising of a list of homogeneous polynomials $\mathcal{F}$, to be fixed. As noted in Section $\ref{subsec:fixedloci}$, this is trivially true if $\mathcal{F} \cap \mathcal{G}_{-} = \emptyset$ and so such cases need not be checked.\\

{\footnotesize
\begin{algorithm}[H]
$m = W.rows$\\
$n = \mathcal{F}.length$\\
let $M$ be a matrix of size $m \times n$\\
\For{$i \leftarrow 0$ \KwTo $n$}{
	set column $i$ of $M$ to be the toric weight vector of $\mathcal{F}[i]$\\
}
	
let $b$ be an array of zeros of size $n$\\
\For{$i \leftarrow 0$ \KwTo $n$}{
	\If{$\mathcal{F}[i] \in \mathcal{G}_{-}$}{
		$b[i] = 1$\\
	}
}

let $s$ be an array of size $n$\\
\For{$i \leftarrow 0$ \KwTo $n$}{
	$s[i] = M_{1i} + \dots + M_{mi}$\\
}

let $Q$ be the set $Q = \{ (q_{1}, \dots, q_{n}) \in \mathbb{Z}^{n}, 0 \le q_{i} < s[i] \}$\\
$r = \textrm{rank(M)}$\\
\For{$q \in Q$}{
	let $M_{aug}$ be the augmented matrix $M|(2q+b)$\\
	$r_{aug} = \textrm{rank}(M_{aug})$\\
	\If{$r == r_{aug}$}{
		\KwRet true
	}
}

\KwRet false

\caption{$\textrm{WEIGHT\_FIXED}(\mathcal{F}, W, \mathcal{G}_{0}, \mathcal{G}_{+}, \mathcal{G}_{-})$}
\end{algorithm}
}

\vskip0.2cm
As discussed in Section \ref{subsec:fixedloci}, the invariant generators are not independent, with some subset of the generators being related by consistency conditions. The following algorithm determines whether or not a given the fixed set determined by the vanishing of a list of homogeneous polynomials $\mathcal{F}$ is consistent. We do this by computing the dimension of an ideal with the nonvanishing invariant polynomials set to 1. \\

{\footnotesize
\begin{algorithm}[H]
$\mathcal{G} = \mathcal{G}_{0} \cup \mathcal{G}_{+} \cup \mathcal{G}_{-}$\\
let $\mathcal{L}$ be an empty list\\
\For{$p \in \mathcal{G}$}{
	\eIf{$p \in \mathcal{F}$}{
		append $p$ to $\mathcal{L}$
	}{
		append $p-1$ to $\mathcal{L}$
	}
}
\KwRet $\textrm{dim}(\langle \mathcal{L} \rangle) \geq 0$

\caption{$\textrm{CONSISTENT}(\mathcal{F},\mathcal{G}_{0}, \mathcal{G}_{+}, \mathcal{G}_{-})$}
\end{algorithm}
}

\vskip0.2cm
Thus, given the sets $\mathcal{G}_{0}, \mathcal{G}_{+}, \mathcal{G}_{-}$ of definite-parity polynomials, we can find all fixed-point loci allowed by the toric weights via the following routine.\\

{\footnotesize
\begin{algorithm}[H]
$\mathcal{G} = \mathcal{G}_{0} \cup \mathcal{G}_{+} \cup \mathcal{G}_{-}$\\
let $\mathcal{L}$ be an empty list\\
let $\mathcal{S}$ be the power set of $\mathcal{G}$\\
\For{$\mathcal{F} \in \mathcal{S}$}{
	
	skip = false\\
	\For{$\mathcal{T} \in \mathcal{L}$}{
		\If{$\mathcal{T} \subset \mathcal{S}$}{
			skip = true\\
			\Break
		}
	}
	\If{skip}{
		\Continue
	}
	
	\If{$\textrm{CONSISTENT}(\mathcal{F}, \mathcal{G}_{0}, \mathcal{G}_{+}, \mathcal{G}_{-})$ \And $\textrm{WEIGHT\_FIXED}(\mathcal{F}, W, \mathcal{G}_{0}, \mathcal{G}_{+}, \mathcal{G}_{-})$}{
		append $\mathcal{F}$ to $\mathcal{L}$\\
	}
}

\KwRet $\mathcal{L}$

\caption{$\textrm{FIXED\_LOCI\_WEIGHTS}(W, \mathcal{G}_{0}, \mathcal{G}_{+}, \mathcal{G}_{-})$}
\end{algorithm}
}

\vskip0.2cm
In order to check whether any given locus in the weighted projective space intersects the ambient space, we first determine the minimal generating sets such that, when each coordinate in a set is non-zero, the Stanley-Reisner conditions are satisfied. We call these sets sectors. In a sense, this finds a minimal list of sets that cover the ambient space. We express the Stanley-Reisner ideal $\mathcal{I}_{SR}$ as a list of lists of integers. For example, if $x_{0}x_{2}x_{3} \in \mathcal{I}_{SR}$, we have a list element $\{0, 2, 3\}$.

Given $\mathcal{I}_{SR}$, we can find the sectors using the following routine. Depending on one's chosen programming language, removing an element from a container mid-loop may invalidate iterators. Thus it may be safer to keep an auxiliary array of the indices of elements to be removed, and then run a separate loop after to remove these elements. \\

{\footnotesize
\begin{algorithm}[H]

$n = \mathcal{I}_{SR}.length$\\
let $\mathcal{S}$ be the power set of $\{0, ..., n-1\}$\\

\tcc{Keep only index sets that contain at least one element \\
     from each list in $\mathcal{I}_{SR}$}
let $\mathcal{T}$ be an empty list\\
\For{$S \in \mathcal{S}$}{
	keep = true\\
	\For{$j \leftarrow 0$ \KwTo $n-1$}{
		\If{$S \cap \mathcal{I}_{SR}[j] == \emptyset$}{
			keep = false\\
			\Break		
		}
	}
	\If{keep}{
		append $S$ to $\mathcal{T}$\\
	}
}

\tcc{Remove any lists that are supersets of others}
\For{$T_{1} \in \mathcal{T}$}{
	\For{$T_{2} \in \mathcal{T}$}{
		\If{$T_{1} \subset T_{2}$}{
			remove $T_{2}$ from $\mathcal{T}$
		}
	}
}

\tcc{Construct the polynomials corresponding to the sectors}
let $\mathcal{U}$ be an empty list\\
\For{$T \in \mathcal{T}$}{
	let $L$ be an empty list\\
	\For{$i \in T$}{
		append $x_{i} - 1$ to $L$
	}
	append $L$ to $\mathcal{U}$\\
}

\KwRet $\mathcal{U}$

\caption{$\textrm{SECTORS}(\mathcal{I}_{SR})$}
\end{algorithm}
}

\vskip0.2cm
The Calabi-Yau hypersurface $X$ is defined by the vanishing of a homogeneous polynomial $P$. In order for $X$ to be invariant under $\sigma$, we must restrict to the subset of moduli space in which $P$ is invariant. This gives us a ``symmetrized'' polynomial $P_{s}$. The requisite steps for creating $P_{s}$ are described in Section \ref{subsec:CY}.

Once the sectors $\mathcal{U}$ and the homogeneous hypersurface polynomial $P_{s}$ have been determined, we can check whether a given fixed locus defined by the vanishing of a list of homogeneous polynomials $\mathcal{F}$ intersects that ambient space, by verifying that it intersects at least one of the sectors. At the same time, we check that the set intersects the hypersurface, defined by $P_{s} = 0$. This is done via the following routine, which also returns the ideal dimensions in the case that the intersection is nonempty, as these will be useful in finding the codimension.\\

{\footnotesize
\begin{algorithm}[H]
\tcc{Check whether or not $\mathcal{F}$ intersects $X$ in at least one sector}
let $D$ be an empty list\\
intersect $\gets$ false\\
\For{$U\in\mathcal{U}$}{
	$\mathcal{I} = \langle U, P_{s}, \mathcal{F} \rangle$\\
	$d = \textrm{dim} \mathcal{I}$\\
	\If{$d \geq 0$}{
		intersect = true\\
	}
	append $d$ to $D$\\
}

\KwRet intersect, $D$

\caption{$\textrm{INTERSECTS\_HYPERSURFACE}(\mathcal{F}, \mathcal{U}, P_{s})$}
\end{algorithm}
}

\vskip0.2cm
Once we have determined that a fixed locus meets the hypersurface, we calculate its codimension using the following routine. Note that the outermost for loop will, at worst, terminate in its final iteration (when $S$ is equal to $\mathcal{F}$). To save calculation time, one can check whether S.length == $\mathcal{F}$.length before the inner \texttt{for} loop, and return the value beforehand if so. We have omitted this check for algorithmic clarity.\\

{\footnotesize
\begin{algorithm}[H]
let $\mathcal{S}$ be the power set of $\mathcal{F}$, as a list\\
sort $\mathcal{S}$ in order of increasing set size\\
\For{$i \leftarrow 0$ \KwTo $\mathcal{S}.length$}{
	same $ \gets $ true\\
	$S = \mathcal{S}[i]$\\	
	
	\For{$j \leftarrow 0$ \KwTo $\mathcal{U}.length$}{
		$\mathcal{I} = \langle \mathcal{U}[j], P_{s}, S, \rangle$\\
		$d_{s} = \textrm{dim} \mathcal{I}$\\
		\If{$d_{s}\neq D[j]$}{
			same = false\\
		}
		\If{same}{
			codim = $9 - 2 * S.length$\\
			\KwRet codim\\
		}
	}
}

\caption{$\textrm{CODIMENSION}(\mathcal{F}, \mathcal{U}, P_{s}, D)$}
\end{algorithm}
}

\vskip0.2cm
The final check is whether the $\sigma$-invariant hypersurface defined by the vanishing of $P_{s}$ is smooth. We do this by checking whether the polynomial and its partial derivatives can all vanishing simultaneously, via the following algorithm.\\

{\footnotesize
\begin{algorithm}[H]
smooth $\gets$ true\\
\For{$U \in \mathcal{U}$}{
	$\mathcal{I}_{smooth} = \langle U, P_{s}, \frac{\partial P_{s}}{\partial x_{1}}, ... , \frac{\partial P_{s}}{\partial x_{k}} \rangle$\\
	\If{$\textrm{dim}(\mathcal{I}_{smooth}) \geq 0$}{
		smooth = false\\
		\Break
	}
}

\KwRet smooth

\caption{$\textrm{SMOOTH}(P_{s}, \mathcal{U})$}
\end{algorithm}
}

\vskip0.2cm
Combining these routines, we can express our full algorithm for determining the fixed-point loci, their codimensions, and whether or not the hypersurface is smooth. The full algorithm is sketched below. The input consists of the involution $\sigma$, the Stanley-Reisner ideal $\mathcal{I}_{SR}$, the hypersurface polynomial $P$ and the weight matrix $W$. The algorithm is given below.\\

{\footnotesize
\begin{algorithm}[H]

$\mathcal{G}_{0}, \mathcal{G}_{+}, \mathcal{G}_{-} = \textrm{INVARIANT\_GENERATORS}(\sigma, W)$\\

$\mathcal{G} = \mathcal{G}_{0} \cup \mathcal{G}_{+} \cup \mathcal{G}_{-}$ \\
$\mathcal{U} = \textrm{SECTORS}(\mathcal{I}_{SR})$\\
let $\mathcal{L}_{\mathcal{F}}, \mathcal{L}_{c}, \mathcal{L}_{s}$ be empty lists\\
let $\mathcal{S}$ be the power set of $G$\\
\For{$\mathcal{F} \in \mathcal{S}$}{
	\If{$\textrm{FIXED\_LOCI\_WEIGHTS}(W, \mathcal{G}_{0}, \mathcal{G}_{+}, \mathcal{G}_{-})$}{
        $X, D \gets \textrm{INTERSECTS\_HYPERSURFACE}(\mathcal{F}, \mathcal{U}, P_{s})$\\	    
	    \If{X}{
            append $\mathcal{F}$ to $\mathcal{L}_{\mathcal{F}}$\\	    
	        append $\textrm{CODIMENSION}(\mathcal{F}, \mathcal{U}, P_{s}, D)$ to $\mathcal{L}_{c}$\\
	        append $\textrm{SMOOTH}(P_{s}, \mathcal{U})$ to $\mathcal{L}_{s}$\\
	    }
		
	}
}
\KwRet $\mathcal{L}_{\mathcal{F}}$, $\mathcal{L}_{c}$, $\mathcal{L}_{s}$

\caption{$\textrm{FIXED\_LOCI}(\sigma, \mathcal{I}_{SR}, P_{s}, W)$}
\end{algorithm}
}

%%%%% Description of database entry
\section{Database Format of Results}

In this appendix we describe the format in which the results of this scan will be stored in the database located at \dburl.  The cohomology of each divisors (Hodge diamond) in defining the Calabi-Yau are also presented.  A thorough description of the general structure and other contents of this website can be found in Section 3 of \cite{Altman:2014bfa}. 

\subsection{Database Fields}

The entry for each involution will contain the following fields:

\begin{itemize}
\item \textbf{Polytope \#}, \textbf{Geometry \#}, \textbf{Triangulation \#}: Identification numbers for the polytope, geometry (within the polytope), and triangulation (within the geometry), inherited from the existing database.
\item \textbf{Involution \#}: An identification number for the involution (within the triangulation)
\item \textbf{h11}, \textbf{h21}: The Hodge numbers $h^{1,1}$ and $h^{2,1}$ of the polytope
\item \textbf{Invol}: The involution, written in terms of its exchanged divisors, in a Mathematica-style list
\item {\bf Geometry-wise Invol}: Whether the involution is Geometry-wise proper involution
%\item \textbf{SR Invol}: Whether the involution keeps the Stanley-Reisner ideal invariant
%\item \textbf{ITens Invol}: Whether the involution keeps the intersection tensor invariant
\item \textbf{Volume Parity}: The parity of the volume form under the involution (0 if the parity is not definite)
%\item \textbf{\# CY Terms}: The number of terms in the hypersurface polynomial
\item \textbf{\# Sym CY Terms}: The number of terms remaining in the symmetrized hypersurface polynomial
%\item \textbf{CY Poly}: A list of the terms, without coefficients, of the hypersurface polynomial
\item \textbf{Sym CY Poly}: A list of the terms, without coefficients, of the symmetrized hypersurface polynomial
\item \textbf{h11+, h11-}: The split Hodge numbers of the orientifold
%\item \textbf{Smooth}: Whether or not the symmetrized hypersurface is smooth
\item \textbf{OPlanes}: A list of the orientifold planes found by the scan. Each orientifold plane will have two fields:
\begin{itemize}
\item \textbf{ODim}: The dimension of the orientifold plane (3, 5, or 7)
\item \textbf{OIdeal}: The (reduced) list of polynomials whose vanishing define the orientifold plane
\end{itemize}
\item {\bf Naive String Vacua}: Whether the orientifold Calabi-Yau admits the naive Type IIB string vacua criteria
\end{itemize}

\subsection{Example Entry}

As an example, we display this information for the example from Section \ref{subsec:nontriv}.

\footnotesize{
\begin{itemize}
\item \textbf{Polytope \#}: 566
\item \textbf{Geometry \#}: 1
\item \textbf{Triangulation \#}: 1
\item \textbf{Involution \#}: 1
\item \textbf{h11}: 4
\item \textbf{h21}: 64
\item \textbf{Invol}: \verb|{D3 -> D6,D6 -> D3,D4 -> D7,D7 -> D4}|
\item {\bf Geometry-wise Invol}: true
%\item \textbf{SR Invol}: true
%\item \textbf{ITens Invol}: true
\item \textbf{Volume Parity}: -1
%\item \textbf{\# CY Terms}: 41
\item \textbf{\# Sym CY Terms}: 48
\item \textbf{Sym CY Poly}: In this case, all of the terms are symmetric, so this is the same as CY Poly. We avoid repeating it for brevity.
\item \textbf{h11+}: 3
\item \textbf{h11-}: 1
%\item \textbf{h21+}: 18
%\item \textbf{h21-}: 16
%\item \textbf{Smooth}: true
\item \textbf{OPlanes}: \begin{verbatim}[
	{ "OIDEAL" : [ "x3*x4-x6*x7" ], "ODIM" : 7 },
	{ "OIDEAL" : [ "x1", "x2", "x5" ], "ODIM" : 3 }
]\end{verbatim}
\item {\bf Naive String Vacua}:  True
\end{itemize}
}

\clearpage
\nocite{*}
\bibliography{ClassifyingCY3}
\bibliographystyle{utphys}

\end{document}